\documentclass[a4paper,UKenglish,cleveref, autoref, thm-restate]{lipics-v2021}
\hideLIPIcs

\usepackage{lowco2}

\usepackage{thmtools}
\usepackage{enumitem}
\usepackage{mnotes}
\newif\ifnotes
\notestrue

\renewcommand\MNOTE[1]{\ifnotes\textcolor{red}{#1}\fi}

\newcommand\THOMAS[1]{\MNOTE{T: #1}}

\newif\ifshortversion
\shortversionfalse

\newif\ifcomposition
\compositionfalse

\newcommand{\compornot}[2]{\ifcomposition #1 \else #2 \fi}

\theoremstyle{plain}

\usepackage[utf8]{inputenc}
\usepackage{amsthm}
\usepackage{amsmath}
\usepackage{amsfonts}
\usepackage{amssymb}
\usepackage{stmaryrd}
\usepackage{xcolor}
\usepackage{bm}
\usepackage{pgf,tikz}
\usepackage{xspace}
\usepackage{float}
\usepackage{ifthen}
\usepackage{mathabx}
\usepackage{centernot}
\usepackage{marginnote}   
\usepackage{ifoddpage}    
\reversemarginpar 
\usetikzlibrary{positioning,automata,shapes,arrows}
\newcommand{\RightMarginNote}[1]{%
\ifshortversion
  \else
  {
  \ifodd\value{page}%
    \normalmarginpar
  \else%
    \reversemarginpar
  \fi
  \marginnote{#1}
  }
\fi
}
\newcommand{\optional}[1]{}

\usepackage{chngcntr}     

\usepackage{mathcommand}
\usepackage[quotation, electronic]{knowledge}

\definecolor{darkblue}{rgb}{.1,0,.75}
\KnowledgeConfigureNotion[darkblue]{notion}
\KnowledgeConfigureNotion{standard notion}

\knowledgeconfigure{diagnose line=true} 

\input{knowledge-tree-algebra.kl}

\usepackage{refcount}

\ExplSyntaxOn
\cs_generate_variant:Nn\str_if_eq:nnTF{xxTF}
\newcommand\plabel[1]{
  \label{#1}
  \IfRefUndefinedExpandable {proof:#1}
    {}
    {\str_if_eq:xxTF
      {\getpagerefnumber{#1}}{\getpagerefnumber{proof:#1}}
        {}
        {\marginpar{\tiny{See~proof~at~page~\pageref{proof:#1}.}}}
     }}
\newenvironment{proofof}[1]{
  \IfRefUndefinedExpandable {#1}
    {\begin{proof}}
    {\str_if_eq:xxTF
      {\getpagerefnumber{#1}}{\getpagerefnumber{proof:#1}}
        {\begin{proof}}
        {\begin{proof}[{Proof~of~\Cref{#1}}]}}
   \AP\label{proof:#1}
  }
  {\end{proof}}
\ExplSyntaxOff

\let\revisiontextcolor\textcolor
\newcommand{\revisionred}{%
  \color{red}%
  \renewcommand{\textcolor}[2]{\revisiontextcolor{red}{##2}}%
}
\newrobustcmd{\redchange}[1]{{\revisionred #1}}

\usetikzlibrary{arrows.meta}
\tikzset{
  singlehead/.style={-{To}},
  doublehead/.style={-{To[sep=.6pt] To}},
  triplehead/.style={-{To[sep=.6pt] To[sep=1pt] To}},
}
\newrobustcmd{\ssmorphism}{\mathrel{\mkern3mu\tikz[baseline=-.55ex]{\draw[-{To[length=3pt,width=3pt]}] (0,0)--(.45,0);}\mkern3mu}}
\newrobustcmd{\sslocsurj}{\mathrel{\mkern3mu\tikz[baseline=-.55ex]{\draw[-{To[length=3pt,width=3pt,sep=.6pt] To[length=3pt,width=3pt]}] (0,0)--(.45,0);}\mkern3mu}}
\newrobustcmd{\ssfolding}{\mathrel{\mkern3mu\tikz[baseline=-.55ex]{\draw[-{To[length=3pt,width=3pt,sep=.2pt] To[length=3pt,width=3pt,sep=.2pt] To[length=3pt,width=3pt]}] (0,0)--(.55,0);}\mkern3mu}}
\knowledgenewrobustcmd{\yieldss}{\cmdkl{\mathrm{ss}}}

\newcommand{\N}{\mathbb{N}}

 \graphicspath{{Pictures/}}

\newcommand\Systems{\kl[\Systems]{\mathrm{Systems}}}
\knowledge\Systems{notion}
\newcommand\Setsystems{\kl[\Setsystems]{\mathrm{Set{-}systems}}}
\knowledge\Setsystems{notion}

\knowledgenewmathcommand\smallprofile{\cmdkl{\mathrm{small{-}profile}}}
\knowledgenewmathcommand\smallrootprofile{\cmdkl{\mathrm{small{-}root{-}profile}}}
\knowledgenewmathcommand\profile{\cmdkl{\mathrm{profile}}}
\knowledgenewmathcommand\rootprofile{\cmdkl{\mathrm{root{-}profile}}}

\knowledgenewcommand{\Vars}{\mathrm{\cmdkl{Vars}}}
\knowledgenewcommand{\rk}{\cmdkl{\mathrm{rk}}}
\knowledgenewcommand{\rksystem}{\cmdkl{\mathrm{rk}}}
\knowledgenewcommand{\hole}{\cmdkl{\square}}
\knowledgenewcommand{\Syst}{\cmdkl{\mathcal{S}}}

\knowledgenewcommand\iss{\cmdkl{\mathrm{iss}}}
\knowledgenewcommand\rss{\cmdkl{\mathrm{rss}}}

\newcommand\V{\kl[\V]{\mathit{V}}}
\knowledge\V{notion}
\knowledgenewcommand\Vroot{\cmdkl{\mathit{Vr}}}
\knowledgenewcommand\Vinit{\cmdkl{\mathit{Vi}}}
\knowledgenewcommand\Label{\cmdkl{\mathit{label}}}
\knowledgenewcommand\Edges{\cmdkl{\mathit{Edges}}}

\knowledgenewcommand\elift[1]{\cmdkl{\widetilde{#1}}}

\knowledgenewcommand\flatten{\cmdkl{\mathrm{flatten}}}
\knowledgenewcommand\sflatten{\cmdkl{\mathrm{flatten}}}

\knowledgenewcommand\syntflatten{\cmdkl{\mathrm{flatten}}}

\knowledgenewcommand\unit{\cmdkl{\mathrm{unit}}}
\let\atomic\unit
\knowledgenewcommand\plant{\cmdkl{\mathrm{plant}}}
\knowledgenewcommand\Plant{\cmdkl{\mathrm{Plant}}}
\knowledgenewcommand\Uproot{\cmdkl{\mathrm{uproot}}}
\knowledgenewcommand\fuproot{\cmdkl{\mathrm{fuproot}}}

\knowledgenewcommand{\yAlg}{\mathcal{\cmdkl{Y}}}

\newcommand{\Alg}{\mathcal{A}}
\newcommand{\YAlg}{\mathcal{Y}}

\knowledgenewcommand{\Yw}{\cmdkl{Y_{\leqslant1}}}

\knowledgenewcommand{\yaleq}{\mathrel{\cmdkl{\sqsubseteq}}}

\knowledgenewcommand{\leqS}{\mathrel{\cmdkl{\leqslant}}}
\knowledgenewcommand{\geqS}{\mathrel{\kl[\leqS]{\geqslant}}}
\knowledgenewcommand{\lS}{\mathrel{\kl[\leqS]{<}}}
\knowledgenewcommand{\gS}{\mathrel{\kl[\leqS]{>}}}

\knowledgenewcommand\lsslift[1]{\cmdkl{\widetilde{#1}}}

\knowledgenewcommand\SyntYield{\mathrm{\cmdkl{Synt}}^{\mathrm{\cmdkl{yield}}}}

\newcommand{\leqSset}{\mathrel{\kl[\leqSset]{\sqsubseteq}}^{\mathrm{\kl[\leqSset]{yield}}}}

\knowledge{\leqSset}{notion}

\knowledgenewcommand\syntclass[1]{\cmdkl{[}#1\cmdkl{]}}
\knowledgenewcommand\synteval{\mathrm{\cmdkl{eval}}^\mathrm{\cmdkl{Synt}}}

\knowledgenewcommand\yAlgSynt{\mathcal{\cmdkl{Y}}^{\mathrm{\cmdkl{Synt}}}}

\knowledgenewcommand\liftin{\mathrel{\cmdkl{\overline\in}}}

\knowledgenewcommand\syntmorph{\mathrm{\cmdkl{synt}}}
\newcommand\setsyntmorph[1]{\kl[\setsyntmorph]{[}#1\kl[\setsyntmorph]{]}^{\kl[\setsyntmorph]{\mathrm{yield}}}}
\knowledge\setsyntmorph{notion}
\knowledgenewcommand\setsyntacc{\cmdkl{F}}

\knowledgenewcommand\Synt{\mathrm{\cmdkl{Syntactic}}}
\newcommand\Setsynt{\mathrm{\kl[\Setsynt]{Syntactic}}^{\kl[\Setsynt]{\mathrm{yield}}}}
\knowledge\Setsynt{notion}

\knowledgenewcommand{\equiS}{\mathrel{\cmdkl{\equiv}}}
\newcommand{\equiSset}{\mathrel{\kl[\equiSset]{\equiv}}^{\mathrm{\kl[\equiSset]{yield}}}}

\knowledgenewcommand\interval[1]{\cmdkl{[}#1\cmdkl{]}}

\knowledgenewcommand\slift[1]{\cmdkl{\overline{#1}}}
\newcommand\slifteval{\mathrm{\kl[\slift]{\overline{eva}l}}}
\knowledgenewcommand\renamerel[1]{\cmdkl{\widehat{#1}}}
\knowledgenewcommand\rename[1]{\cmdkl{\widehat{#1}}}
\knowledgenewcommand\dupname[1]{\cmdkl{\widehat{#1^{-1}}}}
\knowledgenewcommand\Id[1]{\cmdkl{\mathrm{Id_{\interval{#1}}}}}

\knowledgenewcommand\sssum{\mathbin{\cmdkl{+}}}

\let\ssSum\Sssum

\knowledgenewcommand{\LanguageRec}{\cmdkl{\mathrm{Rec}}}

\knowledgenewcommand{\LanguageRes}{\cmdkl{\mathcal L_{\mathrm{res}}}}
\knowledgenewcommand{\LanguageInner}{\cmdkl{\mathcal L_{\mathrm{inner}}}}
\knowledgenewcommand{\LanguageRoot}{\cmdkl{\mathcal L_{\mathrm{root}}}}

\knowledgenewcommand{\Initlanguage}{\cmdkl{\mathrm{InitLanguage}}}
\knowledgenewcommand{\Yields}{\cmdkl{\mathrm{Yields}}}
\knowledgenewcommand{\InitYields}{\cmdkl{\mathrm{InitYields}}}
\knowledgenewcommand{\RootYields}{\cmdkl{\mathrm{RootYields}}}

\newcommand\yieldssubseteq{\mathrel{\kl[\yieldssubseteq]{\subseteq}^{\kl[\yieldssubseteq]{\mathrm{yields}}}}}

\knowledge\yieldssubseteq{notion}
\newcommand\yieldseq{\mathrel{\kl[\yieldseq]{=}^{\kl[\yieldseq]{\mathrm{yields}}}}}
\knowledge\yieldseq{notion}

\knowledgenewcommand{\emptymap}{\cmdkl{\emptyset}}
\knowledgenewcommand{\Resolutions}{\cmdkl{\mathrm{Resolutions}}}
\knowledgenewcommand{\Initresolutions}{\cmdkl{\mathrm{IResolutions}}}
\knowledgenewcommand{\Rootresolutions}{\cmdkl{\mathrm{RResolutions}}}

\knowledgenewcommand{\Directinitresolutions}{\cmdkl{\mathrm{DIResolutions}}}
\knowledgenewcommand{\Directrootresolutions}{\cmdkl{\mathrm{DRRootResolutions}}}

\knowledgenewcommand{\eval}{\mathrm{\textcolor{red}{eval}}}
\knowledgenewcommand{\aeval}{\mathrm{\cmdkl{eval}}}
\knowledgenewcommand{\yaeval}{\mathrm{\cmdkl{eval}}}

\knowledgenewcommand{\Eval}{\mathrm{\cmdkl{Eval}^{\Alg}}}
\knowledgenewcommand{\EvalInit}{\mathrm{\cmdkl{Eval_{init}^{\Alg}}}}
\knowledgenewcommand{\EvalRoot}{\mathrm{\cmdkl{Eval_{root}^\Alg}}}

\knowledgenewcommand{\Acc}{\mathrm{\cmdkl{Acc}}}
\knowledgenewcommand{\Det}{\mathrm{\cmdkl{Det}}}

\knowledgenewcommand{\Context}{\cmdkl{\mathrm{Context}}}

\knowledgenewcommand{\Struct}{\mathrm{\cmdkl{Struct}}}
\knowledgenewcommand{\Dup}{\mathrm{\cmdkl{Dup}}}

\knowledgenewcommand\MRalphabet{\cmdkl{\mathrm{MR}}}

\knowledgenewcommand{\vroot}{\mathit{\cmdkl{root}}}
\knowledgenewcommand{\reset}{\mathit{\cmdkl{reset}}}

\newcommand{\mass}{\scalebox{1.2}{\begin{tikzpicture}
    \draw (0,0) -- (0,.8em);
    
    \draw (-.2em,.6em) -- (0,.7em);
    \draw (-.2em,.5em) -- (0,.6em);
    \draw (-.2em,.4em) -- (0,.5em);
    \draw (-.2em,.3em) -- (0,.4em);
    \draw (-.2em,.2em) -- (0,.3em);
    \draw (-.2em,.1em) -- (0,.2em);
    \draw (-.2em,0em) -- (0,.1em);
    
    \draw (0,.4em) -- (.2em,.4em);
    
\end{tikzpicture}
}
\xspace
}
\newcommand{\initcircle}{
\scalebox{.9}{
\begin{tikzpicture}
    \fill[gray!50, draw=black, line width=.07em] (0,0) circle (.5em);
\end{tikzpicture}
}
}

\newcommand{\meetsquare}{
\scalebox{.9}{
\begin{tikzpicture}
    \draw[fill=gray!50, draw=black, line width=.08em] (0em,0em) -- ++(45:.8em) -- ++(135:.8em) -- ++(225:.8em) -- cycle;
\end{tikzpicture}
}
}

\newrobustcmd\Nats{\mathbb{N}}

\newrobustcmd\alphabet{\kl[\alphabet]{\Sigma}}
\knowledge\alphabet{notion}
\newrobustcmd\alphabetB{\kl[\alphabet]{\Gamma}}

\newrobustcmd\alphabetTS{\kl[\alphabetTS]{\mathbf{Tr}}}
\knowledge\alphabetTS{notion}
\knowledgenewcommand\tsdecode{\cmdkl{\mathrm{decode}}}
\knowledgenewcommand\tsb{\cmdkl{b}}
\newcommand{\T}{\mathcal T}

\newcommand\sinner{{\mathrm{\kl[\sinner]{inner}}}}
\knowledge\sinner{notion}
\newcommand\sroot{{\mathrm{\kl[\sroot]{root}}}}
\knowledge\sroot{notion}

\knowledgenewcommand\alphabetInner{\cmdkl{\Sigma}^{\cmdkl{\mathrm{inner}}}}
\knowledgenewcommand\alphabetRoot{\cmdkl{\Sigma}^{\cmdkl{\mathrm{root}}}}

\newrobustcmd\unfold{\kl[\ unfold]{\mathrm{unfold}}}
\knowledge\unfold{notion}

\newrobustcmd\mto{\mathbin{\kl[\mto]{\to}}}
\knowledge\mto{notion}

\knowledgenewcommand\yameet{\mathbin{\cmdkl{\sqcap}}}
\knowledgenewcommand\yatop{\cmdkl{\top}}

\knowledgenewcommand\yaup{\cmdkl{\Uparrow}}

\knowledgenewcommand\meet{\mathbin{\cmdkl{\sqcap}}}
\DeclareMathOperator\bigmeet{\kl[\yameet]{\bigsqcap}}

\newrobustcmd\powerset{\mathcal{P}}
\knowledge\powerset{notion}

\newrobustcmd\funSystem{\kl[\funSystem]{\mathrm{System}}}
\knowledge\funSystem{notion}

\knowledgenewcommand\rank[1]{\cmdkl{|}#1\cmdkl{|}}
\knowledge\rank{notion}

\knowledgenewcommand\sem[1]{\cmdkl{[\![}#1\cmdkl{]\!]}}
\knowledge\sem{notion}

\knowledgenewcommand\aterms[2]{#1\cmdkl{[}#2\cmdkl{]}}
\knowledge\aterms{notion}

\newcommand{\Pred}{\mathtt{P}}
\newcommand{\aut}{\mathcal{B}}
\newcommand{\D}{\mathcal{D}}
\newcommand{\C}{\mathcal{C}}

\knowledgenewcommand\equivMSO[1][d]{\mathrel{\cmdkl{\equiv_d^{\mathrm{MSO}}}}}

\title{Tree algebras and bisimulation-invariant MSO on finite graphs}

\author{Thomas Colcombet}{CNRS, IRIF, Université Paris Cité, France \and \url{https://www.irif.fr/~colcombe}}{thomas.colcombet@irif.fr}{https://orcid.org/0000-0001-6529-6963}{}

\author{Amina Doumane}{CNRS, LIP, ENS Lyon, France \and \url{http://perso.ens-lyon.fr/amina.doumane}}{amina.doumane@ens-lyon.fr}{}{}
\author{Denis Kuperberg}{CNRS, LIP, ENS Lyon, France \and \url{http://perso.ens-lyon.fr/denis.kuperberg}}{denis.kuperberg@ens-lyon.fr}{https://orcid.org/0000-0001-5406-717X}{ANR ReCiProg}

\authorrunning{T. Colcombet, A. Doumane and D. Kuperberg} 

\Copyright{Thomas Colcombet, Amina Doumane and Denis Kuperberg} 



\category{} 

\relatedversion{https://arxiv.org/abs/2407.12677} 

\nolinenumbers 

\EventEditors{}
\EventNoEds{4}
\EventLongTitle{}
\EventShortTitle{preprint}
\EventAcronym{preprint}
\EventYear{2026}
\EventDate{}
\EventLocation{}
\EventLogo{}
\SeriesVolume{}
\ArticleNo{}

\usepackage{graphicx}
\newcommand{\nocontentsline}[3]{}
\let\origcontentsline\addcontentsline
\newcommand\stoptoc{\let\addcontentsline\nocontentsline}
\newcommand\resumetoc{\let\addcontentsline\origcontentsline}

\usepackage{background}

\begin{document}
\backgroundsetup{placement=center,pages=all,contents={},color=yellow}

\maketitle

\begin{abstract}
We establish that the bisimulation-invariant fragment of MSO over finite transition systems is expressively equivalent in the finite to the modal $\mu$-calculus, a question that had remained open for several decades.

The proof goes by translating the question to an algebraic framework, and showing that the languages of regular trees that are recognised by finitary tree algebras are the regular ones. This corresponds for trees to a weak form of the key translation of Wilke algebras to $\omega$-semigroups over infinite words, and was also a missing piece in the algebraic theory of regular languages of infinite trees for twenty years.
\end{abstract}

\vskip\topmattervskip\baselineskip
\noindent\textcolor{lipicsGray}{\fontsize{9}{12}\sffamily\bfseries
                      \lowcotwo} \ This is a low-CO$_2$ research paper: \lowcotwourl[v1]. This research
was developed, written, submitted and presented without the use of air travel.\\
\noindent Extended version of the conference paper \cite{ColcombetDK25}, published in the proceedings of ICALP 2025.\\
\noindent A Large Language Model was used for finding loopholes in previous versions and coming up with clean fixes.
\noindent The global proof scheme remains the same as in \cite{ColcombetDK25}, but both the presentation and the rigor have been improved.
\noindent A Lean formalization was also performed by an LLM, available at \url{https://laxarchive.org/lax-369799}. It follows the structure of the paper and verifies all claims, including the last remarks on effectivity.

\tableofcontents


\section{Introduction}


A well-known result of Janin and Walukiewicz \cite[Thm 11]{JaninW96} states:
\begin{theorem}[Janin-Walukiewicz]\label{theorem:jw}
	For a property of "transition systems", the following statements are equivalent:
	\begin{itemize}
	\item being "MSO-definable" and "bisimulation-invariant", and
	\item being "$\mu$-calculus-definable"\footnote{In this work, ""$\mu$-calculus"" refers to the standard propositional modal $\mu$-calculus, i.e. the temporal logic constructed from propositions, modalities $\square$ and $\diamond$, boolean connectives, and least and greatest fixpoints.}.
	\end{itemize}
\end{theorem}
A property of "transition systems" is "bisimulation-invariant" if for any two "bisimilar" "transition systems", either both satisfy the property, or neither does.

It is enlightening to view this result in light of the following well-known finite model property: if two "$\mu$-calculus" sentences can be separated by some "transition system", then they can be separated by a finite one. In other words, the semantics of a "$\mu$-calculus" sentence is entirely defined by what it expresses over finite "transition systems". For this reason, the question of whether a version of \Cref{theorem:jw} holds for finite "transition systems" is extremely natural. It remained an important open question in the field for almost three decades.
One contribution of this paper is to provide a positive answer to it:
\begin{theorem}\label{theorem:bisim-invariant-finite}
	For a property of \textbf{finite} "transition systems", the following statements are equivalent:
	\begin{itemize}
	\item being "MSO-definable" and "bisimulation-invariant", and
	\item being "$\mu$-calculus-definable".
	\end{itemize}
\end{theorem}

\begin{example}\label{example:bisim-in-the-finite}
	Let us consider the property of "finite transition systems" having a reachable vertex (from the "initial vertex@@ts" $s_0$) that belongs to a cycle. This property is expressible by the following "MSO" formula:
	\[\varphi=\exists x.\bigl(\mathrm{Path}(s_0,x)\wedge \exists y. \mathrm{Edge}(x,y)\wedge\mathrm{Path}(y,x)\bigr)\]
	where $\mathrm{Path}(x,y)$ is the "MSO" formula\footnote{This is expressible by the following formula, stating that any set $X$ containing $x$ and closed under the edge relation must contain $y$: $\mathrm{Path}(x,y):= \forall X.((x\in X\wedge \forall z,t.((z\in X\wedge \mathrm{Edge}(z,t))\rightarrow t\in X))\rightarrow y\in X)$.} that expresses that there is a (possibly empty) path from vertex $x$ to vertex $y$.
	If the same formula is interpreted over arbitrary "transition systems", the resulting property is not "bisimulation-invariant", as witnessed by the following two "bisimilar" "transition systems":
	\begin{center}
	{
\begingroup%
  \makeatletter%
  \providecommand\color[2][]{%
    \errmessage{(Inkscape) Color is used for the text in Inkscape, but the package 'color.sty' is not loaded}%
    \renewcommand\color[2][]{}%
  }%
  \providecommand\transparent[1]{%
    \errmessage{(Inkscape) Transparency is used (non-zero) for the text in Inkscape, but the package 'transparent.sty' is not loaded}%
    \renewcommand\transparent[1]{}%
  }%
  \providecommand\rotatebox[2]{#2}%
  \newcommand*\fsize{\dimexpr\f@size pt\relax}%
  \newcommand*\lineheight[1]{\fontsize{\fsize}{#1\fsize}\selectfont}%
  \ifx\svgwidth\undefined%
    \setlength{\unitlength}{250.18741157bp}%
    \ifx\svgscale\undefined%
      \relax%
    \else%
      \setlength{\unitlength}{\unitlength * \real{\svgscale}}%
    \fi%
  \else%
    \setlength{\unitlength}{\svgwidth}%
  \fi%
  \global\let\svgwidth\undefined%
  \global\let\svgscale\undefined%
  \makeatother%
  \begin{picture}(1,0.1938808)%
    \lineheight{1}%
    \setlength\tabcolsep{0pt}%
    \put(0,0){\includegraphics[width=\unitlength,page=1]{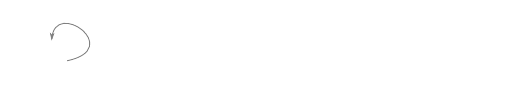}}%
    \put(-0.00298604,0.15941774){\color[rgb]{0,0,0}\makebox(0,0)[lt]{\lineheight{1.25}\smash{\begin{tabular}[t]{l}$T_1$\end{tabular}}}}%
    \put(0,0){\includegraphics[width=\unitlength,page=2]{not_bisim_inf.pdf}}%
    \put(0.08424461,0.067045){\color[rgb]{0,0,0}\makebox(0,0)[lt]{\lineheight{1.25}\smash{\begin{tabular}[t]{l}$a$\end{tabular}}}}%
    \put(0.35281739,0.1665497){\color[rgb]{0,0,0}\makebox(0,0)[lt]{\lineheight{1.25}\smash{\begin{tabular}[t]{l}$T_2$\end{tabular}}}}%
    \put(0,0){\includegraphics[width=\unitlength,page=3]{not_bisim_inf.pdf}}%
    \put(0.43454786,0.0643118){\color[rgb]{0,0,0}\makebox(0,0)[lt]{\lineheight{1.25}\smash{\begin{tabular}[t]{l}$a$\end{tabular}}}}%
    \put(0,0){\includegraphics[width=\unitlength,page=4]{not_bisim_inf.pdf}}%
    \put(0.0869465,0.00848385){\color[rgb]{0,0,0}\makebox(0,0)[lt]{\lineheight{1.25}\smash{\begin{tabular}[t]{l}$s_0$\end{tabular}}}}%
    \put(0.43468608,0.00848385){\color[rgb]{0,0,0}\makebox(0,0)[lt]{\lineheight{1.25}\smash{\begin{tabular}[t]{l}$s_0$\end{tabular}}}}%
    \put(0,0){\includegraphics[width=\unitlength,page=5]{not_bisim_inf.pdf}}%
    \put(0.60242217,0.0643118){\color[rgb]{0,0,0}\makebox(0,0)[lt]{\lineheight{1.25}\smash{\begin{tabular}[t]{l}$a$\end{tabular}}}}%
    \put(0,0){\includegraphics[width=\unitlength,page=6]{not_bisim_inf.pdf}}%
    \put(0.60256039,0.00848385){\color[rgb]{0,0,0}\makebox(0,0)[lt]{\lineheight{1.25}\smash{\begin{tabular}[t]{l}$s_1$\end{tabular}}}}%
    \put(0,0){\includegraphics[width=\unitlength,page=7]{not_bisim_inf.pdf}}%
    \put(0.77029645,0.0643118){\color[rgb]{0,0,0}\makebox(0,0)[lt]{\lineheight{1.25}\smash{\begin{tabular}[t]{l}$a$\end{tabular}}}}%
    \put(0,0){\includegraphics[width=\unitlength,page=8]{not_bisim_inf.pdf}}%
    \put(0.77043467,0.00848385){\color[rgb]{0,0,0}\makebox(0,0)[lt]{\lineheight{1.25}\smash{\begin{tabular}[t]{l}$s_2$\end{tabular}}}}%
    \put(0,0){\includegraphics[width=\unitlength,page=9]{not_bisim_inf.pdf}}%
  \end{picture}%
\endgroup%
}
	\end{center}
	Indeed, the "system@transition system" $T_1$ satisfies $\varphi$, while its "unfolding@full unfolding" $T_2$ does not.
	However, this property is "bisimulation-invariant" over finite "transition systems", where it amounts to the existence of an infinite path from $s_0$, and the latter property is expressible in $\mu$-calculus as $\psi=\nu X.\diamond(X)$.
\end{example}

\medskip

For both \Cref{theorem:jw} and \Cref{theorem:bisim-invariant-finite}, the direction translating "$\mu$-calculus" to "MSO" is the same, and is the easy one. It follows from two well-known facts: (1) "$\mu$-calculus" sentences can be effectively translated into equivalent "MSO-sentences", and (2) properties definable in "$\mu$-calculus" are invariant under "bisimulation".

The difficult direction is to show that "bisimulation-invariant" "MSO-definable" properties can be translated into equivalent "$\mu$-calculus sentences".
In the classical result of Janin and Walukiewicz, the crucial remark is that for every "transition system" there exists a (potentially infinite) "bisimilar" one which is a tree.
Thus, the result amounts to showing that if an "MSO-definable" property of (potentially infinite) trees is "bisimulation-invariant", then it is "$\mu$-calculus" definable. This explains why the inner machinery of their proof relies, in the spirit of the seminal work of Rabin for binary trees~\cite{Rabin68}, on a theory of automata over infinite trees (with unrestricted branching) that correspond in expressiveness to $\mu$-calculus.

This approach cannot be used in our case since there exist finite "transition systems" that are not "bisimilar" to any finite trees (this is the case of the finite "transition system" depicted in \Cref{example:bisim-in-the-finite}).
That is problematic since in \Cref{theorem:bisim-invariant-finite} we are given an MSO-sentence which is "bisimulation-invariant" over finite "transition systems" only, and thus, given a finite "transition system", there is in general no equivalent finite tree on which we could try to analyse the semantics of the MSO-sentence.

We adopt a different strategy. Our proof of the downward implication in \Cref{theorem:bisim-invariant-finite} uses a translation of the problem into an algebraic formalism. From now on, all "transition systems" are assumed to be finite.

The first step is to extend the notion of "transition system" in order to be able to compose them in a sufficiently powerful way.
We use generalised "transition systems", called "systems". Our "systems" are made of ranked letters instead of transitions, and have a finite number of identified variables (that we can understand as exit vertices). Crucially, this means that we introduce more structure into our "systems": the number of outgoing edges from a node is fixed by the label of the node, and these outgoing edges are now ordered. Such "systems" should be understood as pieces that can be plugged in a context. A "system"~$S$ with two free variables and a "context"~$C[\hole_2]$ as well as their composition are illustrated below\footnote{Transitions are always oriented from left to right unless a marking specifies otherwise.}.
\begin{center}
\scalebox{.7}{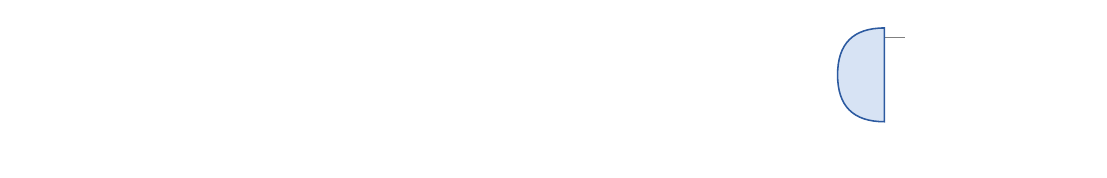}
\end{center}

This approach gives rise to a notion of a multi-sorted "algebra", in which we are considering, for each finite set of variables~$X=\{x_1,\dots,x_k\}$ the set of generalised "transition systems" that use variables in~$X$.
Up to now, this is classical and has been explored in various ways~\cite{BojI09,Blum13}. 
The novelty here is to extend this "algebra" by a kind of powerset construction, where union is interpreted conjunctively. For technical reasons, we will also define two sets of starting vertices: "initial vertices" which are used for composing with other elements, and roots that act as a global starting point in any context.
We then equip these "extended algebras@set-algebras", called "set-algebras", with the usual composition operations, associativity rules, Myhill-Nerode-like equivalence relation, etc.

If a language of "transition systems" is defined by an MSO-sentence, then (by induction on formulas), its "syntactic algebra" is "rankwise-finite", i.e. the sort corresponding to each finite set of variables $X$ is finite.
We then show that its "syntactic set-algebra" is also "rankwise-finite". Finite minimality and upward closure of the "accepting set" allow labels in an accepted "system" to be replaced by "extremal elements". The additional "automaton property" states that extremal elements of positive rank split into extremal unary directions, and that acceptance of the resulting unary/nullary "set-systems" is determined by acceptance along paths.
This in turn allows us to design an automaton model equivalent to the original formula. If the language we started with is "bisimulation-invariant", then this can be seen in the transition structure of the automaton, from which we can finally translate back to the $\mu$-calculus following arguments similar to~\cite{JaninW96}.

Moreover, we can show that this transformation is effective, i.e. from an MSO-formula that is assumed to be "bisimulation-invariant" on finite structures (though this property itself is not decidable), we can compute the wanted $\mu$-calculus formula.
%


%
%

\subsection*{Related work}

\subparagraph*{Bisimulation-invariant logics}
The seminal Hennessy-Milner theorem~\cite{HennessyM80} gives a first characterisation of modal logic via bisimulation-equivalence.

Van Benthem generalised it to first-order logic (FO) in~\cite{VanBenthem84}. The result was then transferred to finite structures~\cite{Rosen97}, and specialised to particular classes of finite structures~\cite{DawarO09,Hansen09}.

In~\cite{MollerR99}, the equivalence between the "bisimulation-invariant" fragment of monadic path logic with CTL* is shown.

An intermediate logic strictly between FO and "MSO", obtained by extending FO with fixed-point operators, is considered in~\cite{PfluegerMK24}, where it is shown that the two-way "bisimulation-invariant" fragment of this logic is equivalent to two-way $\mu$-calculus on finite structures.

Concerning the logic "MSO", as already explained, Janin and Walukiewicz prove the equivalence with $\mu$-calculus on general structures~\cite{JaninW96}. The case of weak "monadic second-order logic" was shown to correspond to continuous $\mu$-calculus in~\cite{CarreiroFVZ14}. On finite structures, this equivalence is shown for structures of bounded Cantor-Bendixson rank in~\cite{BlumWolf20}.
%
%
%
%
%
%

\subparagraph*{\compornot{Composition method and tree algebra}{Tree algebra}}

\compornot{The composition method that we use here was pioneered in~\cite{FefermanV59} for FO and used intensively for "MSO" in~\cite{Shelah75, Rabinovich07}.}{}

There is a long line of research pursuing a well-behaved notion of algebra for regular languages of infinite trees~\cite{Wilke93,Niwinski97,Aczel01,BojanczykW08,Blum11}.

Two-sorted algebras for "regular trees@regular-tree" are defined in~\cite{BojI09}, where the two sorts are closed trees and trees with $1$ output, with an MSO-definable composition operation.
The formalism of $\omega$-hyperclones for tree algebras is proposed in~\cite{Blum13}, and is later shown in~\cite{BojKlin19} to be somewhat ill-behaved with respect to regularity: there are examples of "rankwise-finite" algebras that define non-regular languages of infinite trees. Similarly,~\cite{Blum23} shows that no distributive law exists in this framework. This is due to the fact that these algebras must treat non-regular trees, in the same way that $\omega$-semigroups must be able to evaluate words that are not ultimately periodic.


\subsection*{Structure of the document}


In \Cref{sec:defs} we provide definitions and properties of the objects we will manipulate. We describe in particular "systems" and "set-systems" as well as "morphisms@@ss" between them, and other important tools for the development of this work.
In \Cref{section:algebras}, we define "algebras", and describe the translation from "MSO-definable" to "rankwise-finite" "algebras".
In \Cref{section:set-algebras}, we define "set-algebras", and prove the key result of the section which is that the "syntactic set-algebra" of a language of finite "systems" "recognisable" by a "rankwise-finite" "algebra" is also "rankwise-finite".
In \Cref{sec:core}, we establish extremal replacements for "rankwise-finite" "set-algebras" and prove the "automaton property" for "syntactic set-algebras".
In \Cref{section:bisim-invariant}, we show how this property indeed allows us to prove that any language recognised by a "rankwise-finite" "algebra" is regular in the classical sense. We also adapt this to the "bisimulation-invariant" setting to obtain an equivalent $\mu$-calculus formula.

The document is better read in electronic form since occurrences of notions are hyperlinked to their definitions.


\section{Set-systems and systems}\label{sec:defs}

The goal of this section is to equip finite graphs with additional structure, allowing for an algebraic treatment. We define "set-systems" (containing non-deterministic choices) and "systems" (with no such choices), their "morphisms@@ss", "unfoldings" and "regular trees@regular-tree" as equivalence classes of the "unfolding" relation. We further define "yields" that turn a "set-system" into a "system" by resolving non-deterministic choices. We finally study their compositional properties via "flattening@flattening" and behaviours with respect to "contexts". 

\subsection{Intuitions}
\label{subsection:intuitions}


A ""ranked set""~$\intro*\alphabet$ (or \reintro*"alphabet") is a family of pairwise disjoint sets~$\alphabet_n$ indexed by natural numbers. We write $a\in\alphabet$ for $a\in \alphabet_n$ for some~$n$.
For $a\in \alphabet$, the ""rank"" of $a$, denoted $\intro*\rk(a)$, is the integer $n$ such that $a\in \alphabet_n$. 
\AP A ""map of ranked sets""~$f$ of $\Sigma$ to~$\Gamma$ is a map that preserves the ranks, i.e.\ a family of maps $f_n\colon\Sigma_n\to\Gamma_n$ for all natural numbers~$n$. 
\AP A "ranked set"~$\alphabet$ is ""rankwise-finite"" if~$\alphabet_n$ is finite for all~$n$.
\begin{remark}
	As defined here, "ranked sets" 
	equipped with their "maps@@rs" form a category, which we call the ""category of ranked sets"". 
\end{remark}

\AP We fix an infinite set $\intro*\Vars=\{x_1,x_2,\dots\}$ of ""variables"" ("variables" can be understood to some extent as letters of "rank"~$0$).
For $n\in\Nats$, we denote $\intro*\interval n$ the set~$\{1,\dots,n\}$, and $\reintro*\Vars[n]$ the variables~$\{x_1,\dots,x_n\}$. We write $\intro*\Id n$ for the identity map on~$\interval n$.

\begin{figure}[ht]
	\centering{
	\def\svgwidth{\linewidth}}
	\scalebox{.7}{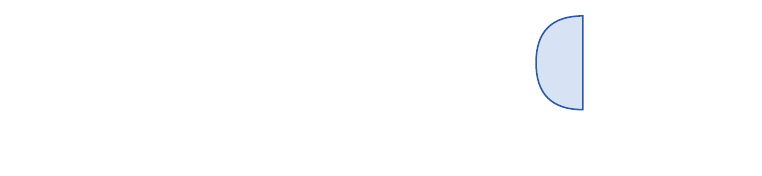}
	\caption{
	 A "$\alphabet$-set-system" $S_1$ and a "$\alphabet$-system"~$S_2$, both over "variables" $\{x_1,x_2\}$ (i.e.\ $S_1$ and $S_2$ are of rank $2$), for $\alphabet$ containing symbols $b_1$ of "rank" $1$, $a_2$ of "rank" $2$, and $c_3$ of "rank" $3$. The topmost outgoing edge of a symbol has "direction"~$1$, the next one~$2$, and so on. The circle $\initcircle$ denotes an "initial vertex", the symbol $\mass$ denotes a "root vertex", and $\meetsquare$ emphasizes the presence of multiple successors in the same direction. Implicitly, edges are directed from left to right unless explicitly using an arrow notation.}\label{fig:system}
\end{figure}

\AP The central notion here is that of "$\alphabet$-set-system". It is illustrated in \Cref{fig:system}.  This is a more general version of the usual notion of "system", which itself abstracts the notion of "transition system".
\begin{itemize}
\item At a high level, "set-systems" consist of finite sets of vertices, to each vertex is attached a symbol of some rank~$k$, and the vertex is then connected, in each of the $k$ directions, to a set of successors. The complete definition also has "initial vertices", "root vertices", and variables. See \Cref{subsection:system-definition}.
\item Note that the `non-determinism' of successors in "set-systems" should not be confused with the possible successor choice in a "transition system" (we shall see that this non-determinism is somehow controlled by an adversarial player).
\item Let us also emphasize that "[set-]systems" are finite. This is a key feature in this work since our ultimate goal is to understand the expressive power of "MSO" over finite "transition systems".
\item "Systems" form the subclass of "set-systems" that are deterministic: there is only one "initial vertex", no "root vertex", there is exactly one successor in each possible direction, and every vertex and every variable is reachable from the "initial vertex". See \kCref{system}.
\item One way to see "systems" is as representing infinite "regular trees@regular infinite tree" over a "ranked alphabet" in a "folded" way. The represented tree is obtained via the usual notion of unravelling. We shall never perform such a complete unravelling since our "systems" have to remain finite. Instead, "systems" are considered modulo (partial) "unfolding", which corresponds to the notion of "morphism@@s" (in a reverse way), and "regular trees@regular infinite tree" correspond to an equivalence class of "systems" under the generated "unfold-equivalence@@ss".
\item "Set-systems" in their general form are of a non-deterministic nature. The intuition here is similar to the case of "systems", but instead of `representing' a unique infinite tree, it would represent a set of "regular trees@regular-tree" called its "yield": the set of "regular trees@regular-tree" that could be obtained by resolving the non-determinism (either from the "initial vertices" or the "root vertices"). See \kCref{yield}.
\item "Set-systems" can also be considered modulo some form of unravelling, also called "unfold-equivalence@@ss". See \kCref(ss){unfold-equivalence}. This equivalence (which is a form of "bisimulation") has the property that two "unfold-equivalent@@ss" "set-systems" have the same "yields" (however, the converse is not true in general).
  \item Another orthogonal aspect of these objects is that "[set-]systems" use "variables" and "initial vertices". These are used as handles for composing several "[set-]systems" into more complex ones. When `composing', the general idea is to connect a variable (which is a descriptor of a possible `exit' of the "system") to an "initial vertex" (which is describing `entry points' of the "system"). This ability to combine "systems" together equips "systems" with an algebraic structure.
  This is formalised using the $\flatten$ operation. See \Cref{subsec:flatten}.
\end{itemize}

\subsection{Set-systems and systems}
\label{subsection:system-definition}


In this section, we define the notions of "systems" (the natural model), and of "set-systems" (which generalise "systems" with a form of non-determinism and a notion of root).

\begin{definition}[set-systems]\AP\label{definition:set-systems}
	\AP Given a "ranked set"~$\alphabet$, a ""$\alphabet$-set-system"" of ""rank@@ss""~$n$ is a tuple $S:=(\V,\intro*\Vinit,\intro*\Vroot,\Label,\Edges)$, composed of
	\begin{itemize}
	\item a finite set of ""vertices"" $\intro*\V$,
	\item a set of ""initial vertices"" $\Vinit\subseteq\V$,
	\item a set of ""root vertices"" $\Vroot\subseteq\V$,
	\item a ""labelling function"" $\intro*\Label: \V\to \alphabet$; in practice, we simply write $S(s)$ for $\Label(s)$,
	\item an ""edge relation"" $\intro*\Edges\subseteq \V\times\N\times (\V\uplus\{x_1,\dots,x_n\})$ consisting of ""edges"" of the form $(v,d,f)$ with $d\in\interval{\rk(S(v))}$; $d$ is called the ""direction"" of the "edge"; if $f$ is a "vertex", then $(v,d,f)$ is called a ""transition edge""; otherwise $f$ is a "variable" and $(v,d,f)$ is called a ""variable edge"". We also denote $\reintro*\Edges(v)$ the set~$\{(d,f)\mid (v,d,f)\in\Edges\}$, and $\reintro*\Edges(v,d)=\{f\mid (v,d,f)\in\Edges\}$.
	\end{itemize}

	\AP
	Given a "set-system"~$S$, we tacitly name its components $(\reintro*\Vinit_S,\reintro*\Vroot_S,\reintro*\Label_S,\reintro*\Edges_S)$.
	\AP

	\AP A "$\alphabet$-set-system" is ""closed@@system""\phantomintro(ss){closed} if it has "rank@@ss"~$0$.
	\AP Two "$\alphabet$-set-systems"~$S,S'$ are of the ""same shape@@ss"" if they differ only on their "labelling", i.e.\ $\V_S=\V_{S'}$, $\Vinit_S=\Vinit_{S'}$, $\Vroot_S=\Vroot_{S'}$, and $\Edges_S=\Edges_{S'}$.

	\AP Given a "map of ranked sets"~$\eta$ from~$\Sigma$ to~$\Gamma$, we denote $\intro*\slift\eta$ the "map@@rs" from "$\alphabet$-set-systems" to "$\Gamma$-set-systems" that sends $S$ to~$\slift\eta(S)=(\V_S,\Vinit_S,\Vroot_S,\eta\circ \Label_S,\Edges_S)$.
\end{definition}
We may drop the mention of the alphabet $\alphabet$, and simply talk about "set-systems".
The variables used in a "[set-]system" of rank~$n$ are $x_1,\dots,x_n$. We sometimes use other variable names such as $x,y,z$ for convenience if there is no ambiguity about the meaning.

\begin{definition}[systems]\AP\label{definition:systems}
	A ""$\alphabet$-system""~$T$ of "rank@@ss"~$n$ is a "$\alphabet$-set-system" such that $\Vinit_T$ is a singleton, $\Vroot_T$ is empty, for all~"vertices"~$v\in\V_T$ and all "directions" $d\in\interval{\rk(T(v))}$, $\Edges_T(v,d)$ is a singleton, and all "vertices" and all "variables"~$x_1,\dots,x_n$ are reachable from the unique "initial vertex" by a path of "edges".
\end{definition}

\begin{remark}[category of set-systems and systems]\AP\label{remark:system-functors}
	For $\alphabet$ fixed, the "$\alphabet$-set-systems" form a "ranked set", denoted $\intro*\Setsystems(\alphabet)$. Furthermore, the lifting of "maps of ranked sets", $f\mapsto \slift{f}$ turns $\Setsystems$ into an endofunctor  of the category of "ranked sets", called the ""set-system functor"".
	The "ranked set"~$\intro*\Systems(\alphabet)$ is defined in the same way. As above, $\Systems$ is an endofunctor of the category of "ranked sets", and more precisely a subfunctor of $\Setsystems$, called the ""system functor"".
\end{remark}

We conclude this section with the introduction of some operations and notations useful for the technical development of this work.
\begin{itemize}
\itemAP The "$\alphabet$-set-system"~$\intro*\Uproot(S)$ is $(\V_S,\Vroot_S,\emptyset,\Label_S,\Edges_S)$. It consists in forgetting the "initial vertices", and transforming "root vertices" into "initial ones@initial vertex". This definition shall be used for describing the semantics of "root vertices" (namely "root yields").
\itemAP The "$\alphabet$-set-system"~$\intro*\plant(S)$ is $(\V_S,\emptyset,\Vroot_S\cup\Vinit_S,\Label_S,\Edges_S)$.	If we interpret "initial vertices" as the cuts of branches of a tree, then planting amounts to putting them in the ground, thus turning them into roots. We shall also apply $\plant$ to sets of "set-systems", with the expected meaning. We denote this pointwise operation by $\intro*\Plant(L):=\{\plant(S)\mid S\in L\}$.
\itemAP Given two "$\alphabet$-set-systems" of same "rank@@ss"~$S$ and~$S'$, then $S \intro*\sssum S'$ is their disjoint union (defined in the natural way).
\itemAP Given a binary relation~$R\subseteq\interval m\times\interval n$, and a "set-system"~$S$ of "rank"~$m$, we denote $\intro*\renamerel R(S)$ the "set-system" which is identical to~$S$, but for the "variable edges" that are defined by $(r,i,x_j)\in \Edges_{\renamerel R(S)}$ if $(k,j)\in R$ and $(r,i,x_k)\in \Edges_S$ for some~$k\in\interval m$.
	As a particular case, for~$\sigma\colon\interval m\to\interval n$, $\intro*\rename\sigma$ turns "set-systems" of "rank@@ss"~$m$ into "set-systems" of "rank@@ss"~$n$, and $\intro*\dupname\sigma$ turns "set-systems" of "rank@@ss"~$n$ into "set-systems" of "rank@@ss"~$m$. We use $\rename\sigma$ as an operation on "systems" only when $\sigma$ is surjective, so every target "variable" is used; $\dupname\sigma$ does not preserve "systems" in general.
\itemAP For the sake of writing examples, we shall sometimes denote "set-systems" as expressions. The main notation is that, given a symbol~$a\in\alphabet$ of "rank"~$n$, and $S_1,\dots,S_n$ that are either "set-systems" of "rank@@ss"~$m$ or variables in~$\Vars[m]$, $a(S_1,\dots,S_n)$ is the "set-system"  obtained as expected: it has a single new "vertex" which is "initial@@vertex" and "labelled@@ss"~$a$, which is connected in "direction"~$d$ to either a "variable"~$y$ if~$S_d=y$, or to all the "initial vertices" of a fresh copy of $S_d$ otherwise ("initial vertices" in~$S_d$ are not initial anymore in the copy). This notation is consistent with what one expects for finite trees: for instance $a_2(x_1,a_2(b,x_2))$ denotes the expected "system" of tree shape.
\end{itemize}

\subsection{Morphisms of set-systems}
\label{subsection:morphisms}

We define "morphisms@@ss", "locally surjective@@ss" "morphisms@@ss", and "foldings" (\Cref{definition:morphism}), establish their closure under composition (\Cref{lemma:morphism-composition}) and compatibility with uprooting and planting (\Cref{statement:uproot-morphism}), and construct pullbacks (\Cref{lemma:pullback}).

The "morphisms@@ss" will be central in the development of this work.
Other notions such as "unfoldings", "unfold-equivalence", and "yields" derive from them.  In fact, we somehow already implicitly used them in the previous section since operations such as the disjoint union are implicitly defined ``up to isomorphism''.

\begin{definition}\AP\label{definition:morphism}
	\AP Given two "$\alphabet$-set-systems" $S$ and $S'$ of same "rank@@ss",  a ""morphism@@ss"" from $S$ to~$S'$ is a map $\eta\colon \V_S\to\V_{S'}$ such that
	\begin{itemize}
	\item $\eta(\Vinit_S)\subseteq \Vinit_{S'}$,
	\item $\eta(\Vroot_S)\subseteq \Vroot_{S'}$, 
	\item $S'(\eta(v))=S(v)$ for all~$v\in \V_S$, and
	\item $\elift\eta(\Edges_S)\subseteq \Edges_{S'}$ where $\intro*\elift\eta$ is the extension to "edges" of~$\eta$, i.e.\ $\elift\eta(v,d,v')=(\eta(v),d,\eta(v'))$, if $v'\in V$, and $\elift\eta(v,d,y)=(\eta(v),d,y)$ for $y\in\Vars$.
	\end{itemize}
	\AP It is ""locally surjective@@ss"" if $\eta(\Vinit_S)=\Vinit_{S'}$, $\eta(\Vroot_S)=\Vroot_{S'}$, and $\elift\eta(\Edges_S(v))=\Edges_{S'}(\eta(v))$ for all~$v\in \V_S$.
	\AP A ""folding@@ss"" is a "locally surjective@@ss" "morphism@@ss" that is also surjective on "vertices", i.e.\ $\eta(\V_S)=\V_{S'}$.
\end{definition}

We use the following arrow notations throughout the document:
\[
  \begin{aligned}
    S\ssmorphism S' &\quad\text{for a "morphism@@ss"},\\
    S\sslocsurj S' &\quad\text{for a "locally surjective@@ss" "morphism@@ss"},\\
    S\ssfolding S' &\quad\text{for a "folding@@ss"}.
  \end{aligned}
\]

A first consistency check in the definition of "morphisms@@ss" is their closure under composition.
\begin{lemma}[composition]\label{lemma:morphism-composition}
	If~$\eta$ is a "morphism of set-systems" from~$S$ to~$S'$, and $\eta'$ is a "morphism of set-systems" from~$S'$ to $S''$, then~$\eta'\circ\eta$ is a "morphism of set-systems" from~$S$ to~$S''$. If~$\eta$ and~$\eta'$ are "locally surjective@@ss" (resp. "foldings@@ss"), then~$\eta'\circ\eta$ is also "locally surjective@@ss" (resp. a "folding@@ss").
\end{lemma}

\begin{remark}\label{remark:system-morphisms}
Every "morphism@@ss" between "systems" is a "folding@@ss": it is "locally surjective@@ss" because successors are unique, and it is surjective on "vertices" because every target "vertex" is reachable from the unique "initial vertex".
\end{remark}

\begin{lemma}[system selection]\label{lemma:system-selection}
Let $\eta\colon S\sslocsurj T$ be a "locally surjective@@ss" "morphism@@ss" from a "set-system" to a "system". There is a "system"~$U$ with an inclusion "morphism@@ss" $\iota\colon U\ssmorphism S$ such that $\eta\circ\iota\colon U\ssfolding T$ is a "folding@@ss".
\end{lemma}
\begin{proof}
Local surjectivity makes $\Vinit_S$ nonempty, $\Vroot_S$ empty, and every set of successors in a permitted "direction" nonempty. Choose one "initial vertex"~$p$ and one outgoing "edge" per permitted "direction" at each "vertex" of~$S$, and restrict to the "vertices" reachable from~$p$, retaining the rank and labels of~$S$. The resulting~$U$ has a unique "initial vertex", no roots, and exactly one successor per permitted "direction". Every path from the "initial vertex" of~$T$ lifts to~$U$, since each chosen "edge" maps to the unique edge of~$T$ in that "direction". As "morphisms@@ss" fix "variables", and all "variables" are reachable in~$U$ via the lifted paths, so~$U$ is a "system". The inclusion~$\iota$ of $U$ into $S$ is a "morphism@@ss", and $\eta\circ\iota$ is a "folding@@ss" by \Cref{remark:system-morphisms}.
\end{proof}

\begin{remark}\label{remark:unfold-systems}
By \Cref{lemma:system-selection}, an "unfolding" of a "system" can be replaced by a "system" "unfolding" of the same target. In the sequel, we shall choose "system" witnesses whenever needed. This can also be done for common "unfoldings" of "systems" (\Cref{lemma:common-system-unfolding}).
\end{remark}

In many proofs we shall concentrate on "initial vertices" only rather than "root vertices".
The following remark explains why many arguments involving "initial vertices" directly transfer to "root vertices".
\begin{remark}\label{statement:uproot-morphism}
	If $\eta$ is a "morphism@@ss" (resp. a "locally surjective@@ss" "morphism@@ss", resp. a "folding@@ss") from~$S$ to $S'$, then it is also a "morphism@@ss" (resp. a "locally surjective@@ss" "morphism@@ss", resp. a "folding@@ss") from~$\Uproot(S)$ to~$\Uproot(S')$ and from~$\plant(S)$ to~$\plant(S')$.
\end{remark}
An important property of "morphisms@@ss" is that they admit pullbacks, as shown now.
\begin{lemma}[pullbacks]\label{lemma:pullback}
	Let $S,S',T$ be "$\alphabet$-set-systems", and $\eta\colon S\ssmorphism T$, $\eta'\colon S'\ssmorphism T$ be "morphisms@@ss". There exist a "$\alphabet$-set-system"~$P$ and  "morphisms@@ss" $\pi\colon P\ssmorphism S$ and $\pi'\colon P\ssmorphism S'$ such that $\eta\circ\pi = \eta'\circ\pi'$, and $P$ is universal with this property, i.e., for all "$\alphabet$-set-systems"~$U$ and "morphisms@@ss" $\tau\colon U\ssmorphism S$ and $\tau'\colon U\ssmorphism S'$, if $\eta\circ\tau = \eta'\circ\tau'$, then there exists a unique "morphism@@ss" $\theta\colon U\ssmorphism P$ such that $\tau = \pi\circ\theta$ and $\tau'=\pi'\circ\theta$. 
	\begin{center}
	\begin{tikzpicture}
		\node (P) at (0,0) {$P$};
		\node (S) at (-2,-1) {$S$};
		\node (S') at (2,-1) {$S'$};
		\node (T) at (0,-2) {$T$};
		\draw[singlehead] (P) -- (S) node[midway,above left]{$\pi$};
		\draw[singlehead] (P) -- (S') node[midway,above right]{$\pi'$};
		\draw[singlehead] (S) -- (T) node[midway,below]{$\eta$};
		\draw[singlehead] (S') -- (T) node[midway,below]{$\eta'$};
	\end{tikzpicture}
	\end{center}	
	Furthermore, if~$\eta$ (resp. $\eta'$) is "locally surjective@@ss", then $\pi'$ (resp. $\pi$) is "locally surjective@@ss". If~$\eta$ (resp. $\eta'$) is a "folding@@ss", then $\pi'$ (resp. $\pi$) is a "folding@@ss".
\end{lemma}

\begin{proof}
	Let~$S,S',T$ be "$\alphabet$-set-systems" and $\eta\colon S\ssmorphism T$, $\eta'\colon S'\ssmorphism T$ be "morphisms@@ss".
	We construct the new "$\alphabet$-set-system" $P = (\V_P,\Vinit_P,\Vroot_P,\Label_P,\Edges_P)$ as follows:
	\begin{itemize}
	\item $\V_P := \{ (v,v')\in \V_S\times \V_{S'}\mid \eta (v)=\eta'(v')\}$,
	\item $\Vinit_P := \{ (v,v')\in \Vinit_S\times \Vinit_{S'}\mid \eta (v)=\eta'(v')\}$,
	\item $\Vroot_P := \{ (v,v')\in \Vroot_S\times \Vroot_{S'}\mid \eta (v)=\eta'(v')\}$,
	\item $P(v,v'):=S(v)$ ($=T(\eta(v))=T(\eta'(v'))=S'(v')$),
	\item there is a "transition edge" $((v,v'),d,(w,w'))\in \Edges_P$ if $(v,d,w)\in \Edges_S$ and $(v',d,w')\in \Edges_{S'}$, and
	\item there is a "variable edge" $((v,v'),d,y)\in \Edges_P$ if $(v,d,y)\in \Edges_S$ and $(v',d,y)\in \Edges_{S'}$.
	\end{itemize}
	We define the "morphisms of set-systems" $\pi\colon P\ssmorphism S$ and~$\pi'\colon P\ssmorphism S'$ to be the projections on the first, and the second component respectively. It is straightforward to check that these are indeed "morphisms@@ss" and that $\eta\circ\pi = \eta'\circ\pi'$.
	
	Let us show that $P$ is universal with this property. Let us consider some "$\alphabet$-set-system"~$U$ together with "morphisms@@ss" $\tau\colon U\ssmorphism S$ and $\tau'\colon U\ssmorphism S'$ such that $\eta\circ\tau = \eta'\circ\tau'$. Define now $\theta\colon U\ssmorphism P$ by $\theta(u)=(\tau(u),\tau'(u))$ for all~$u\in \V_U$. Since $\eta(\tau(u))=\eta'(\tau'(u))$, we indeed have $(\tau(u),\tau'(u))\in \V_P$. Hence $\theta$ is indeed a "morphism@@ss" such that $\tau = \pi\circ\theta$ and $\tau'=\pi'\circ\theta$.
Furthermore, this $\theta$ is unique. Indeed, assume the existence of a "morphism@@ss" $\rho$ from~$U$ to~$P$ such that $\tau = \pi\circ\rho$ and $\tau'= \pi'\circ\rho$, and consider some vertex~$u$ in~$U$. Let $(v,v'):=\rho(u)\in \V_P$. Since $\tau = \pi\circ \rho$ and by definition of~$\pi$ and $\theta$, we have $v=\pi(v,v')=\pi(\rho(u))=\tau(u)$. 
 In the same way, $v'=\tau'(u)$. Overall $\rho(u)=(\tau(u),\tau'(u))$ so $\rho=\theta$.

	Let us finally show that if~$\eta$ is "locally surjective@@ss", then $\pi'$ also is. The interesting case is, for $u\in \V_P$, the surjectivity of $\elift\pi'$ from $\Edges_P(u)$ onto $\Edges_{S'}(\pi'(u))$.
	By definition of $\V_P$, $u=(v,v')$ with $\eta(v)=\eta'(v')$.
	Let $(v',d,w')\in \Edges_{S'}(v')$. Since $\eta'$ is a "morphism@@ss", $(\eta'(v'),d,\eta'(w'))\in \Edges_T(\eta'(v'))$.
	Since $\eta(v)=\eta'(v')$ and using the "local surjectivity@@ss" of~$\eta$, there exists $(v,d,w)\in \Edges_S(v)$ such that $\elift\eta(v,d,w)=(\eta(v),d,\eta'(w'))$. Hence, by definition of~$\Edges_P$, $((v,v'),d,(w,w'))\in \Edges_P(u)$ and is such that $\elift\pi'((v,v'),d,(w,w'))=(v',d,w')$. Hence $\elift\pi'$ is surjective from $\Edges_P(u)$ onto $\Edges_{S'}(\pi'(u))$. The case where $w$ is a variable is handled as well by taking $w'=\eta'(w)=w$.
	If~$\eta$ is a "folding@@ss", then for every $v'\in\V_{S'}$ there is $v\in\V_S$ with $\eta(v)=\eta'(v')$. Thus $(v,v')\in\V_P$ and $\pi'(v,v')=v'$, so $\pi'$ is surjective and hence a "folding@@ss". The assertions for $\eta'$ and $\pi$ follow symmetrically.

\end{proof}

\begin{lemma}[pullbacks of systems]\label{lemma:pullback-systems}
Let $S,S',T$ be "$\alphabet$-systems", let $\eta\colon S\ssmorphism T$ and $\eta'\colon S'\ssmorphism T$ be "morphisms@@ss", and let $P$ be their pullback constructed in \Cref{lemma:pullback}.
The restriction $P_{\mathrm{reach}}$ of~$P$ to the "vertices" reachable from its unique "initial vertex" is a "system", that "folds" to~$S$ and~$S'$ via the restrictions of the "foldings" from $P$. Thus, the pullback property is still valid when restricted to "systems". 
\end{lemma}
\begin{proof}
	Let $i,i'$ be the respective "initial vertices" of~$S,S'$. Both map to the unique "initial vertex" of~$T$, so $\Vinit_P=\{(i,i')\}$; also $\Vroot_P=\emptyset$. At every $(v,v')\in\V_P$ and in each permitted "direction", the unique successors of~$v$ and~$v'$ map to the same successor in~$T$. Thus they are either "vertices" whose pair belongs to~$\V_P$, or the same "variable". Consequently $P$ has exactly one successor in each permitted "direction", and so does its reachable part $P_{\mathrm{reach}}$.

	Every path from~$i$ in~$S$ lifts from~$(i,i')$ by following the same sequence of "directions" in~$S'$: the two paths have the same image in~$T$, so their paired "vertices" belong to~$P$, and whenever one reaches a "variable", both reach the same "variable". All "variables" are therefore reachable in~$P_{\mathrm{reach}}$, which is a "system". Its projections are "morphisms@@ss" between "systems", hence "foldings@@ss" by \Cref{remark:system-morphisms}.

	Finally, let~$U$ be a "system" with "morphisms@@ss" $\tau\colon U\ssmorphism S$ and $\tau'\colon U\ssmorphism S'$ such that $\eta\circ\tau=\eta'\circ\tau'$. By \Cref{lemma:pullback}, these induce a unique "morphism@@ss" $\theta\colon U\ssmorphism P$. Every "vertex" of~$U$ is reachable from its unique "initial vertex", and $\theta$ maps such paths to paths from~$(i,i')$, so its image lies in~$P_{\mathrm{reach}}$. It therefore factors uniquely through~$P_{\mathrm{reach}}$, proving the universal property among "systems".
\end{proof}

\begin{remark}\label{remark:pullback-reachability}
The full "set-system" pullback~$P$ need not be a "system", even when $S,S',T$ are "systems". For example, let $S=S'$ be the "closed@@system" "system" with two "vertices" $r,s$, both labelled by a symbol~$a$ of "rank"~$1$, initial set $\{r\}$, and "edges" $(r,1,s)$ and $(s,1,r)$. Let $T$ be the one-vertex $a$-labelled loop, and let $\eta=\eta'$ be the constant map. Then $\V_P=\{r,s\}^2$, but only $(r,r)$ and $(s,s)$ are reachable from its "initial vertex" $(r,r)$. Thus the restriction to reachable "vertices" is necessary.
\end{remark}

Our "morphisms@@ss" are defined from a given "rank@@ss" to the same one. The following lemma is a tool for transforming "morphisms@@ss" between "set-systems" of a "rank@@ss" into "morphisms@@ss" between "set-systems" of another "rank@@ss".
\begin{lemma}\label{lemma:dup-morphism}
	Given a "set-system"~$S$ of "rank@@ss"~$m$, a "set-system"~$S'$ of rank~$n$, and a map~$\sigma\colon\interval m\to\interval n$, then, for a map~$\rho$ from "vertices" of~$S$ to "vertices" of~$S'$, the following properties are equivalent
	\begin{itemize}
	\item $\rho$ is a "morphism@@ss" from $\rename\sigma(S)$ to~$S'$, and 
	\item $\rho$ is a "morphism@@ss" from $S$ to $\dupname\sigma(S')$.
	\end{itemize}
\end{lemma}
\begin{proof}
	Assume  $\rho$ is a "morphism@@ss" from $\rename\sigma(S)$ to~$S'$.
	Only "variable edges" differ between $\rename\sigma(S)$ and~$S$ (resp. $S'$ and $\dupname\sigma(S')$), so we only have to check the condition for "variable edges".
	Consider a "variable edge"~$(s,i,x_j)\in \Edges_S$.
	This implies that $(s,i,x_{\sigma(j)})\in \Edges_{\rename\sigma(S)}$.
	Hence, by "morphism@@ss", $(\rho(s),i,x_{\sigma(j)})\in \Edges_{S'}$.
	And thus $(\rho(s),i,x_j)\in \Edges_{\dupname\sigma(S')}$.

	Conversely, assume  $\rho$ is a "morphism@@ss" from $S$ to $\dupname\sigma(S')$.
	Consider a "variable edge"~$(s,i,x_j)\in \Edges_{\rename\sigma(S)}$.
	By definition of $\rename\sigma$, there is a "variable edge"~$(s,i,x_k)\in \Edges_S$ with $j=\sigma(k)$.
	Since~$\rho$ is a "morphism@@ss", $(\rho(s),i,x_k)\in \Edges_{\dupname\sigma(S')}$.
	Hence, by definition of~$\dupname\sigma$ there is a "variable edge"~$(\rho(s),i,x_j)\in \Edges_{S'}$. 
\end{proof}

\def\IGNORE{
\begin{lemma}\THOMAS{Ne semble pas utilisé.}
	Let~$S$ be a "set-system" of "rank"~$m$, $S'$ a "set-system" of "rank"~$n$ and a map $\sigma\colon\interval m\to\interval n$ such that there is a "morphism@@ss"~$\rho$ from $S'$ to $\rename\sigma(S)$. Then, there exists a "set-system"~$S''$ of "rank"~$n$ such that $\rename\sigma(S'')=S'$ (and hence $S''$ and $S'$ have the same "vertices") and $\rho$ is a "morphism@@ss" from~$S''$ to $S$.
\end{lemma}
\begin{proof}
	Let us define $S''$ to be identical to~$S'$ but for "variable edges" which are defined as follows.
	There is a "variable edge" $(s,i,x_j)\in \Edges_{S''}$ if and only if $(s,i,x_{\sigma(j)})\in \Edges_{S'}$ and $(\rho(s),i,x_j)\in \Edges_S$.

	Let us prove that $\rename\sigma(S'')=S'$. It is obvious in all respects by for the "variable edges". It is also clear from the definition that if $(s,i,x_j)\in \Edges_{S''}$ the $(s,i,x_{\sigma(j)})\in \Edges_{S'}$, i.e.\ that $\Edges_{\rename\sigma(S'')}\subseteq \Edges_{S'}$. Conversely, let~$(s,i,x_k)\in \Edges_{S'}$, then (by "morphism@@ss") $(\rho(s),i,x_k)\in \Edges_{\rename\sigma(S)}$, hence (by definition of~$\rename\sigma$) there exists $(\rho(s,i,x_j))\in \Edges_S$ with $\sigma(j)=k$, and (by definition of~$S'$) we get $(s,i,x_j)\in \Edges_{S''}$. It follows that $(s,i,x_k)\in \Edges_{\rename\sigma(S'')}$, and hence $\Edges_{S'}\subseteq \Edges_{\rename\sigma(S'')}$.
\end{proof}}

\subsection{Unfoldings and regular-trees}

We define "unfoldings" using "foldings" (\Cref{definition:set-unfold}), prove that "unfold-equivalence" is an equivalence relation (\Cref{lemma:unfold-equivalence-ss}), and define "regular-trees" and their relabelling functor (\Cref{definition:regular-trees}). We compare these classes with classical "fully unfolded" trees.


The notion of "unfolding@@ss" of "set-systems" that we now introduce derives from the notion of "foldings@@ss".

\begin{definition}[unfolding and unfold-equivalence]\label{definition:set-unfold}
\AP If $S$ and $S'$ are "$\alphabet$-set-systems" of same "rank@@ss", then $S'$ is an ""unfolding@@ss"" of~$S$ if there exists a "folding@@ss" $S'\ssfolding S$. 
Two "$\alphabet$-set-systems" that have a common "unfolding@@ss" are called ""unfold-equivalent@@ss"".
\end{definition}
\AP By \Cref{remark:system-morphisms}, a "system"~$T$ is an "unfolding@@ss" of a "system"~$T'$ if and only if there is a "morphism@@s" $T\ssmorphism T'$.

The first lemma shows that this terminology of equivalence is suitably chosen.
\begin{lemma}\label{lemma:unfold-equivalence-ss}
	The binary "unfold-equivalence@@ss" relation between "$\alphabet$-set-systems" is indeed an equivalence relation.
\end{lemma}
\begin{proof}
	Reflexivity and symmetry are straightforward. For transitivity, assume $S$ is a common "unfolding@@ss" of $T,T'$, and $S'$ is a common "unfolding@@ss" of $T',T''$. By \Cref{lemma:pullback}, the "foldings@@ss" from $S$ and $S'$ to $T'$ have a pullback $P$, whose projections to $S$ and $S'$ are "foldings@@ss". By composition of "foldings@@ss" (\Cref{lemma:morphism-composition}), $P$ is a common "unfolding@@ss" of $T,T''$. The relation is therefore transitive. In picture:
	\begin{center}
	\begin{tikzpicture}
		\node (P) at (0,0) {$P$};
		\node (S) at (-1,-1) {$S$};
		\node (S') at (1,-1) {$S'$};
		\node (T) at (-2,-2) {$T$};
		\node (T') at (0,-2) {$T'$};
		\node (T'') at (2,-2) {$T''$};
		\draw[triplehead] (P) -- (S);
		\draw[triplehead] (P) -- (S');
		\draw[triplehead] (S) -- (T);
		\draw[triplehead] (S) -- (T');
		\draw[triplehead] (S') -- (T');
		\draw[triplehead] (S') -- (T'');
	\end{tikzpicture}
	\end{center}
\end{proof}
\begin{lemma}[common system unfolding]\label{lemma:common-system-unfolding}
If two "$\alphabet$-systems" $T,T'$ are "unfold-equivalent@@ss", then there exists a "$\alphabet$-system"~$U$ that is an "unfolding@@ss" of both.
\end{lemma}
\begin{proof}
Let $P$ be a common "set-system" "unfolding@@ss", with "foldings@@ss" $\eta\colon P\ssfolding T$ and $\eta'\colon P\ssfolding T'$. Apply \Cref{lemma:system-selection} to~$\eta$, obtaining a "system"~$U$ and an inclusion $\iota\colon U\ssmorphism P$ such that $\eta\circ\iota\colon U\ssfolding T$. The other composite $\eta'\circ\iota\colon U\ssmorphism T'$ is a "folding@@ss" by \Cref{remark:system-morphisms}. Thus~$U$ is a common "system" "unfolding@@ss".
\end{proof}

\begin{definition}[regular tree]\AP\label{definition:regular-trees}
	A ""$\alphabet$-regular-tree"" is an "unfold-equivalence" class of "$\alphabet$-systems".
\end{definition}

\begin{figure}[h]
	\centering{\def\svgwidth{\linewidth}}\scalebox{.7}{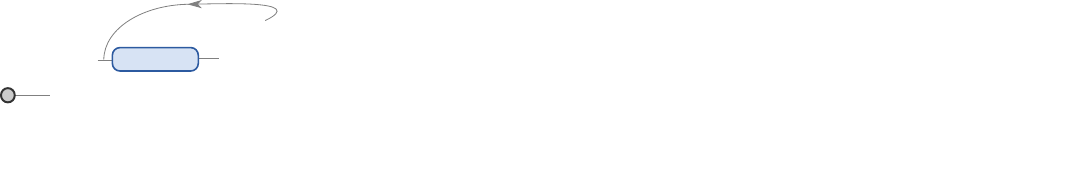}
	\caption{$S_1$ and $S_2$ are "unfold-equivalent", $F$ is a common "folding" of both, and $U$ a common "unfolding" of both.}\label{fig:unfold}\label{figure:unfoldings}
\end{figure}

This is illustrated in \Cref{fig:unfold}.
\begin{remark}
	Note that if one ""fully unfolded"" a "system" (as is classically done), one would obtain an infinite tree which is ``regular in the classical sense'' (a finite number of subtrees)\phantomintro{regular infinite tree}, and that all "regular trees@regular infinite tree" (in the classical sense) can be obtained as the "full unfolding" of some "system".
	It is also easy to check that two "systems" are "unfold-equivalent" if and only if they have isomorphic "full unfoldings".
	Hence, the map which associates to a "regular tree@regular infinite tree" (in the classical sense) the set of "systems" that "fully unfold" to it is a bijection between "regular trees@regular infinite tree" (in the classical sense) and "regular-trees" as defined in this work.
\end{remark}

\begin{remark}
Relabelling along a "map of ranked sets" induces a map on "regular-trees", since it preserves "foldings@@ss". The resulting endofunctor of the "category of ranked sets" is called the ""regular-tree functor"".
\end{remark}


Let us conclude this section with a technical lemma.
\begin{lemma}\label{statement:system-sigma-lift}
	Let~$\sigma\colon\interval m\to\interval n$ be surjective.
	Let $T$ be a "system" of "rank"~$m$ and $T'$ be a "system" of "rank"~$n$ which is an "unfolding" of $\rename\sigma(T)$.
	There exists an "unfolding" $T''$ of~$T$ such that $\rename\sigma(T'')=T'$.
\end{lemma}
\begin{proof}
	Let~$\eta$ be the "morphism@@ss" from $T'$ to~$\rename\sigma(T)$.
	We define~$T''$ to be identical to~$T'$, but for the fact that there is a "variable edge"~$(v,d,y)\in\Edges_{T''}$ if and only if there is a "variable edge"~$(\eta(v),d,y)$ in~$T$. One easily checks that (1) $T''$ is a "system", (2) $\eta$ is a "morphism@@ss" from~$T''$ to~$T$, and (3) $\rename\sigma(T'')=T'$.
\end{proof}

\subsection{Yields of set-systems and yield-equivalence}
\label{subsection:resolutions}

We define "init yields" and "root yields", and "yield-equivalence" (\Cref{definition:yields}). We prove invariance under "unfolding" of the set of yielded "systems" (\Cref{lemma:yield-unfold-invariance}), and study properties of "yields" with respect to "morphisms".

We have seen so far that "systems" and "set-systems" can be dealt with modulo "unfold-equivalence".
We shall now see how "set-systems" can be understood as ``finite non-deterministic machines that would non-deterministically produce "systems"''.
Again, these notions derive from "morphisms@@ss". The notion of "init yield" corresponds to a "system" produced starting from some "initial vertex", and "root yield" to a "system" produced from a "root vertex".
Two "set-systems" are "yield-equivalent" if they agree on their "yields".
An important tool developed in this section is that the existence of a "morphism@@ss" between "set-systems" entails the inclusion of their sets of "yields", and their equality if it is a "locally surjective@@ss" "morphism of set-systems" (see \Cref{statement:locsurj->res-eq}).

\begin{figure}[ht]
\begin{center}
\begin{tikzpicture}
\node (U) at (0,0) {$U$};
\node (T) at (0,-1.2) {$T$};
\node (US) at (3,0) {$\rename\sigma(U)$};
\node (S) at (3,-1.2) {$S$};
\draw[triplehead] (U) -- (T);
\draw[singlehead] (US) -- (S);
\draw[->,densely dotted] (U) -- (US) node[pos=.5,above]{$\rename\sigma$};

\node (eq) at (4.5,-0.6) {$\Longleftrightarrow$};
\node (descr) at (4.5,-2) {$(T,\sigma)$ is an "init-yield" of $S$};
\node (U2) at (6,0) {$U$};
\node (T2) at (6,-1.2) {$T$};
\node (S2) at (9,0) {$\dupname\sigma(S)$};
\draw[triplehead] (U2) -- (T2);
\draw[singlehead] (U2) -- (S2);
\end{tikzpicture}
\end{center}
\caption{An "init-yield" $(T,\sigma)$ of~$S$ is witnessed by a "system"~$U$, a "folding@@ss" $U\ssfolding T$, and a "morphism@@ss" $\rename\sigma(U)\ssmorphism S$, equivalently $U\ssmorphism\dupname\sigma(S)$ by \Cref{lemma:dup-morphism}.}\label{fig:def-yield}
\end{figure}

\begin{definition}\AP\label{definition:yields}
Let~$S$ be a "set-system" of "rank@@ss"~$n$.
An ""init-yield"" of~$S$ is a pair $(T,\sigma)$, where $T$ is a "system" of "rank@@ss"~$m$ and $\sigma\colon\interval m\to\interval n$, such that there exist a "system" $U$ that is an "unfolding@@ss" of $T$ and a "morphism@@ss" $\rename\sigma(U)\ssmorphism S$ (or equivalently by \Cref{lemma:dup-morphism}, a "morphism@@ss" $U\ssmorphism\dupname\sigma(S)$).
A ""root-yield"" of~$S$ is an "init-yield" of~$\Uproot(S)$, recorded separately as a "root-yield".
A ""yield"" is either an "init-yield" or a "root-yield".
The "set-system" associated\footnote{This is a bit cumbersome, but this choice was made to preserve the property that "yields" are actual "systems" (without "root"), together with a renaming map.} with a "yield"~$(T,\sigma)$ is denoted $\intro*\yieldss(T,\sigma)$ and is defined, with the initial or root status of the pair retained, by
\[
  \yieldss(T,\sigma):=
  \begin{cases}
    \rename\sigma(T) & \text{if $(T,\sigma)$ is an "init-yield"},\\
    \plant(\rename\sigma(T)) & \text{if $(T,\sigma)$ is a "root-yield"}.
  \end{cases}
\]
 For a family~$F$ of "yields", write $\yieldss(F):=\{\yieldss(q)\mid q\in F\}$; this notation always retains the initial or root status of each member.

\AP We denote the sets of "init-yields" and "root-yields" by $\intro*\InitYields(S)$ and $\intro*\RootYields(S)$ respectively, and their disjoint union by $\intro*\Yields(S)$. Thus a pair may occur in both families, with its initial or root status retained.
For "set-systems"~$S,S'$ of the same "rank@@ss", write $S\intro*\yieldssubseteq S'$ if both families for~$S$ are included in the corresponding families for~$S'$.
They are ""yield-equivalent"", written $S\intro*\yieldseq S'$, if both inclusions are equalities.
\end{definition}

The definition of "yields" include a renaming map~$\sigma$, see \Cref{example:direct-resolutions-not-compositional} to see why that is necessary.
The map~$\sigma$ need be neither injective nor surjective: different variables of~$T$ may represent the same variable of~$S$, and some variables of~$S$ may be unused. Although $T$ uses every one of its own variables, $\rename\sigma(T)$ is in general only a "set-system".
Whenever a "yield"~$q=(T,\sigma)$ is inserted into a sum or a "set-context", we use the rank-$n$ "set-system" $\yieldss(T,\sigma)=\yieldss(q)$ from \Cref{definition:yields}.
For a "closed@@system"~$S$, necessarily $m=0$ and $\sigma$ is the empty map. In this case $\yieldss(T,\sigma)$ is $T$ for an "init-yield" and $\plant(T)$ for a "root-yield". When treating closed "yields" as "set-systems", we always mean this explicit conversion.

\begin{lemma}[unfold-invariance of yields]\label{lemma:yield-unfold-invariance}
Both families of "yields" are closed under "unfold-equivalence" of the "system" component, with the variable map $\sigma$ fixed.
\end{lemma}
\begin{proof}
Let $(T,\sigma)$ be an "init-yield" of~$S$, witnessed by a "system"~$U$, a "folding@@ss" $U\ssfolding T$, and a "morphism@@ss" $\beta\colon U\ssmorphism\dupname\sigma(S)$. If~$T'$ is "unfold-equivalent" to~$T$, then $U$ and~$T'$ are "unfold-equivalent" by \Cref{lemma:unfold-equivalence-ss}. By \Cref{lemma:common-system-unfolding}, choose a "system"~$P$ with "foldings@@ss" $\pi\colon P\ssfolding U$ and $P\ssfolding T'$. The composite $\beta\circ\pi\colon P\ssmorphism\dupname\sigma(S)$ then makes~$P$ a witness for $(T',\sigma)\in\InitYields(S)$. This shows the statement for "init-yields". Apply the same argument to~$\Uproot(S)$ for "root-yields".
\end{proof}

\begin{lemma}\label{statement:locsurj->res-eq}
	If there is a "morphism of $\alphabet$-set-systems" from~$S$ to~$S'$, then~$S \yieldssubseteq S'$. If furthermore it is "locally surjective@@ss" then $S \yieldseq S'$.
\end{lemma}
\begin{proofof}{statement:locsurj->res-eq}
Let~$\eta\colon S\ssmorphism S'$ be a "morphism@@ss". A witness $U\ssfolding T$, $\rename\sigma(U)\ssmorphism S$ for an "init-yield" $(T,\sigma)$ of~$S$ also witnesses that pair for~$S'$, by composition with~$\eta$.

Suppose now that~$\eta$ is "locally surjective@@ss". To prove the reverse inclusion, take $(T,\sigma)\in\InitYields(S')$, with a "system" witness~$U$, a "folding@@ss" $\alpha\colon U\ssfolding T$, and a "morphism@@ss" $\beta\colon U\ssmorphism\dupname\sigma(S')$. Variable renaming preserves local surjectivity, so~$\eta$ induces a "locally surjective@@ss" "morphism@@ss" $\eta_\sigma\colon\dupname\sigma(S)\sslocsurj\dupname\sigma(S')$.

By \Cref{lemma:pullback}, there is a pullback $P$ of~$\eta_\sigma$ and~$\beta$ verifying the following commuting diagram:

\begin{center}
\begin{tikzpicture}
\node (P) at (0,0) {$P$};
\node (S) at (4,0) {$\dupname\sigma(S)$};
\node (U) at (0,-1.5) {$U$};
\node (Sp) at (4,-1.5) {$\dupname\sigma(S')$};
\draw[singlehead] (P) -- (S) node[midway,above] {$q$};
\draw[doublehead] (P) -- (U) node[midway,left] {$p$};
\draw[doublehead] (S) -- (Sp) node[midway,right] {$\eta_\sigma$};
\draw[singlehead] (U) -- (Sp) node[midway,below] {$\beta$};
\end{tikzpicture}
\end{center}

Here~$p$ is "locally surjective@@ss" because~$\eta_\sigma$ is. Since~$U$ is a "system", \Cref{lemma:system-selection} applied to~$p$ gives a "system"~$D$ and an inclusion "morphism@@ss" $\iota\colon D\ssmorphism P$ such that $\gamma:=p\circ\iota\colon D\ssfolding U$ is a "folding@@ss". Thus the following diagram commutes.
\begin{center}
\begin{tikzpicture}
\node (D) at (-2,1.5) {$D$};
\node (P) at (0,0) {$P$};
\node (S) at (4,0) {$\dupname\sigma(S)$};
\node (U) at (0,-2) {$U$};
\node (Sp) at (4,-2) {$\dupname\sigma(S')$};
\node (T) at (-2,-3.5) {$T$};
\draw[singlehead] (D) -- (P) node[midway,above right] {$\iota$};
\draw[triplehead] (D) -- (U) node[midway,left] {$\gamma$};
\draw[singlehead] (P) -- (S) node[midway,above] {$q$};
\draw[doublehead] (P) -- (U) node[midway,right] {$p$};
\draw[doublehead] (S) -- (Sp) node[midway,right] {$\eta_\sigma$};
\draw[singlehead] (U) -- (Sp) node[midway,below] {$\beta$};
\draw[triplehead] (U) -- (T) node[midway,below right] {$\alpha$};
\draw[triplehead,densely dotted] (D) -- (T) node[midway,left] {$\alpha\circ\gamma$};
\draw[singlehead,densely dotted] (D) to[bend left=20] node[midway,above] {$q\circ\iota$} (S);
\end{tikzpicture}
\end{center}

The dotted composed "morphisms@@ss" witness that $(T,\sigma)$ is an "init-yield" of~$S$, as desired. The argument for "root-yields" is identical, applied to $\Uproot(S)$ and $\Uproot(S')$.
\end{proofof}
\begin{corollary}\label{cor:unfold-yield}
	"Set-systems" that are "unfold-equivalent@@ss" are "yield-equivalent".
\end{corollary}
The converse is not true, as witnessed by the following "set-systems" $S$ and $S'$, that are "yield-equivalent" but not "unfold-equivalent@@ss":

\begin{center}
\scalebox{.9}{
\begingroup%
  \makeatletter%
  \providecommand\color[2][]{%
    \errmessage{(Inkscape) Color is used for the text in Inkscape, but the package 'color.sty' is not loaded}%
    \renewcommand\color[2][]{}%
  }%
  \providecommand\transparent[1]{%
    \errmessage{(Inkscape) Transparency is used (non-zero) for the text in Inkscape, but the package 'transparent.sty' is not loaded}%
    \renewcommand\transparent[1]{}%
  }%
  \providecommand\rotatebox[2]{#2}%
  \newcommand*\fsize{\dimexpr\f@size pt\relax}%
  \newcommand*\lineheight[1]{\fontsize{\fsize}{#1\fsize}\selectfont}%
  \ifx\svgwidth\undefined%
    \setlength{\unitlength}{364.72040131bp}%
    \ifx\svgscale\undefined%
      \relax%
    \else%
      \setlength{\unitlength}{\unitlength * \real{\svgscale}}%
    \fi%
  \else%
    \setlength{\unitlength}{\svgwidth}%
  \fi%
  \global\let\svgwidth\undefined%
  \global\let\svgscale\undefined%
  \makeatother%
  \begin{picture}(1,0.12937598)%
    \lineheight{1}%
    \setlength\tabcolsep{0pt}%
    \put(0,0){\includegraphics[width=\unitlength,page=1]{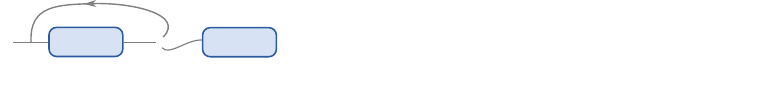}}%
    \put(0.10081785,0.06712278){\color[rgb]{0,0,0}\makebox(0,0)[lt]{\lineheight{1.25}\smash{\begin{tabular}[t]{l}$a_1$\end{tabular}}}}%
    \put(0,0){\includegraphics[width=\unitlength,page=2]{yield-not-unfold.pdf}}%
    \put(0.16185001,0.00490822){\makebox(0,0)[t]{\lineheight{1.25}\smash{\begin{tabular}[t]{c}$S$\end{tabular}}}}%
    \put(0,0){\includegraphics[width=\unitlength,page=3]{yield-not-unfold.pdf}}%
    \put(0.63595014,0.00570291){\makebox(0,0)[t]{\lineheight{1.25}\smash{\begin{tabular}[t]{c}$S'$\end{tabular}}}}%
    \put(0,0){\includegraphics[width=\unitlength,page=4]{yield-not-unfold.pdf}}%
    \put(0.30806863,0.06709812){\color[rgb]{0,0,0}\makebox(0,0)[lt]{\lineheight{1.25}\smash{\begin{tabular}[t]{l}$b_0$\end{tabular}}}}%
    \put(0.94348758,0.06709812){\color[rgb]{0,0,0}\makebox(0,0)[lt]{\lineheight{1.25}\smash{\begin{tabular}[t]{l}$b_0$\end{tabular}}}}%
    \put(0.58070689,0.06712266){\color[rgb]{0,0,0}\makebox(0,0)[lt]{\lineheight{1.25}\smash{\begin{tabular}[t]{l}$a_1$\end{tabular}}}}%
    \put(0.77964445,0.06712266){\color[rgb]{0,0,0}\makebox(0,0)[lt]{\lineheight{1.25}\smash{\begin{tabular}[t]{l}$a_1$\end{tabular}}}}%
  \end{picture}%
\endgroup%
}
\end{center}
Indeed, the presence of an $a$-node without an edge towards an $a$-node in $S'$ witnesses the non-"unfold-equivalence" of the two "systems". Their "yields" are in both cases simply lines of $a$ ending with a $b$-node.

\begin{lemma}\label{statement:ue=te}
Two "systems" of the same "rank@@ss" are "unfold-equivalent@@ss" if and only if they are "yield-equivalent".
\end{lemma}
\begin{proof}
Let $T,T'$ be "systems" of the same "rank@@ss". If they have a common "unfolding@@ss", \Cref{statement:locsurj->res-eq} applied to its two "foldings@@ss" gives equality of their "yields".

Conversely, if $T,T'$ are "yield-equivalent", then $(T,\Id{\rk(T)})$, witnessed by~$T$ itself, is an "init-yield" of~$T'$. By \Cref{definition:yields}, there are a "system"~$U$, a "folding@@ss" $U\ssfolding T$, and a "morphism@@ss" $U\ssmorphism T'$. The latter is a "folding@@ss" by \Cref{remark:system-morphisms}, so~$U$ is a common "unfolding@@ss" of~$T,T'$.
\end{proof}

\begin{example}
	In \Cref{fig:system}, $(S_2,\Id 2)$ is an "init-yield" of~$S_1$, witnessed by~$S_2$ itself.
\end{example}

See \Cref{example:direct-resolutions-not-compositional} for a case where $U$ is different from $T$, and why this is useful.

\subsection{Flattening}
\label{subsec:flatten}


We describe in this section the operations used to construct complex "set-systems" out of simpler ones.
In particular, the notions of context and application of a context to a "system" derive from it.
The general idea is to perform the disjoint union of "systems", connecting variables of subsystems to "initial vertices" of others according to a pattern.
The operation for this is $\flatten$ that we define below. 

It is better explained through an example first.
\begin{example}\label{example:flatten}
	The following illustration shows a "set-system" of "set-systems" and its "flattening". The subsystems are drawn inside the dashed boxes, and boxes are organised in an outer "system" structure called the pattern.
	Direction names for edges going out of a node $v$ will always be implicit, the topmost one being $1$, then $2$, and so on, ending with $\rk(S(v))$ for the bottommost.
	\begin{center}
		\scalebox{.7}{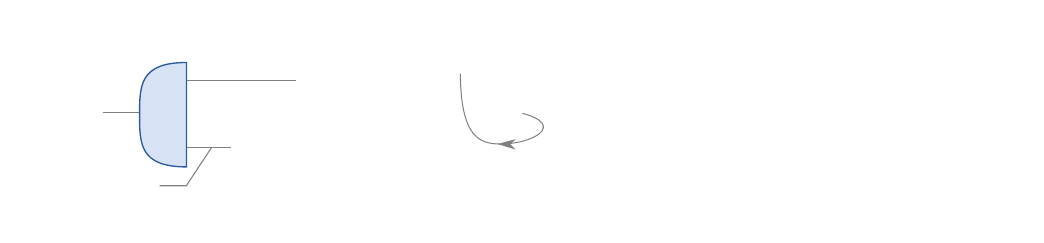}
	\end{center}
\end{example}

	As is shown in \Cref{example:flatten}, the $\flatten$ operation glues the subsystems together, keeping their internal structures. If an "edge" of a subsystem ends in a variable~$x_d$, then it is connected instead to all the "initial vertices" of the subsystems reachable by following "direction"~$d$ in the pattern, as well as to the variables reachable by following direction~$d$.  After "flattening", the "initial vertices" are the "initial vertices" of the subsystems that were themselves "initial vertices" in the pattern.
	"Root vertices" are treated very differently. "Root vertices" from subsystems remain "root vertices" during the process, while "initial vertices" of "root@@vertex" subsystems are promoted to "roots@@vertex".

	\AP Formally, we already noted that "$\alphabet$-set-systems" do form a "ranked set". Hence, the notion of "($\alphabet$-set-system)-set-systems" is meaningful. The $\flatten$ operation maps "($\alphabet$-set-system)-set-systems" to "$\alphabet$-set-systems", and is defined as follows.
\begin{definition}[flattening]\AP\label{definition:flattening}
	Given a "($\alphabet$-set-system)-set-system"~$S=(\V,\Vinit,\Vroot,\Label,\Edges)$ with $S(v)=:(\V_v,\Vinit_v,\Vroot_v,\Label_v,\Edges_v)$, we define $\intro*\flatten(S):=(\V',\Vinit',\Vroot',\Label',\Edges')$ where:\phantomintro{flattening}
	\begin{itemize}
	\item $\V' := \{(v,w)\mid v\in \V, w\in \V_v\}$,
	\item $\Vinit' :=\{(v,w)\mid v\in \Vinit,\ w\in \Vinit_v\}$,
	\item $\Vroot' := \{(v,w)\mid v\in \V,\ w\in \Vroot_v\}\cup\{(v,w)\mid v\in\Vroot,\ w\in\Vinit_v\}$,
	\item $\flatten(S)(v,w):=S(v)(w)$,
	\itemAP The edge relation is the union of the following three relations:\phantomintro(flatten){intra-subsystem}\phantomintro(flatten){inter-subsystem}\phantomintro(flatten){variable}

    \begin{align*}
      E_{\mathrm{intra}}&:=
        \{((v,w),d,(v,w'))\mid v\in\V,\ (w,d,w')\in\Edges_v\},\\
      E_{\mathrm{inter}}&:=
        \{((v,w),d,(v',w'))\mid (v,e,v')\in\Edges,\ (w,d,x_e)\in\Edges_v,\ w'\in\Vinit_{v'}\},\\
      E_{\mathrm{var}}&:=
        \{((v,w),d,y)\mid (v,e,y)\in\Edges,\  y\in\Vars,\ (w,d,x_e)\in\Edges_v\},\\
      \Edges'&:=E_{\mathrm{intra}}\cup E_{\mathrm{inter}}\cup E_{\mathrm{var}}.
    \end{align*}
    These are respectively the \reintro(flatten){intra-subsystem}, \reintro(flatten){inter-subsystem}, and \reintro(flatten){variable} cases.
\end{itemize}
\end{definition}
We can also use $\flatten$ for "systems" as justified by the following lemma.
\begin{lemma}
	\label{lemma:flatten-systems}The "flattening" of a "($\alphabet$-system)-system" is a "$\alphabet$-system".
\end{lemma}
\begin{proof}
	The unique "initial vertex", absence of "root vertices", and unique successor in each "direction" follow from the definition. Every "variable" of each subsystem is reachable from its "initial vertex", so every edge of the outer "system" can be realised by a path through that subsystem. Since all outer "vertices" and "variables" are reachable, all subsystem entries and all outer "variables" are reachable after "flattening". Reachability of every inner "vertex" then gives reachability of every flattened "vertex".
\end{proof}

Let us also provide two simple results stating that "morphisms@@ss" behave well with respect to "flattening".
\begin{lemma}\label{lemma:inner-morphism-ok}
	Let $S,S'$ be two "($\alphabet$-set-system)-set-systems" of  "same shape@@ss" such that there is a "set-system morphism" $\eta_v$ from~$S(v)$ to $S'(v)$ for all~$v\in \V_S$.
	Then $\eta$ defined as $\eta(v,w)=(v,\eta_v(w))$ is a "morphism@@ss" from~$\flatten(S)$ to~$\flatten(S')$.
	Furthermore, if~$\eta_v$ is "locally surjective@@ss" (resp. a "folding@@ss") for all~$v$, then $\eta$ is "locally surjective@@ss" (resp. a "folding@@ss").
\end{lemma}
\begin{proof}
	Simple verification. In the "folding@@ss" case, every target "vertex" $(v,w')$ has a preimage $(v,w)$ because $\eta_v$ is surjective.
\end{proof}

\begin{lemma}\label{lemma:outer-morphism-ok}
	Let~$S,S'$ be two "($\alphabet$-set-system)-set-systems", and $\eta$ be a "morphism@@ss" from $S$ to $S'$, then $\eta'$ defined by $\eta'(v,w)=(\eta(v),w)$ is a "morphism@@ss" from~$\flatten(S)$ to $\flatten(S')$. Furthermore, if~$\eta$ is a "folding@@ss", then $\eta'$ is a "folding@@ss".
\end{lemma}
\begin{proof}
	The "morphism@@ss" conditions follow directly from \Cref{definition:flattening}. Suppose now that~$\eta$ is a "folding@@ss". Surjectivity on outer "vertices" makes~$\eta'$ surjective on "vertices" and covers all roots inherited from labels. Local surjectivity on the outer initial and root sets covers the flattened initial vertices and the roots induced by outer root vertices. Finally, local surjectivity on outer edges, together with equality of corresponding labels, gives local surjectivity on flattened edges. Thus~$\eta'$ is a "folding@@ss".
\end{proof}

\AP We also define the $\intro*\atomic$  operation, which, given $a\in\alphabet$ builds a "$\alphabet$-system"~$\atomic(a)$ of "rank" $\rk(a)$ in a natural way.
\AP Formally, $\atomic(a)=(\{*\},\{*\},\emptyset,\Label,\Edges)$, where $\atomic(a)(*)=a$, and $\Edges=\{(*,i,x_i)\mid i\in\interval{\rk(a)}\}$. We will sometimes omit the $\atomic$ function to avoid notational clutter.

The following lemma shows that the identities defining a monad are satisfied.

\begin{lemma}\label{lemma:monad-identities}
	For any "(($\alphabet$-set-system)-set-system)-set-system"~$S_3$ and "$\alphabet$-set-system"~$S_1$, we have
	\begin{align*}
		\flatten(\flatten(S_3)) &= \flatten(\slift\flatten(S_3))\ ,&
		\flatten(\slift \atomic(S_1)) &= S_1
		\\&&\text{and}\quad
		\flatten(\atomic(S_1)) &= S_1\ .
	\end{align*}
\end{lemma}
\begin{proof}
Write a vertex of~$S_3$ as~$u$, a vertex of the "set-system"~$S_3(u)$ as~$v$, and a vertex of its label~$S_3(u)(v)$ as~$w$. The vertices on the two sides of the associativity identity are respectively $((u,v),w)$ and $(u,(v,w))$. Identify them with the same triple $(u,v,w)$.
Its label is $S_3(u)(v)(w)$ in both constructions. It is initial exactly when all three components are initial at their respective levels. It is a root exactly when $w$ is a root, or $v$ is a root and $w$ is initial, or $u$ is a root and both $v,w$ are initial. Thus the initial and root sets coincide.

An edge either stays inside the innermost subsystem, follows one of its "variable edges" and then a transition of the middle subsystem, or follows "variable edges" at both inner levels and then a transition or "variable edge" of the outer "system". In each case \Cref{definition:flattening} gives the same target and direction, independently of which pair of levels is flattened first. This proves associativity under the canonical identification of vertices.

For $\flatten(\slift\atomic(S_1))$, each vertex~$v$ is replaced by the sole vertex of $\atomic(S_1(v))$; the map $(v,*)\mapsto v$ preserves its label, its initial and root status, and every edge. For $\flatten(\atomic(S_1))$, the map $(*,v)\mapsto v$ has the same properties: the outer atomic edges simply restore the "variable edges" of~$S_1$.
\end{proof}

\begin{remark}
	\label{remark:monad-set-systems}What we have shown so far is that
	the operations $\flatten$ and $\atomic$ are natural transformations that equip the "set-system functor" with a monad structure.
	We call it the ""monad of set-systems"".
	For "systems", it defines similarly the ""monad of systems"".
\end{remark}

We end this section with a technical tool: the $\Uproot$ operation was used for defining $\RootYields$. We introduce here the companion operation, $\fuproot$, that allows us to reason about $\Uproot$ in the context of $\flatten$, see \Cref{statement:fuproot}.

\AP The operation $\intro*\fuproot$ transforms a "(set-system)-set-system"~$S$ into a  "(set-system)-set-system"~$\fuproot(S)=S'$ defined as follows:
\begin{itemize}
\item $\V_{S'}=\{0,1\}\times \V_S$,
\item $\Vinit_{S'}=\{0\}\times\V_S \cup \{1\}\times\Vroot_S$,
\item $\Vroot_{S'}=\emptyset$,
\item $S'(0,s)=\Uproot(S(s))$ and $S'(1,s)=\mathrm{noroot}(S(s))$, where $\mathrm{noroot}(S(s))$ is the "set-system" obtained from $S(s)$ by setting the set of roots to $\emptyset$.
\item $((m,s),d,(m',s'))$ is a "transition edge" in $S'$ if~$(s,d,s')\in\Edges_S$ and $m'=1$,
\item $((m,s),d,y)$ is a "variable edge" in $S'$ if~$(s,d,y)\in\Edges_S$.
\end{itemize}
The operation $\fuproot$ is justified by the following lemma.
\begin{lemma}\label{statement:fuproot}
	For all "(set-system)-set-system"~$S$, there exists a "folding@@ss" from~$\flatten(\fuproot(S))$ to~$\Uproot(\flatten(S))$.
\end{lemma}
\begin{proof}
We define the "folding@@ss" $\eta$ from $\flatten(\fuproot(S))$ to $\Uproot(\flatten(S))$ by forgetting the $\{0,1\}$-component: if $(b,v)\in \{0,1\}\times V_S$, and $w\in V_{S(v)}$, we have $\eta(((b,v),w))=(v,w)$. It is straightforward to check that this "morphism@@ss" is "locally surjective@@ss". It is surjective on "vertices" since every $(v,w)$ is the image of $((0,v),w)$.
\end{proof}

\subsection{Unfold-equivalence is a congruence}


The two following lemmas show that $\flatten$ behaves well with respect to "unfold-equivalence@@ss" (it is a congruence of the monad of "set-systems").

\begin{lemma}\label{statement:ue-is-congruence-2}
	Let~$S,S'$ be "unfold-equivalent@@ss" "($\alphabet$-set-system)-set-systems",
	then $\flatten(S)$ and $\flatten(S')$ are "unfold-equivalent@@ss".
\end{lemma}
\begin{proof}
	By definition of "unfold-equivalence@@ss", this means that $S$ and $S'$ both "unfold@@ss" to some~$S''$.
	By using twice \Cref{lemma:outer-morphism-ok}, we get that both $\flatten(S)$ and $\flatten(S')$ "unfold" to~$\flatten(S'')$.
	Hence~$\flatten(S)$ and $\flatten(S')$ are "unfold-equivalent@@ss".
\end{proof}

\begin{lemma}\label{statement:ue-is-congruence}
	Let~$S,S'$ be "($\alphabet$-set-system)-set-systems" of "same shape@@ss" such that $S(s)$ is "unfold-equivalent@@ss" to $S'(s)$ for all "vertices"~$s$,
	then $\flatten(S)$ and $\flatten(S')$ are "unfold-equivalent@@ss".
\end{lemma}
\begin{proof}
	Let us construct a third "($\alphabet$-set-system)-set-system"~$P$ that has the "shape@@ss" of $S$ and $S'$ and such that $P(s)$ is an "unfolding@@ss" of both~$S(s)$ and $S'(s)$ for all "vertices"~$s$ (it exists since $S(s)$ and $S'(s)$ are "unfold-equivalent@@ss").
	By using twice \Cref{lemma:inner-morphism-ok}, we get that $\flatten(P)$ is an "unfolding@@ss" of both~$\flatten(S)$ and~$\flatten(S')$.
	Hence, $\flatten(S)$ and $\flatten(S')$ are "unfold-equivalent@@ss".
\end{proof}

\begin{remark}\label{remark:quotient-monads}
By \Cref{statement:ue-is-congruence-2,statement:ue-is-congruence}, the operations $\flatten$ and $\atomic$ can be considered on classes of "set-systems" modulo "unfold-equivalence@@ss", equipping the quotient of the "set-system functor" with a monad structure. Restricting to "systems" equips the "regular-tree functor" with this structure, called the ""monad of regular-trees"".
\end{remark}

\subsection{Set-contexts, contexts, and their decomposition}\label{subsection:contexts}


A special form of composition, which is derived from $\flatten$ is the notion of "set-contexts", which are "set-systems" with a "hole" that can be filled by a "set-system".
We also explain in this section how "set-contexts" can be decomposed into "pieces@@context" which can be recomposed into an "unfold-equivalent@@ss" "set-context" (see \Cref{lemma:context-decomposition}).
  
\begin{figure}[h]
	\centering{\def\svgwidth{\linewidth}}
	\scalebox{.7}{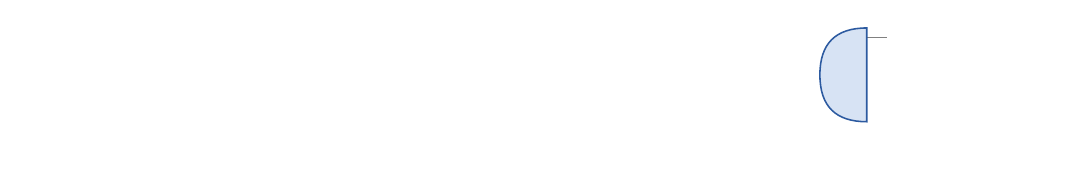}
	\caption{A "context" $C[\hole_2]$, a "system"~$S$, and the result of combining them into $C[S]$.}
	\label{fig:context}
\end{figure}

\AP A ""$\alphabet$-set-context"" with a ""$k$-hole"" is a "$(\alphabet\uplus\{\hole_k\})$-set-system", in which $\intro*\hole_k$ is a new symbol of "rank" $k$, called the ""hole symbol"", and which "labels" exactly one "vertex", called the ""hole vertex"". This is illustrated in \Cref{fig:context}. 
"Set-contexts" are denoted $C[\hole_k]$, $D[\hole_k]$, \dots\, or sometimes simply $C,D,\dots$, the hole being left implicit.
A "set-context" is ""closed@@context"" if it is of "rank@@ss"~$0$. \phantomintro{closed set-context}A "set-context" will be called a ""context"" if it is a "$(\alphabet\uplus\{\hole_k\})$-system".

\AP Given a "$\alphabet$-set-context"~$C[\hole_k]$ and a "$\alphabet$-set-system"~$S$ of "rank@@ss"~$k$, $C[S]$ denotes the "$\alphabet$-set-system" obtained by substituting $S$ for the "hole". This is illustrated in \Cref{fig:context}.
Formally, this amounts to construct the "(set-system)-set-system" $S'$ of the "same shape@@ss" as~$C$ such that $S'(h)=S$ for $h$ the "hole vertex" of~$C$, and $S'(s)=\atomic(C(s))$ for all other "vertices"~$s\in\V_C$. We then define~$C[S]$ to be~$\flatten(S')$.

Note in particular that this substitution follows the ``rules of "flattening"'', which means that (1) if the "hole vertex" is "initial@@vertex" in~$C[\hole_k]$, then all "initial vertices" of~$S$ remain "initial@@vertex" in $C[S]$, (2) all the "root vertices" of~$S$ remain "root vertices" in~$C[S]$, and (3) if the "hole vertex" is "root@@vertex" in~$C[\hole_k]$, then all "initial vertices" of~$S$ become "root vertices" in $C[S]$.

Depending on the situation, we shall use $\flatten$ or "[set-]contexts". The following technical lemma gives a recipe for transferring properties from one presentation to another. 
\begin{lemma}\label{statement:context-to-flatten}
	For all reflexive relations~${\mathcal R}$ between "set-systems" of same "rank@@ss", and all reflexive and transitive relations ${\mathcal R'}$ between "set-systems" of same "rank@@ss", the following assertions are equivalent:
	\begin{itemize}
	\item For all "set-contexts"~$C[\hole_m]$, and all~$S,S'$ "set-systems" of "rank@@ss"~$m$ such that $S\mathrel{{\mathcal R}}S'$, we have $C[S]\mathrel{{\mathcal R'}} C[S']$.
	\item For all "(set-system)-set-systems"~$S,S'$ of the "same shape@@ss" such that $S(s)\mathrel{{\mathcal R}} S'(s)$ for all "vertices"~$s\in\V_S$, we have $\flatten(S)\mathrel{{\mathcal R'}}\flatten(S')$.
	\end{itemize}
	The same holds for "contexts" and "systems" instead of "set-contexts" and "set-systems".
\end{lemma}
\begin{proof}
	The upward direction is straightforward, simply using the definition of~$C[S]$ as a "flattening".

	For the downward direction, let us first introduce a notation.
	Let~$S$ be some "(set-system)-set-system", $s\in\V_S$, and $k=\rk(S(s))$.
	Define~$S/s[\hole_{k}]$ to be the "set-context" $\flatten(S[s\leftarrow \atomic(\hole_{k})])$.
	In other words, this is $\flatten(S)$ in which the whole part that should have arisen from~$S(s)$ is a "hole" of suitable rank.
	With this definition, we naturally have that $\flatten(S)$ and $S/s[S(s)]$ are isomorphic.
	
	Let us now assume that the first item holds, and that~$S$ and $S'$ are of the "same shape@@ss" and such that $S(s)\mathrel{{\mathcal R}} S'(s)$ for all~$s\in\V_S$.
	Let us fix an enumeration $s_1,\dots,s_n$ of~$\V_S$.
	Define~$S_i$ for all~$i\in\{0,\dots, n\}$ as a "(set-system)-set-system" that has the "shape@@ss" of~$S$, and such that $S_i(s_j)=S'(s_j)$ if~$j\leq i$, and $S_i(s_j)=S(s_j)$ otherwise. Clearly, we have~$S_0=S$, and $S_n=S'$. 

	Now, for all~$1\leq i\leq n$, we have that $S_{i-1}/s_i=S_i/s_i$.
	Hence
	\begin{align*}
		\flatten(S_{i-1})=S_{i-1}/s_i[S_{i-1}(s_i)]\mathrel{{\mathcal R'}} S_{i-1}/s_i[S_{i}(s_i)]=S_i/s_i[S_{i}(s_i)] = \flatten(S_i)\ ,
	\end{align*}
	which by induction yields~$\flatten(S)\mathrel{{\mathcal R'}} \flatten(S')$.
\end{proof}

In the following, we will use ""point"" to refer to either a "vertex" or a "variable" of a "set-system".

To account for a "hole" that is itself "initial@@vertex" or "root@@vertex", or is one of its own successors, we use the following extension only for the pieces from which a "set-context" is constructed.
The phenomenon is analogous\footnote{In some sense it is a generalisation of the treatment of $\varepsilon$ in the context of infinite words.} to accounting for the empty word $\varepsilon$ in the context of infinite words: it is sometimes necessary, but is treated as a special case to avoid considering $\varepsilon^\omega$.

\begin{definition}[extended set-systems of rank~$1$]\AP\label{definition:extended-set-systems}
	\AP An ""extended set-system"" of "rank@@ss"~$1$ over~$\alphabet$ has the same components as an ordinary "$\alphabet$-set-system" of "rank@@ss"~$1$, except that its sets of ""initial points"" and ""root points"" may contain the "variable"~$x_1$:
	\[
		\Vinit_S,\Vroot_S\subseteq\V_S\uplus\{x_1\}.
	\]
	An ordinary "set-system" is the special case in which neither set contains~$x_1$. 
\end{definition}

In other words, the "extended set-system" $\varepsilon=(\emptyset,\{x_1\},\emptyset,\emptyset,\emptyset)$ represents an immediate exit through~$x_1$, and $\varepsilon\sssum S$ adds this initial possibility to an ordinary "set-system"~$S$. The analogous root possibility is represented by $x_1\in\Vroot_S$. Disjoint union extends by taking the union of the initial and root sets.

\AP Given $n+1$ "extended set-systems"~$S_0,S_1,\dots,S_n$ of "rank@@ss"~$1$ over~$\alphabet$, define $\intro*\Context(S_0,S_1,\dots,S_n)$ to be the following "closed@@system" "$\alphabet$-set-context" with a "$n$-hole":
	\begin{center}
		\scalebox{.7}{
\begingroup%
  \makeatletter%
  \providecommand\color[2][]{%
    \errmessage{(Inkscape) Color is used for the text in Inkscape, but the package 'color.sty' is not loaded}%
    \renewcommand\color[2][]{}%
  }%
  \providecommand\transparent[1]{%
    \errmessage{(Inkscape) Transparency is used (non-zero) for the text in Inkscape, but the package 'transparent.sty' is not loaded}%
    \renewcommand\transparent[1]{}%
  }%
  \providecommand\rotatebox[2]{#2}%
  \newcommand*\fsize{\dimexpr\f@size pt\relax}%
  \newcommand*\lineheight[1]{\fontsize{\fsize}{#1\fsize}\selectfont}%
  \ifx\svgwidth\undefined%
    \setlength{\unitlength}{173.27171746bp}%
    \ifx\svgscale\undefined%
      \relax%
    \else%
      \setlength{\unitlength}{\unitlength * \real{\svgscale}}%
    \fi%
  \else%
    \setlength{\unitlength}{\svgwidth}%
  \fi%
  \global\let\svgwidth\undefined%
  \global\let\svgscale\undefined%
  \makeatother%
  \begin{picture}(1,0.49652913)%
    \lineheight{1}%
    \setlength\tabcolsep{0pt}%
    \put(0,0){\includegraphics[width=\unitlength,page=1]{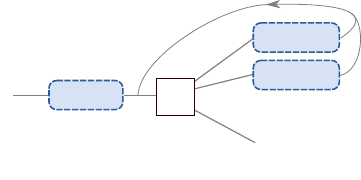}}%
    \put(0.21122453,0.21597132){\color[rgb]{0,0,0}\makebox(0,0)[lt]{\lineheight{1.25}\smash{\begin{tabular}[t]{l}$S_0$\end{tabular}}}}%
    \put(0,0){\includegraphics[width=\unitlength,page=2]{nfcontext.pdf}}%
    \put(0.79974469,0.0875688){\color[rgb]{0,0,0}\makebox(0,0)[lt]{\lineheight{1.25}\smash{\begin{tabular}[t]{l}$S_n$\end{tabular}}}}%
    \put(0,0){\includegraphics[width=\unitlength,page=3]{nfcontext.pdf}}%
    \put(0.79974469,0.26936262){\color[rgb]{0,0,0}\makebox(0,0)[lt]{\lineheight{1.25}\smash{\begin{tabular}[t]{l}$S_2$\end{tabular}}}}%
    \put(0.79974469,0.3732458){\color[rgb]{0,0,0}\makebox(0,0)[lt]{\lineheight{1.25}\smash{\begin{tabular}[t]{l}$S_1$\end{tabular}}}}%
  \end{picture}%
\endgroup%
}
	\end{center}
Formally, take disjoint copies of the "vertices" of $S_0,\dots,S_n$ and add a fresh "vertex"~$h$ "labelled"~$\hole_n$. The resulting "set-context"~$C$ has:
\begin{itemize}
\item the "initial vertices" of~$S_0$, together with~$h$ if $x_1\in\Vinit_{S_0}$;
\item all the "root vertices" of~$S_0,\dots,S_n$, together with~$h$ if $x_1\in\Vroot_{S_i}$ for at least one~$i$;
\item all "transition edges" of the pieces, and an edge $(v,d,h)$ in place of every "variable edge" $(v,d,x_1)$ of a piece;
\item for each $i\in\interval n$, an edge $(h,i,v)$ to every "initial vertex"~$v$ of~$S_i$, and the edge $(h,i,h)$ if $x_1\in\Vinit_{S_i}$.
\end{itemize}
Here "initial vertices" and "root vertices" of a piece refer to actual nodes in~$\V_{S_i}$; the occurrences of~$x_1$ are handled separately as specified above. In particular, $C$ is an ordinary "set-context" of "rank@@ss"~$0$.

\AP Given a "closed@@context" "set-context"~$C[\hole_n]$ of "hole vertex"~$h$, define its ""pieces@@context"" $P_0,P_1,\dots,P_n$ as follows:	
\begin{itemize}
\item For every $i\in\{0,\dots,n\}$, $P_i$ is an "extended set-system" of "rank@@ss"~$1$ whose "vertices" are $\V_C\setminus\{h\}$, with the same "labelling" as~$C$. Its "edges" are those of~$C$ that do not start at~$h$, with every edge $(v,d,h)$ replaced by the "variable edge" $(v,d,x_1)$. Its "root points" are $(\Vroot_C\setminus\{h\})\cup\{x_1\mid h\in\Vroot_C\}$.
\item The "initial points" of~$P_0$ are $(\Vinit_C\setminus\{h\})\cup\{x_1\mid h\in\Vinit_C\}$.
\item For $i\in\interval n$, the "initial points" of~$P_i$ are $\{v\in\V_C\setminus\{h\}\mid(h,i,v)\in\Edges_C\}\cup\{x_1\mid(h,i,h)\in\Edges_C\}$.
\end{itemize}
\begin{lemma}[decomposition and recomposition of set-contexts]\label{lemma:context-decomposition}
	Let~$(P_0,P_1,\dots,P_n)$ be the "pieces@@context" of a "closed set-context"~$C[\hole_n]$, then $C$ and $\Context(P_0,P_1,\dots,P_n)$ are "unfold-equivalent@@ss".
\end{lemma}
\begin{proof}
	The map from~$\Context(P_0,P_1,\dots,P_n)$ to~$C$ sends its "hole vertex" to~$h$ and each copied "vertex" to its original in~$C$. It is surjective because~$P_0$ contains every "vertex" of~$C$ other than~$h$. The definition of the "pieces@@context" preserves the initial and root sets, including the possible membership of~$h$, and gives exactly the outgoing "edges" of~$C$ at each image "vertex". Thus this map is "locally surjective@@ss" and hence a "folding@@ss". By \Cref{definition:set-unfold}, $C$ and $\Context(P_0,P_1,\dots,P_n)$ are "unfold-equivalent@@ss".
\end{proof}

\subsection{Yields and flattening}
\label{subsection:flatten-resolution}

We start with an example showing why the definition of "yields" allow for a renaming of variables, for sound composition with "set-contexts" and "flattening". We then study the general relationship between "yields" and "flattening".

The variable maps in "yields" allow different occurrences of the same exit to be connected independently when composing "set-systems". Here is why this needed:
\begin{example}\label{example:direct-resolutions-not-compositional}
Let $C[\hole_1]=\hole_1(b\sssum c)$ and $S=a_2(x_1,x_1)$, with $b,c$ of rank~$0$ and $a_2$ of rank~$2$.
\begin{center}
\scalebox{.8}{
\begingroup%
  \makeatletter%
  \providecommand\color[2][]{%
    \errmessage{(Inkscape) Color is used for the text in Inkscape, but the package 'color.sty' is not loaded}%
    \renewcommand\color[2][]{}%
  }%
  \providecommand\transparent[1]{%
    \errmessage{(Inkscape) Transparency is used (non-zero) for the text in Inkscape, but the package 'transparent.sty' is not loaded}%
    \renewcommand\transparent[1]{}%
  }%
  \providecommand\rotatebox[2]{#2}%
  \newcommand*\fsize{\dimexpr\f@size pt\relax}%
  \newcommand*\lineheight[1]{\fontsize{\fsize}{#1\fsize}\selectfont}%
  \ifx\svgwidth\undefined%
    \setlength{\unitlength}{403.44305564bp}%
    \ifx\svgscale\undefined%
      \relax%
    \else%
      \setlength{\unitlength}{\unitlength * \real{\svgscale}}%
    \fi%
  \else%
    \setlength{\unitlength}{\svgwidth}%
  \fi%
  \global\let\svgwidth\undefined%
  \global\let\svgscale\undefined%
  \makeatother%
  \begin{picture}(1,0.13552828)%
    \lineheight{1}%
    \setlength\tabcolsep{0pt}%
    \put(0,0){\includegraphics[width=\unitlength,page=1]{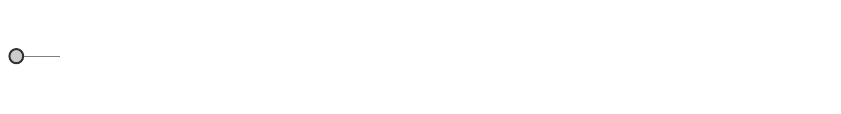}}%
    \put(-0.00185173,0.09648627){\color[rgb]{0,0,0}\makebox(0,0)[lt]{\lineheight{1.25}\smash{\begin{tabular}[t]{l}$C[\hole_1]$\end{tabular}}}}%
    \put(0,0){\includegraphics[width=\unitlength,page=2]{counter_ex_compose.pdf}}%
    \put(0.47368405,0.06121466){\color[rgb]{0,0,0}\makebox(0,0)[lt]{\lineheight{1.25}\smash{\begin{tabular}[t]{l}$a_2$\end{tabular}}}}%
    \put(0,0){\includegraphics[width=\unitlength,page=3]{counter_ex_compose.pdf}}%
    \put(0.53108111,0.10380589){\color[rgb]{0,0,0}\makebox(0,0)[lt]{\lineheight{1.25}\smash{\begin{tabular}[t]{l}$x_1$\end{tabular}}}}%
    \put(0.3886127,0.09355691){\color[rgb]{0,0,0}\makebox(0,0)[lt]{\lineheight{1.25}\smash{\begin{tabular}[t]{l}$S$\end{tabular}}}}%
    \put(0,0){\includegraphics[width=\unitlength,page=4]{counter_ex_compose.pdf}}%
    \put(0.24701287,0.08645342){\color[rgb]{0,0,0}\makebox(0,0)[lt]{\lineheight{1.25}\smash{\begin{tabular}[t]{l}$b$\end{tabular}}}}%
    \put(0,0){\includegraphics[width=\unitlength,page=5]{counter_ex_compose.pdf}}%
    \put(0.24701287,0.03581032){\color[rgb]{0,0,0}\makebox(0,0)[lt]{\lineheight{1.25}\smash{\begin{tabular}[t]{l}$c$\end{tabular}}}}%
    \put(0,0){\includegraphics[width=\unitlength,page=6]{counter_ex_compose.pdf}}%
    \put(0.53108111,0.02200994){\color[rgb]{0,0,0}\makebox(0,0)[lt]{\lineheight{1.25}\smash{\begin{tabular}[t]{l}$x_1$\end{tabular}}}}%
    \put(0,0){\includegraphics[width=\unitlength,page=7]{counter_ex_compose.pdf}}%
    \put(0.74881608,0.06121466){\color[rgb]{0,0,0}\makebox(0,0)[lt]{\lineheight{1.25}\smash{\begin{tabular}[t]{l}$a_2$\end{tabular}}}}%
    \put(0,0){\includegraphics[width=\unitlength,page=8]{counter_ex_compose.pdf}}%
    \put(0.66374469,0.09355691){\color[rgb]{0,0,0}\makebox(0,0)[lt]{\lineheight{1.25}\smash{\begin{tabular}[t]{l}$C[S]$\end{tabular}}}}%
    \put(0,0){\includegraphics[width=\unitlength,page=9]{counter_ex_compose.pdf}}%
    \put(0.95552509,0.09980925){\color[rgb]{0,0,0}\makebox(0,0)[lt]{\lineheight{1.25}\smash{\begin{tabular}[t]{l}$b$\end{tabular}}}}%
    \put(0,0){\includegraphics[width=\unitlength,page=10]{counter_ex_compose.pdf}}%
    \put(0.95748264,0.01630672){\color[rgb]{0,0,0}\makebox(0,0)[lt]{\lineheight{1.25}\smash{\begin{tabular}[t]{l}$c$\end{tabular}}}}%
    \put(0,0){\includegraphics[width=\unitlength,page=11]{counter_ex_compose.pdf}}%
  \end{picture}%
\endgroup%
}
\end{center}
Choosing either $b$ or $c$ as a single successor of the hole produces $a_2(b,b)$ or $a_2(c,c)$, but misses the closed "yields" $a_2(b,c)$ and $a_2(c,b)$ of~$C[S]$.
The "init-yield" $(a_2(x_1,x_2),\sigma)$ of~$S$, where $\sigma(1)=\sigma(2)=1$, keeps its two exits separate, so that they can be connected to~$b$ and~$c$ independently. 
\end{example}

The preceding examples show that thanks to the renaming maps in "yields", the "yields" of a composite "set-system" can be obtained by composing the "yields" of the components.

The following lemmas describe how "init-yields" indeed commute with the "flattening" operation. \Cref{lemma:yield-composition} states that if a "($\alphabet$-system)-system" $T$ has labels that are "yields" of a "($\alphabet$-set-system)-set-system" $S$, then $\flatten(T)$ is a "yield" of $\flatten(S)$.

\begin{lemma}[composition of yields]\label{lemma:yield-decomposition}\label{lemma:yield-composition}
Let~$S$ be a "($\alphabet$-set-system)-set-system" of rank~$n$, let~$T$ be a "($\alphabet$-system)-system" of rank~$m$, and let $\tau\colon\interval m\to\interval n$.
Suppose there is a map $\delta\colon\V_T\to\V_S$ and there are maps $\sigma_t\colon\interval{\rk(T(t))}\to\interval{\rk(S(\delta(t)))}$ such that:
\begin{enumerate}
\item $\delta(\Vinit_T)\subseteq\Vinit_S$;
\item $(T(t),\sigma_t)\in\InitYields(S(\delta(t)))$ for every~$t\in\V_T$;
\item $(\delta(t),\sigma_t(i),\delta(t'))\in\Edges_S$ for every "transition edge" $(t,i,t')\in\Edges_T$;
\item $(\delta(t),\sigma_t(i),x_{\tau(j)})\in\Edges_S$ for every "variable edge" $(t,i,x_j)\in\Edges_T$.
\end{enumerate}
Then $(\flatten(T),\tau)\in\InitYields(\flatten(S))$.
\end{lemma}
\begin{proof}
For each~$t$, choose a witness consisting of a "system"~$U_t$, a "folding@@ss" $\gamma_t:U_t\ssfolding T(t)$, and a "morphism@@ss" $\iota_t:\rename{\sigma_t}(U_t)\ssmorphism S(\delta(t))$. 
Let $T_U$ be the "system-system" obtained from $T$ by replacing the label~$T(t)$ by~$U_t$ at every vertex $t$ of $T$. Its labels have the required ranks because $\gamma_t$ is a "folding". By \Cref{lemma:flatten-systems,lemma:inner-morphism-ok}, $\flatten(T_U)$ is a "system", and
\[
  (t,u)\longmapsto(t,\gamma_t(u))
  \quad\text{defines a "folding@@ss"}\quad
  \flatten(T_U)\ssfolding\flatten(T).
\]
Define
\[
  \eta:\rename\tau(\flatten(T_U))\ssmorphism\flatten(S),
  \qquad \eta(t,u)=(\delta(t),\iota_t(u)),
\]
This map preserves labels. We show that $\eta$ is a "morphism@@ss". An "initial vertex" has an initial outer component~$t$, whose image is initial by condition~1, and an initial inner component, whose image is initial because $\iota_t$ is a "morphism@@ss". There are no "root vertices" in the source.

For edges inside~$U_t$, preservation follows from~$\iota_t$ and the intra-subsystem rule of \Cref{definition:flattening}. An edge passing from~$U_t$ through its variable~$x_i$ to an "initial vertex" of~$U_{t'}$ maps to an inter-subsystem edge: $\iota_t$ sends the exit to~$x_{\sigma_t(i)}$, condition~3 supplies the outer edge $(\delta(t),\sigma_t(i),\delta(t'))$, and $\iota_{t'}$ preserves "initial vertices". Finally, if the outer edge is $(t,i,x_j)$, condition~4 supplies the edge to~$x_{\tau(j)}$ needed after renaming by~$\tau$. Thus~$\eta$ is a "morphism@@ss". Together with the "folding" $\flatten(T_U)\ssfolding\flatten(T)$, it witnesses $(\flatten(T),\tau)\in\InitYields(\flatten(S))$ by \Cref{definition:yields}.
\end{proof}

In the other direction, \Cref{lemma:yield-decomposition-converse} shows that if $S$ is a "($\alphabet$-set-system)-set-system" and $R$ is a "yield" of $S$, then $R$ unfolds to $\flatten(T)$ where $T$ is a "($\alphabet$-system)-system" whose labels are "yields" of the labels of $S$.

\begin{lemma}[decomposition of yields]\label{lemma:yield-decomposition-converse}
Let~$S$ be a "($\alphabet$-set-system)-set-system" of rank~$n$ and $(R,\tau)\in\InitYields(\flatten(S))$, where $\tau:\interval m\to\interval n$.
There are a "($\alphabet$-system)-system"~$T$ of rank~$m$, a map~$\delta$, and maps~$(\sigma_t)_t$ satisfying the four conditions of \Cref{lemma:yield-composition}, together with a "folding@@ss" $\flatten(T)\ssfolding R$. Moreover, for every~$t$, the witness for $(T(t),\sigma_t)\in\InitYields(S(\delta(t)))$ is $(T(t),\sigma_t)$ itself, i.e. there is a "morphism@@ss" $\rename{\sigma_t}(T(t))\ssmorphism S(\delta(t))$.
\end{lemma}

\begin{proof}
Choose a witness "system"~$U$, a "folding@@ss" $\gamma:U\ssfolding R$, and a "morphism@@ss" $\eta:\rename\tau(U)\ssmorphism\flatten(S)$. Write $\eta(u)=(s(u),v(u))$.
For every edge of~$U$, fix one witness for its image, referring to the types of edges in $\flatten(S)$ according to \Cref{definition:flattening}. A "transition edge" $(u,d,u')$ is assigned either an intra-subsystem witness, with $s(u)=s(u')$ and $(v(u),d,v(u'))\in\Edges_{S(s(u))}$, or an inter-subsystem witness, with a chosen exit~$e$ such that
\[
 (v(u),d,x_e)\in\Edges_{S(s(u))},\quad
 (s(u),e,s(u'))\in\Edges_S,\quad v(u')\in\Vinit_{S(s(u'))}.
\]
For a "variable edge" $(u,d,x_j)$, fix an exit~$e$ with
\[
 (v(u),d,x_e)\in\Edges_{S(s(u))},\qquad
 (s(u),e,x_{\tau(j)})\in\Edges_S.
\]
Such choices exist because~$\eta$ is a "morphism@@ss".

Let~$u_0$ be the "initial vertex" of~$U$, and let~$E$ consist of~$u_0$ and the targets of the edges assigned inter-subsystem witnesses. For each~$t\in E$, construct a "system"~$Q_t$ whose vertices are copies of precisely those vertices reachable from~$t$ along the chosen intra-subsystem edges. Its "initial vertex" is the copy of~$t$, and it has no roots. Keep its intra-subsystem edges. Replace each remaining outgoing edge at a retained vertex by a distinct "variable edge", numbering these new variables $x_1,\ldots,x_{k_t}$. All vertices and all these variables are reachable, and every direction still has exactly one successor, so~$Q_t$ is a "system".
The outer component~$s(u)$ is constant on these vertices. Set $\delta(t)=s(t)$, and let $\sigma_t(i)$ be the exit~$e$ chosen for the edge replaced by~$x_i$. The map sending the copy of~$u$ to~$v(u)$ is a "morphism@@ss" $\rename{\sigma_t}(Q_t)\ssmorphism S(\delta(t))$: retained edges use the intra-subsystem witnesses, new "variable edges" use their chosen exits, and the "initial vertex" maps to an "initial vertex" of the local subsystem. For~$u_0$ this follows from initial preservation by~$\eta$, and for any other entry from its inter-subsystem witness. Thus $(Q_t,\sigma_t)$ is an "init-yield", witnessed by~$Q_t$ and its identity "folding".

Construct~$T$ with vertex set~$E$, "initial vertex"~$u_0$, no roots, and label~$Q_t$ at~$t$. For each exit~$x_i$ of~$Q_t$, put an outer edge $(t,i,u')$ if the cut edge ended at an entry~$u'$, or $(t,i,x_j)$ if it ended at the original variable~$x_j$ of~$U$. These give exactly one successor in each direction. Every path in~$U$ is a concatenation of intra-subsystem segments and chosen crossing edges, so every entry and every original variable is reachable in~$T$. Hence~$T$ is a "system-system" of rank~$m$. The initial preservation of~$\eta$ gives condition~1, the preceding local "morphisms@@ss" give condition~2, and the chosen inter-subsystem and variable witnesses give conditions~3 and~4.

Finally, define $\pi:\flatten(T)\ssfolding U$ by sending a copied vertex $(t,u)$ to its original~$u$. An intra-subsystem edge is unchanged, and crossing an exit restores exactly the edge that was cut, so~$\pi$ is a "morphism@@ss". By \Cref{lemma:flatten-systems}, its source is a "system", so~$\pi$ is a "folding@@ss" by \Cref{remark:system-morphisms}. Composing with $\gamma$ gives the required "folding" $\flatten(T)\ssfolding R$.
\end{proof}

We finally use the two preceding lemmas to show that the operation of combining "set-systems" can be considered up to "yield-equivalence".

\begin{corollary}["yield-equivalence" is a congruence for the "flattening" functor]\label{statement:re-is-congruence}
Let~$S,S'$ be "($\alphabet$-set-system)-set-systems" of the "same shape@@ss" such that $S(s)\yieldssubseteq S'(s)$ (resp. $S(s)\yieldseq S'(s)$) for every~$s$. Then $\flatten(S)\yieldssubseteq\flatten(S')$ (resp. $\flatten(S)\yieldseq\flatten(S')$).
\end{corollary}
\begin{proof}
Decompose an "init-yield" $(R,\tau)$ of~$\flatten(S)$ using \Cref{lemma:yield-decomposition-converse}. Every local "yield" $(T(t),\sigma_t)$ is also an "init-yield" of~$S'(\delta(t))$. By \Cref{lemma:yield-composition}, the composition assertion gives $(\flatten(T),\tau)\in\InitYields(\flatten(S'))$. The "folding" $\flatten(T)\ssfolding R$ then gives $(R,\tau)\in\InitYields(\flatten(S'))$ by \Cref{lemma:yield-unfold-invariance}.
For "root-yields" apply the same argument to $\fuproot(S)$ and $\fuproot(S')$, using \Cref{statement:fuproot}. At a vertex $(0,s)$ the local "init-yields" are the "root-yields" of~$S(s)$; at $(1,s)$ they are its "init-yields". Thus, the required local inclusions hold in both cases. Equality follows by symmetry.
\end{proof}
\begin{remark}\label{remark:monad-ss-modulo-yields}
The endofunctor of "ranked sets" that builds "yield-equivalence" classes of "set-systems", equipped with the $\flatten$ and $\atomic$ natural transformations, is not a monad.
It was shown in particular in~\cite{Blum23} that it cannot be described using a distributive law with the powerset monad.
\end{remark}

We retain the following finite-witness property of "yields".
\begin{corollary}\label{corollary:yield-continuous}
For every "set-context"~$C[\hole_n]$ and "set-system"~$S$ of rank~$n$,
\[
\Yields(C[S])=\bigcup_{\substack{F\subseteq\Yields(S)\\F\text{ finite}}}
\Yields\left(C\left[\ssSum_{q\in F}\yieldss(q)\right]\right),
\]
where $\yieldss(q)$ is as defined in \Cref{definition:yields}.
\end{corollary}
\begin{proof}
Decompose a "yield" of~$C[S]$ by \Cref{lemma:yield-decomposition-converse}, using \Cref{statement:fuproot} for a "root-yield". Only finitely many local "yields" of~$S$ occur at copies of the hole; let~$F$ collect them, keeping their initial or root status. Each local pair remains a "yield" of the corresponding summand, witnessed by its "system" component itself. Composition therefore recovers the original "yield" in $C[\ssSum_{q\in F}\yieldss(q)]$.
Conversely, $\yieldss(q)\yieldssubseteq S$ for every $q\in\Yields(S)$: a witness~$U\ssfolding T$ for $q=(T,\sigma)$ gives an "unfolding@@ss" $\yieldss(U,\sigma)$ of~$\yieldss(q)$ and a "morphism@@ss" $\yieldss(U,\sigma)\ssmorphism S$, so \Cref{statement:locsurj->res-eq} applies. Applying \Cref{statement:re-is-congruence} to the outer "set-system" with one "initial vertex"~$v_q$ for each $q\in F$, no roots, and the edges $(v_q,i,x_i)$ for all $q\in F$ and $i\in\interval n$ gives
\[
\ssSum_{q\in F}\yieldss(q)
\yieldssubseteq\ssSum_{q\in F}S
\yieldssubseteq S,
\]
where the last inclusion follows from \Cref{statement:locsurj->res-eq} applied to the "morphism@@ss" identifying the copies of~$S$.
Using again \Cref{statement:re-is-congruence} for substitution into $C$ allows us to conclude.
\end{proof}

\section{Algebras}\label{section:algebras}

We introduce the notion of "algebra" in \Cref{subsection:definition-algebras},
and explain how "algebras" can be used to "recognise" "languages of systems" in \Cref{subsection:recognition}.

\subsection{Algebras}
\label{subsection:definition-algebras}

We give the "algebra" axioms, describe their "morphisms@@algebra", and define "free algebras", and "regular-tree algebras" (\Cref{definition:algebras}). We then compare evaluation through "contexts" and "flattening" (\Cref{remark:context-to-flatten}), and introduce the neutral element of "extended algebras" (\Cref{definition:extended-algebras}). \Cref{ex:odd,ex:aexists-alg} show a "non-unfold-invariant" "algebra" and an "unfold-invariant" one.

The following definitions follow the standard approach via monads.
\begin{definition}[algebras]\label{definition:algebras}
	An ""algebra"" $\Alg$ is a "ranked set" $A$ together with a "map of ranked sets" $\intro*\aeval$ (called the ""evaluation@@algebra"") from "$A$-systems" to $A$, such that
	\begin{itemize}
	\item $\aeval(\atomic(a))=a$, for all~$a\in A$,
	\item $\aeval(\slifteval(S))=\aeval(\flatten(S))$ for all "($A$-system)-systems"~$S$.
	\end{itemize} 
	\AP A ""morphism@@algebra"" from an "algebra" $\Alg=(A,\aeval)$ to an "algebra"~$\Alg'=(A',\aeval')$ is a "map@@rs"~$\rho$ from~$A$ to~$A'$
	such that $\aeval'(\slift\rho(S))=\rho(\aeval(S))$ for all "$A$-systems"~$S$.
	
	\smallskip
	\noindent \AP An "algebra" is ""unfold-invariant"" if $\aeval(S) = \aeval(S')$ for all "unfold-equivalent" "$A$-systems"~$S,S'$.
	\AP "Unfold-invariant" "algebras" are called ""regular-tree algebras"".
	
	\smallskip
	\noindent \AP The "ranked set" of "$\alphabet$-systems" equipped with $\flatten$ as "evaluation" is an "algebra" called the ""free algebra generated by~$\alphabet$"", or the \reintro*"$\alphabet$-free-algebra". 
\end{definition}
\begin{remark}\label{remark:context-to-flatten}
	A consequence of \Cref{statement:context-to-flatten} is that, in the definition of "algebras", one could safely exchange the second identity for 
	\begin{itemize}
	\item $\aeval(C[S])=\aeval(C[\aeval(S)])$ for all "$A$-contexts"~$C[\hole_m]$ and all "$A$-systems"~$S$ of "rank@@ss"~$m$.
	\end{itemize}
\end{remark}

\begin{definition}[extended algebras]\AP\label{definition:extended-algebras}
The ""extended algebra"" associated with an "algebra" $\Alg=(A,\aeval)$ is its formal enlargement $\Alg^+$, with
\[
A^+_1=A_1\uplus\{\varepsilon\},\qquad A^+_n=A_n\quad(n\neq1).
\]
The new ""neutral element""~$\varepsilon$ is used as a "rank@@ss"-$1$ context piece. It represents the vertex-free "extended set-system"
\[
\varepsilon=(\emptyset,\{x_1\},\emptyset,\emptyset,\emptyset).
\]
Thus~$\varepsilon$ provides an immediate exit: substituting a "system"~$S$ at that exit gives~$S$.
The extended evaluation assigns~$\varepsilon$ to this piece and agrees with~$\aeval$ on ordinary "$A$-systems". For an expression assembled from these pieces, first expand occurrences of~$\varepsilon$ as immediate exits, and apply~$\aeval$ when the result is an ordinary "$A$-system". This specifies evaluation at the admissible uses of the formal extension; it does not assign values to cycles consisting only of~$\varepsilon$.
\end{definition}
A "morphism@@algebra" extends to these formal enlargements by fixing the "neutral element". If~$\Alg$ is "rankwise-finite", so is its "extended algebra", since only one element has been added in rank~$1$.


\begin{example}["non-unfold-invariant algebra"]\label{ex:odd}
	\newcommand{\even}{\mathrm{Even}}
	\newcommand{\odd}{\mathrm{Odd}}
	Consider~$\Alg:=(A,\aeval)$ defined by~$A_n=\{\even,\odd\}$ for all~$n\in\Nats$, and for all "$A$-systems"~$S$, let $\aeval(S)$ be~$\odd$ if the number of vertices $s$ such that $S(s)=\odd$ is odd, and $\even$ otherwise.
	It is easy to show that~$\Alg$ is an "algebra".
	However, it is not a "regular-tree algebra" since it is not "unfold-invariant", as shown by the following example, where all vertices are labelled $\odd$:
	\begin{center}
	\scalebox{.7}{
\begingroup%
  \makeatletter%
  \providecommand\color[2][]{%
    \errmessage{(Inkscape) Color is used for the text in Inkscape, but the package 'color.sty' is not loaded}%
    \renewcommand\color[2][]{}%
  }%
  \providecommand\transparent[1]{%
    \errmessage{(Inkscape) Transparency is used (non-zero) for the text in Inkscape, but the package 'transparent.sty' is not loaded}%
    \renewcommand\transparent[1]{}%
  }%
  \providecommand\rotatebox[2]{#2}%
  \newcommand*\fsize{\dimexpr\f@size pt\relax}%
  \newcommand*\lineheight[1]{\fontsize{\fsize}{#1\fsize}\selectfont}%
  \ifx\svgwidth\undefined%
    \setlength{\unitlength}{469.70091248bp}%
    \ifx\svgscale\undefined%
      \relax%
    \else%
      \setlength{\unitlength}{\unitlength * \real{\svgscale}}%
    \fi%
  \else%
    \setlength{\unitlength}{\svgwidth}%
  \fi%
  \global\let\svgwidth\undefined%
  \global\let\svgscale\undefined%
  \makeatother%
  \begin{picture}(1,0.20383409)%
    \lineheight{1}%
    \setlength\tabcolsep{0pt}%
    \put(0,0){\includegraphics[width=\unitlength,page=1]{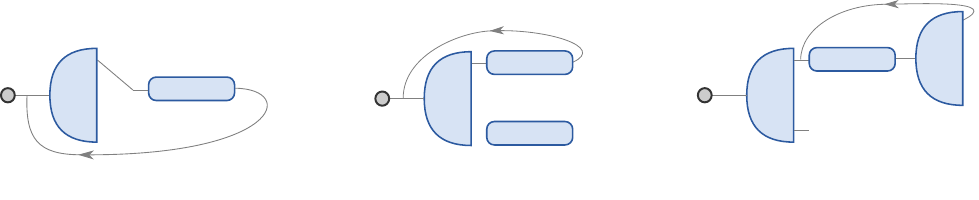}}%
    \put(0.09531301,0.01394889){\color[rgb]{0,0,0}\makebox(0,0)[lt]{\lineheight{1.25}\smash{\begin{tabular}[t]{l}{\Large$\even$}\end{tabular}}}}%
    \put(0,0){\includegraphics[width=\unitlength,page=2]{non-unfold-invariant-algebra.pdf}}%
    \put(0.82034589,0.00576899){\color[rgb]{0,0,0}\makebox(0,0)[lt]{\lineheight{1.25}\smash{\begin{tabular}[t]{l}{\Large$\even$}\end{tabular}}}}%
    \put(0.4818749,0.0050906){\color[rgb]{0,0,0}\makebox(0,0)[lt]{\lineheight{1.25}\smash{\begin{tabular}[t]{l}{\Large$\odd$}\end{tabular}}}}%
  \end{picture}%
\endgroup%
}
	\end{center}
	These three "systems" are "unfold-equivalent", but their evaluations in the "algebra" $\Alg$ are not all equal.
\end{example}
\begin{example}["unfold-invariant algebra"]\label{ex:aexists-alg}
	Consider $\Alg=(A,\aeval)$ defined by $A_n=\{0,1\}$ for every~$n$, and $\aeval(T)=1$ if some "vertex" of~$T$ is labelled~$1$, and $0$ otherwise. The unit identity is immediate, and the "flattening" identity follows because the "vertices" of a "flattening" are the disjoint union of the vertices of its subsystems.
	This "algebra" is "unfold-invariant", since a "folding@@ss" preserves the presence or absence of a vertex labelled~$1$.
\end{example}


\subsection{Languages and their recognition by algebras}
\label{subsection:recognition}

We define "languages of systems" and "regular trees@regular-tree", recognition by an "algebra", and syntactic "algebras" (\Cref{definition:syntactic-algebra}). \Cref{ex:reco-exists-lang,ex:reco-all-paths,example:synt-algebra} illustrate these definitions. Relabelling operations on languages preserve "unfold-invariance" as specified in \Cref{lemma:regular-tree-languages}; \Cref{statement:recognition-closure} establishes closure of "rankwise-finite" recognisability under Boolean operations, "cylindrification", and projection.

We call a ""language of systems"" (or simply a \reintro*"language") over an alphabet~$\alphabet$ a set of "closed@@system" "$\alphabet$-systems".
 A ""language of $\alphabet$-regular-trees"" is a "language of $\alphabet$-systems"~$L$ which is invariant under "unfold-equivalence" (i.e.\ such that if~$T,T'$ are "unfold-equivalent" and $T\in L$, then $T'\in L$, for all "$\alphabet$-systems" $T,T'$). 
  
\AP Let $\rho$ be a "morphism@@algebra" (called the ""recognising morphism"") from the "$\alphabet$-free algebra" to an "algebra"~$\Alg$,  and $P\subseteq A_0$ (called the ""accepting set""), the ""language recognised"" by $(\Alg,\rho,P)$ is the set of "closed@@system" "$\alphabet$-systems" defined as\phantomintro\LanguageRec
\begin{align*}
	\reintro*\LanguageRec(\Alg,\rho,P)&:=\{T\mid \text{$T$ "closed@@system" "$\alphabet$-system" such that~$\rho(T)\in P$}\}.
\end{align*}

We say that $L$ is "recognised" by $\Alg$ if it is recognised by $(\Alg,\rho,P)$ for some $\rho$ and $P$.

This is the standard definition of recognition, as for word languages. The only subtlety is that we focus on "closed systems", which is reflected in the fact that $P\subseteq A_0$. 
As in the case of word languages, we can think of $\rho$ as simply mapping each symbol of~$\alphabet$ to an element of the "algebra" (of same rank here), and then extending this mapping to all "closed@@system" "$\alphabet$-systems" by the axioms of "morphisms@@algebra", via $\aeval$.

\begin{lemma}\label{lemma:regular-tree-languages}
	The "languages of systems" "recognised" by "unfold-invariant" "algebras" are "languages of regular-trees". 
\end{lemma}
\begin{proof}
	Let~$T$ and $T'$ be "unfold-equivalent" "$\alphabet$-systems".
	Then $\slift\rho(T)$ and $\slift\rho(T')$ are "unfold-equivalent", and hence $\rho(T)=\aeval(\slift\rho(T))=\aeval(\slift\rho(T'))=\rho(T')$.
	Thus $T\in L$ if and only if~$T'\in L$, which means that $L$ is a "language of regular-trees".
\end{proof}

Quite naturally, we shall be interested in "languages" that are "recognisable" by "rankwise-finite" "algebras", i.e.\ "algebras" $\Alg=(A,\aeval)$ such that $A_n$ is finite for all~$n\in\Nats$.
\begin{example}[presence of a symbol]\label{ex:reco-exists-lang}
	Consider some "ranked alphabet"~$\alphabet$, and a set of letters $R\subseteq\alphabet$.
	Let also $\Alg$ be the "algebra" from \Cref{ex:aexists-alg}, and define $\rho$ as the unique "algebra morphism" from the "$\alphabet$-free algebra" to $\Alg$ such that 
	\begin{align*}
	\rho(a(x_1,\dots,x_k))=\begin{cases}
					1&\text{if}~a\in R\\
					0&\text{otherwise.}
					\end{cases}
	\end{align*}
	Thus $\rho(T)=1$ exactly when~$T$ contains a symbol from~$R$. The "language recognised" by $(\Alg,\rho,\{1\})$ consists of the "closed@@system" "$\alphabet$-systems" that contain a symbol from~$R$. Since this "language" is preserved under "unfold-equivalence", it is a "language of regular-trees".
	Similarly, $(\Alg,\rho,\{0\})$ "recognises" the "closed@@system" "$\alphabet$-systems" containing no symbol of~$R$.
\end{example}

\begin{example}[a symbol of $R$ on every path]\label{ex:reco-all-paths}
	Let~$R\subseteq\alphabet$, and let~$L$ consist of the "closed@@system" "$\alphabet$-systems" in which every ""branch"" contains a label from~$R$. Here a "branch" is a maximal path from the "initial vertex", it is either infinite or ends at a vertex of rank~$0$.

	Define an "algebra" $\Alg=(A,\aeval)$ with $A_n=\powerset(\interval n)\uplus\{\bot\}$. The intuition behind a subset $J\subseteq\powerset(\interval n)$ is that each element of $J$ represents a direction where we still have to verify that all branches from there contain a symbol from $R$. The element $\bot$ marks the failure of the condition. Thus, accepting "systems" are those which are mapped to $\emptyset$. \AP In an "$A$-system", a ""controlled path"" starts at the "initial vertex" and follows only edges whose direction belongs to the subset labelling their source; no edge is followed from a vertex labelled~$\bot$. For an "$A$-system"~$T$ of rank~$n$, put $\aeval(T)=\bot$ if a "controlled path" reaches~$\bot$, or if there is an infinite "controlled path". Otherwise, put
	\[
		\aeval(T)=\{i\in\interval n\mid\text{a "controlled path" reaches }x_i\}\ .
	\]
	A vertex labelled by the empty subset stops a "controlled path" successfully. The unit identity follows immediately. For the "flattening" identity, a "controlled path" decomposes into paths inside successive subsystems: a subsystem contributes the exits reachable by such paths, or~$\bot$ if it admits a failing "branch". This proves the "flattening" identity. These path properties are preserved under "unfolding", so~$\Alg$ is "unfold-invariant".

	The "recognising morphism"~$\rho$ is determined, for $a$ of rank~$k$, by
	\[
		\rho(\atomic(a))=
		\begin{cases}
		\emptyset&\text{if }a\in R,\\
		\interval k&\text{if }a\notin R\text{ and }k>0,\\
		\bot&\text{if }a\notin R\text{ and }k=0.
		\end{cases}
	\]
	For a "system"~$T$ of rank~$n$, $\rho(T)=\bot$ if there is a "branch" avoiding~$R$. Otherwise, $\rho(T)$ records the variables reachable along paths avoiding~$R$. Thus $(\Alg,\rho,\{\emptyset\})$ "recognises"~$L$.
\end{example}

\begin{definition}\label{definition:syntactic-algebra}
	\AP The ""syntactic algebra"" of a language identifies "systems" of the same rank that no "closed@@context" "context" distinguishes by membership in that language.
	That is, $\Alg$ is the "syntactic algebra" of a "language"~$L$ if $(\Alg,\rho,P)$ recognizes $L$ for some $\rho$ surjective and $P\subseteq A_0$, and for all "$\alphabet$-systems" $T,T'$ of the same rank $m$, we have $\rho(T)=\rho(T')$ if and only if for all "closed@@context" "$\alphabet$-contexts" $C[\hole_m]$, we have $C[T]\in L$ if and only if $C[T']\in L$.
\end{definition} 

\begin{example}\label{example:synt-algebra}
	For $\alphabet=\{b_2,r_1,r_0,b_0\}$ and $R=\{r_1,r_0\}$, the "algebra" from \Cref{ex:reco-all-paths} is precisely the "syntactic algebra" of~$L$. Every subset $J\subseteq\interval n$ is realised by a $b_2$-tree using every variable, with occurrences of $x_i$ placed below an $r_1$-vertex exactly when $i\notin J$; for $n=0$, use~$r_0$. The value~$\bot$ is realised by adding a "branch" ending in~$b_0$ (or by~$b_0$ itself for $n=0$). Distinct subsets are separated by a "context" with the hole initial, connecting a variable in their symmetric difference to~$b_0$ and all other variables to~$r_0$. Connecting every variable to~$r_0$ separates~$\bot$ from every subset. Hence the syntactic carrier in rank~$n$ is $\powerset(\interval n)\uplus\{\bot\}$.
\end{example}

\AP Let us now briefly describe the important closure operations enjoyed by the "languages" "recognisable" by "rankwise-finite" "algebras". These will be the justification of the translation from "monadic second-order logic" to "recognisability" by "rankwise-finite" "algebras" (see \Cref{subsection:MSO}).
\AP For this, let~$h$ be  a "map@@rs" from a "ranked set"~$\alphabet$ to a "ranked set"~$\alphabetB$. Then $\slift h$ is a map from
"$\alphabet$-systems" to~"$\alphabetB$-systems", that preserves the structure of the "systems". We define the following operations:
\begin{itemize}
\itemAP For~$L$ a "language of $\alphabetB$-systems", its ""cylindrification"" by~$h$ is the "language" of "$\alphabet$-systems"~$\slift h^{-1}(L)$.
\itemAP For~$L$ a "language of $\alphabet$-systems", its ""projection"" by~$h$ is the "language" of "$\alphabetB$-systems"~$\slift h(L)$.
\end{itemize}
\begin{remark}
	Note that "languages of regular-trees" are closed under boolean connectives and "cylindrification". 
	However, this is not the case for "projection" as illustrated now. 
	Consider the "ranked alphabet"~$\alphabet$ consisting of letters $a_2,b,c$ of respective "ranks"~$2,0,0$, and the singleton "language"~$L=\{a_2(b,c)\}$.
	It is a "language of $\alphabet$-regular-trees". Consider now the map~$h$ which maps~$a_2$ to itself, $b$ to $b$ and~$c$ to~$b$.
	Then $a_2(b,b)\in \slift h(L)$. Notice now that $a_2(b,b)$ is "unfold-equivalent" to~$S_1(b)$, in which $S_1=\rename{\sigma_{2\to 1}}(a_2)$ where $\sigma_{2\to 1}$ is the constant map from~$\interval 2$ to~$\interval 1$.
	Informally, $S_1(b)$ is a version of~$a_2(b,b)$ in which the two $b$-subtrees have been merged.
	Now $S_1(b)$ has only two antecedents by $\slift h$, namely $S_1(b)$ and $S_1(c)$, and none belongs to~$L$, so $S_1(b)\notin \slift h(L)$. This means that $\slift h(L)$ is not a "language of regular-trees", as it is not closed under "unfold-equivalence": it separates $S_1(b)$ from $a_2(b,b)$.
\end{remark}

In the above remark, the problem stems from the invariance under "unfold-equivalence" which is required for "languages of regular trees".
The following statement is not subject to this difficulty, since it considers "languages of systems" in general.
\begin{lemma}\label{statement:recognition-closure}
	"Languages" "recognisable" by "rankwise-finite" "algebras" are closed under boolean connectives, "cylindrification", and "projection".
\end{lemma}
\begin{proof}
	All the operations follow from generic algebraic arguments that are independent of the specific algebraic theory, except for "projection".
	For the "projection", it suffices to perform a rankwise powerset of the "algebra" "recognising"~$L$.
	Indeed, let us assume that $L$ is "recognised" by $(\Alg,\rho,P)$, and let $\Alg'=(A',\aeval')$ be the "algebra" defined by $A'_n=\powerset(A_n)$ for all~$n\in\Nats$, and $\aeval'(S)=\{\aeval(T)\mid T\text{ has the same shape as $S$ and }T(v)\in S(v)\text{ for all }v\in \V_S\}$, for all "$A'$-systems"~$S$.
	Then $\Alg'$ is a "rankwise-finite" "algebra". Let $\rho'$ be the "morphism of algebras" from the "$\alphabetB$-free algebra" to $\Alg'$ defined by $\rho'(a')=\{\rho(a)\mid h(a)=a'\}$. Finally, let $P'=\{X\subseteq A_0\mid X\cap P\neq\emptyset\}$.
	It now suffices to notice that for any closed "$\alphabetB$-system" $T'$ we have $\rho'(T')=\{\rho(T)\mid \slift h(T)=T'\}$, to conclude that $\LanguageRec(\Alg',\rho',P')=\slift h(L)$.
\end{proof}

\section{Set-algebras}\label{section:set-algebras}

We introduce in \Cref{subsection:set-algebras} a richer notion of "algebras", called "set-algebras". These are the "algebras" arising from "set-systems" considered modulo "unfold-equivalence".
In particular, we introduce the notion of "syntactic set-algebra" of a "language of regular-trees" in \Cref{subsection:syntactic-set-algebra}, and use it in \Cref{subsection:small-profile} to establish that the "syntactic set-algebras" of "languages of regular-trees" "recognised" by "rankwise-finite" "algebras" are also "rankwise-finite" (see \Cref{lemma:algebra-to-set-algebra}).

\subsection{Set-algebras: definition and first properties}
\label{subsection:set-algebras}

The definition of a "set-algebra" combines evaluation of "set-systems", invariance under "unfold-equivalence@@ss", and an order compatible with "morphisms@@ss".
Since "set-systems" are equipped with a form of non-determinism, "set-algebras" are naturally endowed with an order~$\yaleq$ that provides an inf-semi-lattice structure. Note that the non-deterministic sum will be denoted~$\yameet$ at the level of "set-algebras", i.e.\ as an infimum rather than a supremum; this is a design choice that corresponds to the fact that we see in this work this non-determinism as ``adversarial'', i.e.\ controlled by an opponent.
\begin{definition}\label{definition:set-algebras}
	\AP A ""set-algebra"" $\intro*\yAlg$  is a "ranked set" $Y$ together with a "map of ranked sets" $\intro*\yaeval$ from "$Y$-set-systems" to $Y$\phantomintro(ya){evaluation map}, such that\phantomintro\yaleq
	\begin{itemize}
	\item $\yaeval(\atomic(a))=a$, for all~$a\in Y$,
	\item $\yaeval(\slifteval(S))=\yaeval(\flatten(S))$ for all "($Y$-set-system)-set-systems"~$S$, and
	\item  for all "unfold-equivalent@@ss" "$Y$-set-systems"~$S,S'$ of same "rank@@ss",  $\yaeval(S)= \yaeval(S')$.
	\item (notation) for $a,b\in Y_n$, we set $a\intro*\yameet b:= \yaeval(a\sssum b)$, and $a\reintro*\yaleq b$ to hold if $a\yameet b = a$. 
	\item ("morphism@@ss") if there is a "morphism@@ss" of "set-systems" from $S$ to $S'$, then $\yaeval(S')\yaleq\yaeval(S)$.
	\end{itemize}
	\noindent \AP For $a\in Y_n$, let $a\intro*\yaup$ be $\{b\in Y_n\mid a\yaleq b\}$.
	
	\noindent \AP A ""morphism@@ya"" from the "set-algebra" $\yAlg=(Y,\yaeval)$ to the "set-algebra"~$\yAlg'=(Y',\yaeval')$ is a "map of ranked sets"~$\rho$ from $Y$ to $Y'$ such that $\yaeval'(\slift\rho(S))=\rho(\yaeval(S))$ for all~"$Y$-set-systems"~$S$.
\end{definition}

The first two items in \Cref{definition:set-algebras} state that "set-algebras" are algebras\footnote{In the categorical sense of Eilenberg-Moore algebras, not in the sense of \Cref{definition:algebras}} for the "monad of set-systems".
The third one states that "set-algebras" are invariant under "unfold-equivalence@@ss".
The "morphism@@ss" axiom states that $\yaleq$ corresponds to the pre-order induced by "morphisms@@ss" (beware of the direction).

As is classical with such a definition, $\yaleq$ happens to be an order, and the operation $\yameet$ is in fact an infimum for it. This is the subject of the following statement.
\begin{lemma}\label{lemma:meet-is-meet}
	The operation $\yameet$ is associative, commutative, and idempotent.
	The relation $\yaleq$ is an order, and
	$\yameet$ computes the infimum with respect to~$\yaleq$.
\end{lemma}
\begin{proof}
	Since $a\sssum b$ and $b\sssum a$ are "unfold-equivalent@@ss", $a\yameet b=b\yameet a$.
	Similarly, $a\sssum(b\sssum c)$ being "unfold-equivalent@@ss" to $(a\sssum b)\sssum c$, $\yameet$ is associative.
	Finally, $a\sssum a$ is "unfold-equivalent@@ss" to~$a$, and thus $\yameet$ is idempotent.
	Transitivity of $\yaleq$: Assume $a\yaleq b$ and $b\yaleq c$, then $a\yameet c=(a\yameet b)\yameet c=a\yameet(b\yameet c)=a\yameet b = a$.
	Hence $a\yaleq c$.
	Reflexivity of $\yaleq$: Since $a\yameet a=a$, $a\yaleq a$.
	Antisymmetry of $\yaleq$: Assume $a\yaleq b$ and $b\yaleq a$, then $a = a\yameet b = b\yameet a = b$.
	Let us show that $a\yameet b\yaleq a$. This comes from the fact that $(a\yameet b)\yameet a=a\yameet b$, and hence $a\yameet b\yaleq a$. 
	Finally, let us assume that $c\yaleq a$ and $c\yaleq b$, then $c\yameet(a\yameet b)=(c\yameet a)\yameet(c\yameet b)=c\yameet c=c$. Hence $c\yaleq a\yameet b$.
\end{proof}
\begin{definition}[extended set-algebras]\AP\label{definition:extended-set-algebras}
Let~$\yAlg=(Y,\yaeval)$ be a "set-algebra", and put $\top_1:=\yaeval(\emptyset_1)$, where $\emptyset_1$ is the empty rank-$1$ "set-system". The ""extended set-algebra""~$\yAlg^+$ has
\[
Y^+_1=Y_1\times\powerset(\{\varepsilon,\mathsf r\}),
\qquad Y^+_n=Y_n\quad(n\neq1).
\]
We identify $a\in Y_1$ with $(a,\emptyset)$. The "neutral element" $\varepsilon=(\top_1,\{\varepsilon\})$ represents the immediate exit from \Cref{definition:extended-algebras}. The ""root element"" $\mathsf r=(\top_1,\{\mathsf r\})$ represents the vertex-free "extended set-system"
\[
\mathsf r=(\emptyset,\emptyset,\{x_1\},\emptyset,\emptyset).
\]
Substituting a "set-system"~$S$ at its exit gives~$\plant(S)$.
The pair $(a,E)$ records an ordinary part of value~$a$ together with the immediate initial and root exits listed in~$E$. In particular, formal sums with the new elements are retained. Their meet is
\[
(a,E)\yameet(b,F)=(a\yameet b,E\cup F).
\]
Thus the induced order satisfies $(a,E)\yaleq(b,F)$ exactly when $a\yaleq b$ and $E\supseteq F$.

For a rank-$1$ "extended set-system"~$P$ over~$Y$, let~$P^\circ$ be obtained by removing~$x_1$ from its initial and root sets, and let
\[
E_P=\{\varepsilon\mid x_1\in\Vinit_P\}\cup\{\mathsf r\mid x_1\in\Vroot_P\}.
\]
Define its extended evaluation by $\yaeval^+(P):=(\yaeval(P^\circ),E_P)$.
To use $(a,E)$ as a context piece, take $\atomic(a)$ together with the vertex-free pieces listed in~$E$, using disjoint union. For $a=\top_1$, the ordinary part may equivalently be represented by~$\emptyset_1$. Expand all such pieces before evaluating a completed expression, and use~$\yaeval$ when the expansion is an ordinary "$Y$-set-system". As for "extended algebras", the evaluation function is not total on compositions of "extended set-systems": an $\varepsilon$-loop or $\mathsf r$-loop will not evaluate.
\end{definition}
A "morphism@@ya" $\rho$ extends by $(a,E)\mapsto(\rho(a),E)$, i.e. fixing the two markers. A "rankwise-finite" "set-algebra" has a "rankwise-finite" formal extension, with $|Y^+_1|=4|Y_1|$.

\begin{lemma}\label{lemma:extended-context-evaluation}
Let~$P_0,\dots,P_k$ be rank-$1$ "extended set-systems" over~$Y$, and let~$a\in Y_k$. With the interpretation of extended values as pieces from \Cref{definition:extended-set-algebras},
\[
\yaeval\bigl(\Context(P_0,\dots,P_k)[a]\bigr)
=
\yaeval\bigl(\Context(\yaeval^+(P_0),\dots,\yaeval^+(P_k))[a]\bigr).
\]
\end{lemma}
\begin{proof}
Write each piece as its ordinary part~$P_i^\circ$ together with the markers in~$E_{P_i}$. In the construction of~$\Context$, the marker $\varepsilon\in E_{P_0}$ makes the hole initial, a marker $\mathsf r\in E_{P_i}$ in any piece makes it root, and $\varepsilon\in E_{P_i}$ for $i>0$ adds a self-loop in direction~$i$. Keep these connections fixed. The remaining construction is a "flattening" with the ordinary parts~$P_i^\circ$ as labels. The evaluation axiom allows each such label to be replaced by~$\yaeval(P_i^\circ)$. This gives the equality. The same axiom justifies using either~$\emptyset_1$ or~$\atomic(\top_1)$ for an ordinary part of value~$\top_1$.
\end{proof}

\begin{remark}["set-algebras" are "regular-tree algebras"]
	Since "systems" are particular cases of "set-systems", a "set-algebra" can map a value to any "system", and so it is in particular an "algebra".
	Furthermore, since "unfold-equivalence@@ss" over "systems" is consistent with its generalisation over "set-systems", and we required "unfold-invariance" for all "set-algebras", we obtain that "set-algebras" seen as "algebras" are "unfold-invariant@@a", and hence "regular-tree algebras".
	This downgrading of "set-algebras" to "regular-tree algebras" is made implicitly in the rest of this work.
\end{remark}

\subsection{Syntactic set-algebra for a language of regular-trees}
\label{subsection:syntactic-set-algebra}

We now introduce the syntactic yield-congruence. This is a variation on the idea of "syntactic algebra", tailored for "set-algebras", and giving a special role to the "root vertices".

\AP Here, we consider a "language~$L$ of regular-trees", and shall use the notation~$\Plant(L)$ for the set  $\{\plant(T)\mid T\in L\}$, i.e.\ the same elements except that "initial vertices" have been turned into "root vertices". 
In the following definition we use properties of the form $\yieldss(\Yields(S))\subseteq L\cup\Plant(L)$.
Here $S$ is closed, and we explicitly convert each initial or "root yield" to its associated "set-system" using the function $\yieldss$ from \Cref{definition:yields}. Unravelling the definitions, it means that all "systems" produced starting from either an "initial vertex" or a "root vertex" have to belong to~$L$. It would be equivalent to state it as, for instance, $\Plant(\yieldss(\Yields(S)))\subseteq\Plant(L)$.

\begin{definition}\label{definition:syntorder}
	Let $\alphabet$ be a "ranked set", and $L$ be a "language of regular-trees".
	The ""yield-syntactic pre-order of~$L$"" between two "$\alphabet$-set-systems"~$S$ and $S'$ of same "rank@@ss"~$n$, denoted $S\intro*\leqSset_L S'$,  holds if for all "closed@@ss" "$\alphabet$-set-contexts"~$C[\hole_n]$,
	\begin{align*}
		\yieldss(\Yields(C[S]))\subseteq L\cup\Plant(L)\quad \text{implies}\quad \yieldss(\Yields(C[S']))\subseteq L\cup\Plant(L)\text{.}
	\end{align*}
	\AP The ""$L$-yield-syntactic equivalence"" is defined as $S \intro*\equiSset_L S'$ if $S\leqSset_L S'$ and $S'\leqSset_L S$. 
	\AP $\intro*\setsyntmorph S_L$ denotes the "$L$-yield-syntactic equivalence class" of~$S$.
	Let $\intro*\SyntYield_L$ be the set of $\equiSset_L$-classes of "set-systems".
\end{definition}

Notice that contrary to the definition of $\yieldssubseteq$, where $S\yieldssubseteq S'$ means that $S$ has ``less "yields"'' than $S'$, here, $S\leqSset_L S'$ means that $S$ has intuitively ``more "yields"'' than $S'$. Indeed, if $S$ has more "yields" than $S'$, and if all "set-systems" in $\yieldss(\Yields(C[S]))$ are in~$L\cup\Plant(L)$, then it is enough to conclude that all "set-systems" in $\yieldss(\Yields(C[S']))$ are in $L\cup\Plant(L)$ as well.
\newcommand{\inclass}{\mathit{init}}
\newcommand{\roclass}{\mathit{root}}
\newcommand{\inro}{\mathit{inro}}
\newcommand{\leaf}{\mathit{leaf}}
\begin{example}\label{ex:syntyield}
Let $\alphabet=\{a,b,c\}$ be a "ranked set" with $a$ of "rank@@ss"~$2$ and $b,c$ of "rank@@ss"~$0$, and let $L$ be the "language of $\alphabet$-regular-trees" consisting of all "regular-trees" containing at least one occurrence of~$b$. This is a particular case of the family of languages given in \Cref{ex:reco-exists-lang}.
We will describe the syntactic equivalence classes of "set-systems" of "rank@@ss"~$1$. 
For $S$ a "$\alphabet$-set-system" of "rank@@ss"~$1$, we define the following properties:
\begin{itemize}
	\item $I_0$: some "closed@@system" "init-yield" of $S$ does not contain $b$.
	\item $I_1$: not $I_0$ and some "init-yield" of $S$ does not contain $b$.
	\item $I_2$: all "init-yields" of $S$ contain $b$.
\end{itemize}
Properties $R_0,R_1,R_2$ are defined similarly for "root-yields". 
For example, here is a "set-system" (left) satisfying $R_1$ and $I_1$, as witnessed by the representatives of a "root-yield" (middle) and an "init-yield" (right):
\begin{center}
\scalebox{.8}{
\begingroup%
  \makeatletter%
  \providecommand\color[2][]{%
    \errmessage{(Inkscape) Color is used for the text in Inkscape, but the package 'color.sty' is not loaded}%
    \renewcommand\color[2][]{}%
  }%
  \providecommand\transparent[1]{%
    \errmessage{(Inkscape) Transparency is used (non-zero) for the text in Inkscape, but the package 'transparent.sty' is not loaded}%
    \renewcommand\transparent[1]{}%
  }%
  \providecommand\rotatebox[2]{#2}%
  \newcommand*\fsize{\dimexpr\f@size pt\relax}%
  \newcommand*\lineheight[1]{\fontsize{\fsize}{#1\fsize}\selectfont}%
  \ifx\svgwidth\undefined%
    \setlength{\unitlength}{457.87516436bp}%
    \ifx\svgscale\undefined%
      \relax%
    \else%
      \setlength{\unitlength}{\unitlength * \real{\svgscale}}%
    \fi%
  \else%
    \setlength{\unitlength}{\svgwidth}%
  \fi%
  \global\let\svgwidth\undefined%
  \global\let\svgscale\undefined%
  \makeatother%
  \begin{picture}(1,0.26307042)%
    \lineheight{1}%
    \setlength\tabcolsep{0pt}%
    \put(0,0){\includegraphics[width=\unitlength,page=1]{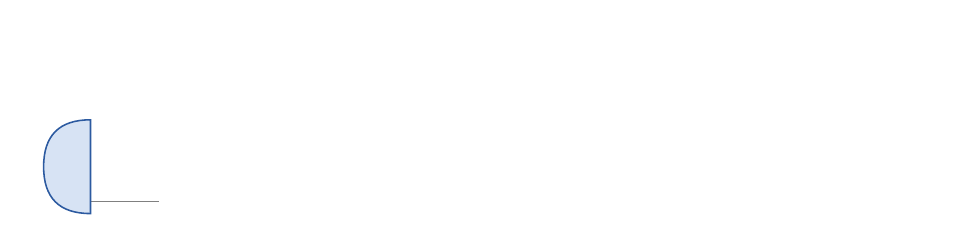}}%
    \put(0.06608799,0.08173179){\color[rgb]{0,0,0}\makebox(0,0)[lt]{\lineheight{1.25}\smash{\begin{tabular}[t]{l}$a$\end{tabular}}}}%
    \put(0,0){\includegraphics[width=\unitlength,page=2]{example_yield_class.pdf}}%
    \put(0.25337479,0.07520684){\color[rgb]{0,0,0}\makebox(0,0)[lt]{\lineheight{1.25}\smash{\begin{tabular}[t]{l}$b$\end{tabular}}}}%
    \put(0,0){\includegraphics[width=\unitlength,page=3]{example_yield_class.pdf}}%
    \put(0.25322415,0.15017483){\color[rgb]{0,0,0}\makebox(0,0)[lt]{\lineheight{1.25}\smash{\begin{tabular}[t]{l}$x_1$\end{tabular}}}}%
    \put(0,0){\includegraphics[width=\unitlength,page=4]{example_yield_class.pdf}}%
    \put(0.16138961,0.18561245){\color[rgb]{0,0,0}\makebox(0,0)[lt]{\lineheight{1.25}\smash{\begin{tabular}[t]{l}$a$\end{tabular}}}}%
    \put(0,0){\includegraphics[width=\unitlength,page=5]{example_yield_class.pdf}}%
    \put(0.57380435,0.12969984){\color[rgb]{0,0,0}\makebox(0,0)[lt]{\lineheight{1.25}\smash{\begin{tabular}[t]{l}$x_1$\end{tabular}}}}%
    \put(0,0){\includegraphics[width=\unitlength,page=6]{example_yield_class.pdf}}%
    \put(0.48196979,0.16513746){\color[rgb]{0,0,0}\makebox(0,0)[lt]{\lineheight{1.25}\smash{\begin{tabular}[t]{l}$a$\end{tabular}}}}%
    \put(0,0){\includegraphics[width=\unitlength,page=7]{example_yield_class.pdf}}%
    \put(0.75291945,0.14436541){\color[rgb]{0,0,0}\makebox(0,0)[lt]{\lineheight{1.25}\smash{\begin{tabular}[t]{l}$a$\end{tabular}}}}%
    \put(0,0){\includegraphics[width=\unitlength,page=8]{example_yield_class.pdf}}%
    \put(0.94824567,0.14377845){\color[rgb]{0,0,0}\makebox(0,0)[lt]{\lineheight{1.25}\smash{\begin{tabular}[t]{l}$x_1$\end{tabular}}}}%
    \put(0,0){\includegraphics[width=\unitlength,page=9]{example_yield_class.pdf}}%
    \put(0.85641115,0.17921608){\color[rgb]{0,0,0}\makebox(0,0)[lt]{\lineheight{1.25}\smash{\begin{tabular}[t]{l}$a$\end{tabular}}}}%
  \end{picture}%
\endgroup%
}
\end{center}

We can now describe the $7$ syntactic equivalence classes of $\SyntYield_L$, as well as their order, by noting e.g. $[R_2,I_2]$ the class of "$\alphabet$-set-systems" of "rank@@ss"~$1$ satisfying $R_2$ and $I_2$. An arrow $\mathcal C\to \mathcal C'$ means $\mathcal C \leqSset \mathcal C'$, i.e. for all $S\in \mathcal C$ and $S'\in \mathcal C'$, $S\leqSset_L S'$.

\begin{center}
\begin{tikzpicture}[scale=0.8]
\node (R0) at (0,0) {$[R_0]$};
\node (R1I0) at (2,0) {$[R_1,I_0]$};
\node (R1I1) at (4,1) {$[R_1,I_1]$};
\node (R2I0) at (4,-1) {$[R_2,I_0]$};
\node (R1I2) at (6,2) {$[R_1,I_2]$};
\node (R2I1) at (6,0) {$[R_2,I_1]$};
\node (R2I2) at (8,1) {$[R_2,I_2]$};
\path[->] (R0) edge (R1I0)
 (R1I0) edge (R1I1)
 (R1I0) edge (R2I0)
 (R1I1) edge (R1I2)
 (R1I1) edge (R2I1)
 (R2I0) edge (R2I1)
 (R1I2) edge (R2I2)
 (R2I1) edge (R2I2);
\end{tikzpicture}
\end{center}

The "set-context" $\hole_1(b)$ witnesses $[R_1,I_1]\not\leqSset_L [R_2,I_0]$.
The "set-context" $a(b,\hole_1(c))$ witnesses $[R_2,I_0]\not\leqSset_L [R_1,I_1]$.

The class $[R_1,I_2]$ is strictly above $[R_1,I_1]$ and incomparable with $[R_2,I_1]$. To witness both $[R_1,I_2]\not\leqSset_L [R_1,I_1]$ and $[R_1,I_2]\not\leqSset_L [R_2,I_1]$, take the "closed set-context" consisting of one "initial@@vertex" hole vertex~$h$ and the self-loop edge $(h,1,h)$. For $[R_1,I_2]$, every "init-yield" of the inserted "set-system" contains~$b$, and any $b$-free "root-yield" uses the exit, which the context feeds into an initial component guaranteeing~$b$. In either of the other two classes, a $b$-free "init-yield" can repeat through the loop indefinitely, producing a closed $b$-free yield. Conversely, $a(b,\hole_1(c))$ witnesses $[R_2,I_1]\not\leqSset_L [R_1,I_2]$, since the "root-yields" of the inserted "set-system" are tested independently of the outer~$b$.

\end{example}

We now need some preparatory lemmas to proceed.
The first one states that $\leqSset_L$ induces an inf-semi-lattice structure.
\begin{lemma}\plabel{lemma:leqSset-and-sum}
	$S\leqSset_L S'\sssum S''$ holds if and only if both $S\leqSset_L S'$ and $S\leqSset_L S''$ hold.
\end{lemma}
\begin{proofof}{lemma:leqSset-and-sum}
	The direction from left to right is obvious, as $\Yields(C[S'])\subseteq \Yields(C[S'\sssum S''])$ and $\Yields(C[S''])\subseteq \Yields(C[S'\sssum S''])$ for all "closed $\alphabet$-set-context"~$C[\hole_n]$.
	
	For the other direction, we claim first that $S\leqSset_L S'$ implies $S\leqSset_L S\sssum S'$.
	For this, consider some "closed $\alphabet$-set-context"~$C[\hole_n]$ such that $\yieldss(\Yields(C[S]))\subseteq L\cup\Plant(L)$.
	Define now $D[\hole_n]$ to be $C[S\sssum\hole_n]$.
	Since $S\sssum S$ "folds" to~$S$, $D[S]=C[S\sssum S]$ has the same "yields" as~$C[S]$ by \Cref{cor:unfold-yield,statement:re-is-congruence}. Hence $\yieldss(\Yields(D[S]))\subseteq L\cup\Plant(L)$. Applying $S\leqSset_L S'$ to~$D$ now gives $\yieldss(\Yields(D[S']))\subseteq L\cup\Plant(L)$. As $D[S']=C[S\sssum S']$, the claim follows.

	Assuming $S\leqSset_L S'$ and $S\leqSset_L S''$, apply the claim first to $S\leqSset_L S'$ and then to $S\sssum S'\leqSset_L S\leqSset_L S''$. The natural inclusion of $S'\sssum S''$ into $S\sssum S'\sssum S''$ gives the last inequality in the following chain:
	\begin{align*}
	S\leqSset_L S\sssum S' \leqSset_L S\sssum S'\sssum S'' \leqSset_L S'\sssum S''\ .
	\end{align*}
\end{proofof}

The second one states that $\equiSset_L$ is a congruence relation.
\begin{lemma}\plabel{lemma:synt-is-congruence}
	For~$S,S'$ "($\alphabet$-set-system)-set-systems" of the "same shape@@ss" such that $S(v)\leqSset_L S'(v)$ for all "vertices"~$v$ of~$S$, then $\flatten(S)\leqSset_L\flatten(S')$.
\end{lemma}
\begin{proofof}{lemma:synt-is-congruence}
	By \Cref{statement:context-to-flatten}, we just have to show the result for "set-contexts", i.e.\ to show that for all "$\alphabet$-set-context"~$C[\hole_k]$ of "rank@@ss"~$n$, and all "$\alphabet$-set-systems"~$S\leqSset_L S'$ of "rank@@ss"~$k$, $C[S]\leqSset_L C[S']$.

	Let $C[\hole_k]$, $S,S'$ be as above, and let~$D[\hole_n]$ be some "closed $\alphabet$-set-context". We define the "set-context" $D\circ C[\hole_k]:=D[C[\hole_k]]$.
	Assuming $\yieldss(\Yields((D\circ C)[S]))\subseteq L\cup\Plant(L)$, and since $S\leqSset_L S'$, we have $\yieldss(\Yields((D\circ C)[S']))\subseteq L\cup\Plant(L)$.
	This is true for all $D$, so this establishes $C[S]\leqSset_L C[S']$.
\end{proofof}

\AP It follows that the operation $\flatten$ can be regarded as acting on $\equiSset_L$-equivalence classes.
Let us denote~$\intro*\syntflatten_L$ the $\flatten$ operation applied to $\equiSset_L$-equivalence classes, that is $\syntflatten_L$ takes a "$\SyntYield_L$-set-system"~$S$, and associates to it an element of $\SyntYield_L$ defined as $\syntflatten_L(S):=\setsyntmorph{\flatten(S')}_L$, where $S'$ is any "($\alphabet$-set-system)-set-system" of same shape as $S$ such that $S'(v)\in S(v)$ for all "vertices"~$v$ of~$S$. This is well-defined by \Cref{lemma:synt-is-congruence}.
\begin{lemma}\label{lemma:syntorder}
	Given a "language of $\alphabet$-regular-trees"~$L$,  $(\SyntYield_L,
	\syntflatten_L)$ is a "set-algebra" for which $\yaleq$ coincides with $\leqSset_L$.
\end{lemma}
\begin{proof}
	Our first objective is to establish that $\yaleq$ and $\leqSset_L$ coincide. Indeed, for all "$\alphabet$-set-systems" $S,S'$ of "rank@@ss"~$n$,
	\begin{align*}
	&\setsyntmorph S_L \yaleq \setsyntmorph {S'}_L\\
	\text{if and only if}\quad&
	\setsyntmorph S_L\yameet \setsyntmorph{S'}_L= \setsyntmorph {S}_L\ ,\tag{definition of $\yaleq$}\\
	\text{if and only if}\quad&
	\setsyntmorph{S\sssum S'}_L= \setsyntmorph {S}_L\ ,\tag{definition of $\yameet$}\\
	\text{if and only if}\quad&
	S\sssum S' \equiSset_L  S\ ,\tag{definition of $\setsyntmorph {-}_L$}\\
	\text{if and only if}\quad&
	S \leqSset_L  S\sssum S'\ ,  \tag{since $S\sssum S' \leqSset_L  S$}\\
	\text{if and only if}\quad&
	S \leqSset_L  S'\ . \tag{using \Cref{lemma:leqSset-and-sum}}
	\end{align*}

	We can now complete the proof that $(\SyntYield_L,\syntflatten_L)$ is a "set-algebra". The first two axioms follow from \Cref{lemma:synt-is-congruence}.
For base-alphabet "set-systems", a "morphism@@ss" $S\ssmorphism S'$ gives "morphisms@@ss" $C[S]\ssmorphism C[S']$ in every "closed set-context". Hence, by \Cref{statement:locsurj->res-eq}, we have $\Yields(C[S])\subseteq\Yields(C[S'])$, so $S'\leqSset_L S$.
For a "morphism@@ss" of $\SyntYield_L$-set-systems, choose one base-alphabet representative for each label. Substitution and "flattening" lift the "morphism@@ss" to the resulting base-alphabet "set-systems", giving the required evaluation inequality. A "folding" likewise lifts to a "folding"; \Cref{cor:unfold-yield,statement:re-is-congruence} then gives equality of their syntactic classes. This proves the "unfold-invariance" axiom as well.
\end{proof}

Let us call the "set-algebra" $(\SyntYield_L,\syntflatten_L)$ the ""syntactic set-algebra"" of~$L$, noted~$\intro*\yAlgSynt_L$.

\begin{lemma}\label{lemma:synt-yield-order}
For "$\alphabet$-set-systems"~$S,S'$ of the same rank, $S\yieldssubseteq S'$ implies $S'\leqSset_L S$, and hence $\setsyntmorph{S'}_L\yaleq\setsyntmorph S_L$.
\end{lemma}
\begin{proof}
By \Cref{statement:re-is-congruence}, every "closed set-context"~$C$ satisfies $\Yields(C[S])\subseteq\Yields(C[S'])$. Thus acceptance of all "yields" of~$C[S']$ implies acceptance of all "yields" of~$C[S]$. Apply the definition of~$\leqSset_L$ and \Cref{lemma:syntorder}.
\end{proof}

\begin{lemma}\label{lemma:synt-recognition}
	For $L$ a "language of $\alphabet$-regular-trees", the "algebra" $\yAlgSynt_L$ "recognises" $L$. Moreover, the corresponding "accepting set" $P\subseteq(\SyntYield_L)_0$ is $\yaleq$-upward closed and closed under $\yameet$.
\end{lemma}
\begin{proof}
	Put
	\[
	P:=\{\setsyntmorph{S}_L\mid S\text{ is closed and }\yieldss(\Yields(S))\subseteq L\cup\Plant(L)\}.
	\]
	Taking the "set-context" consisting of a single initial $0$-hole in \Cref{definition:syntorder} shows that this acceptance condition is invariant under $\equiSset_L$ and preserved under $\leqSset_L$. Thus $P$ is well-defined and $\yaleq$-upward closed by \Cref{lemma:syntorder}. It is closed under $\yameet$ because $\Yields(S\sssum S')=\Yields(S)\cup\Yields(S')$.
	For every "closed@@system" "$\alphabet$-system"~$T$, the set $\yieldss(\Yields(T))$ consists exactly of the "systems" "unfold-equivalent@@ss" to~$T$, by \Cref{statement:ue=te}. Since $L$ is invariant under this relation, $T\in L$ if and only if $\setsyntmorph T_L\in P$. Hence the canonical "morphism@@algebra" $T\mapsto\setsyntmorph T_L$ "recognises" $L$ with "accepting set"~$P$.
\end{proof}

\subsection{Rankwise-finite set-algebras}
\label{subsection:small-profile}

The goal of this section is to establish the following lemma.
\begin{lemma}\label{lemma:algebra-to-set-algebra}
	If a "language of regular-trees"~$L$ is "recognised" by an "unfold-invariant" "rankwise-finite" "algebra", then its "syntactic set-algebra" is also "rankwise-finite".
\end{lemma}
We fix from now on an "unfold-invariant" "rankwise-finite" "algebra"~$\Alg=(A,\aeval)$ that "recognises" a "language of $\alphabet$-regular-trees"~$L$. Let~$\rho$ be the "recognising morphism".

The proof of \Cref{lemma:algebra-to-set-algebra} consists in defining an equivalence relation over "$\alphabet$-set-systems" for each rank, called the "small-$A$-profile-equivalence". It happens to be of finite index for each rank (this is straightforward) and at the same time to refine the "yield-syntactic congruence" for~$L$ (this is the more involved part proved in \Cref{statement:spe-refines-ssc}). \Cref{lemma:algebra-to-set-algebra} immediately follows. Let us begin with introducing this notion of "small-$A$-profile-equivalence".

Put $b_A:=|A_0|+|A_1|+1$. The summands correspond to "closed@@system" "systems", rank-$1$ "systems", and the "extended system" $\varepsilon$ respectively.

\AP A map $\sigma\colon\interval m\to\interval n$ is ""$A$-small@@map"", or simply \reintro*"small@@map", if $|\sigma^{-1}(i)|\leqslant b_A$ for every~$i\in\interval n$.
\AP The ""profile"", ""root-profile"", ""small-profile"" and ""small-root-profile"" of a "$\alphabet$-set-system"~$S$ are\phantomintro{\profile}\phantomintro{\rootprofile}\phantomintro{\smallprofile}\phantomintro{\smallrootprofile}
\begin{align*}
\reintro*\profile(S)&:=\{(\rho(T),\sigma)\mid (T,\sigma)\in\InitYields(S)\},\\
\reintro*\rootprofile(S)&:=\profile(\Uproot(S)),\\
\reintro*\smallprofile(S)&:=\{(a,\sigma)\in\profile(S)\mid\sigma\text{ is "small@@map"}\},\\
\reintro*\smallrootprofile(S)&:=\smallprofile(\Uproot(S)).
\end{align*}
A "profile" records the algebraic value of a "yield" together with its variable map. The "small-profile" bounds the number of local variables representing each variable of~$S$.
Two "set-systems" of the same rank are ""profile-equivalent"" if their "profiles" and "root-profiles" agree, and ""small-profile-equivalent"" if their "small-profiles" and "small-root-profiles" agree.
Since~$\rho$ is "unfold-invariant", every element $(a,\sigma)$ of $\profile(S)$ can be witnessed by a "yield" $(U,\sigma)$ such that there is a "morphism@@ss" $\rename\sigma(U)\ssmorphism S$.
\begin{remark}[profiles of closed set-systems]
For a closed "set-system", every variable map is the empty map $\intro*\emptymap\colon\interval0\to\interval0$, which is small. Thus "profile-equivalence" and "small-profile-equivalence" coincide at rank~$0$.
\end{remark}
\begin{remark}\label{statement:prof-refines-language}
For "profile-equivalent" closed "set-systems"~$S,S'$, we have
\begin{align*}
  \yieldss(\InitYields(S))\subseteq L
    &\quad\Longleftrightarrow\quad \yieldss(\InitYields(S'))\subseteq L,\\
  \yieldss(\RootYields(S))\subseteq\Plant(L)
    &\quad\Longleftrightarrow\quad \yieldss(\RootYields(S'))\subseteq\Plant(L).
\end{align*}

\end{remark}

We first establish the congruence property for full "profiles", before imposing the finite bound.
\begin{lemma}[profile-equivalence is a congruence]\label{statement:pe-congruence}
If~$S,S'$ are "($\alphabet$-set-system)-set-systems" of the "same shape@@ss", with "profile-equivalent" labels at every vertex, then $\flatten(S)$ and $\flatten(S')$ are "profile-equivalent".
\end{lemma}
\begin{proofof}{statement:pe-congruence}
Take $(a,\tau)\in\profile(\flatten(S))$, represented by a "yield" $(R,\tau)$ with $\rho(R)=a$.
By \Cref{lemma:yield-decomposition-converse}, choose $T,\delta,(\sigma_t)_t$ satisfying the four conditions of \Cref{lemma:yield-composition}, with a "folding" $\flatten(T)\ssfolding R$. In particular $\rho(\flatten(T))=a$.
For each~$t$, "profile-equivalence" provides $(R_t,\sigma_t)\in\InitYields(S'(\delta(t)))$ with $\rho(R_t)=\rho(T(t))$. Replace the label at~$t$ by~$R_t$ to obtain a "system-system"~$T'$. \Cref{lemma:yield-composition} gives $(\flatten(T'),\tau)\in\InitYields(\flatten(S'))$, while the "algebra" axioms give
\[
\rho(\flatten(T'))=\aeval(\slift\rho(T'))=\aeval(\slift\rho(T))=\rho(\flatten(T))=a.
\]
Thus, $(a,\tau)\in\profile(\flatten(S'))$.
By symmetry, we obtain the equality of "profiles". Apply the same argument to $\fuproot(S)$ and $\fuproot(S')$ for equality of "root-profiles".
\end{proofof}
\begin{corollary}\label{statement:pe-refines-ssc}
"Profile-equivalence@profile-equivalence" refines the "yield-syntactic equivalence".
\end{corollary}
\begin{proof}
Apply \Cref{statement:pe-congruence} in every "closed set-context" and use \Cref{statement:prof-refines-language}.
\end{proof}

\begin{example}[profile-equivalence need not have rankwise-finite index]\label{ex:profile-not-finite}
Use the "algebra" $\Alg$ from \Cref{ex:aexists-alg} recognising the presence of nullary symbol~$b$, with $A_n=\{0_n,1_n\}$. Over an alphabet containing $a_1,a_2,b$ of the indicated ranks, put $T_1=a_1(x_1)$ and $T_{n+1}=a_2(T_n,x_1)$, so that $T_n$ is comb-shaped with $n$ leaves pointing to $x_1$.
A "yield" $(R,\sigma_m)$ of~$T_n$, where $\sigma_m\colon\interval m\to\interval1$ is the constant map, uses all $m$ variables among its $n$ leaf positions. Thus $1\leqslant m\leqslant n$. Conversely every such~$m$ is possible by assigning the leaf positions surjectively to~$\interval m$. No such "system" contains~$b$, so its algebraic value is $0_m\in A_m$. Therefore
\[
\profile(T_n)=\{(0_m,\sigma_m)\mid1\leqslant m\leqslant n\}.
\]
These "profiles" are pairwise distinct for different values of $n$.
\end{example}

We now want to bound the number of antecedents of each variable through the renaming map, without changing the algebraic value of a closed "yield". The next lemma performs one such reduction directly on a context.
\begin{lemma}\plabel{lemma:context-smallification}
Let~$T$ be a "$\alphabet$-system" of rank~$m$, let $C[\hole_m]$ be a closed "$\alphabet$-context", and let $\sigma\colon\interval m\to\interval n$.
There exists $m'\leq m$ and a surjection $\tau\colon\interval m\to\interval{m'}$ together with an injection $\tau'\colon\interval{m'}\to\interval m$ such that $\sigma\circ\tau'\circ\tau=\sigma$, the map $\sigma\circ\tau'$ is small, and
\[
\rho(C[T])=\rho(C'[\rename\tau(T)]).
\]
Where~$C'[\hole_{m'}]$ is obtained from~$C$ as follows: the direction $j$-successor of the hole of $C'$ is defined to be the $\tau'(j)$-successor of the hole of $C$. We then restrict $C'$ to the vertices reachable from the "initial vertex".
\end{lemma}
\begin{proofof}{lemma:context-smallification}
Let~$h$ be the hole vertex of~$C$, and let~$P_0,P_1,\dots,P_m$ be the "pieces" of $C$. For each $i$, let $Q_i$ be the reachable part of $P_i$, removing the variable $x_1$ when unused, and let $q_i=\rho(Q_i)$, an element of $\{\varepsilon\}\uplus A_0\uplus A_1$. There are at most~$b_A$ possibilities for the value of $q_i$.
Define $i\sim j$ by $\sigma(i)=\sigma(j)$ and $q_i=q_j$. Let~$\tau$ enumerate the equivalence classes and choose a representative~$\tau'(j)$ of each class. Then $\sigma\tau'\tau=\sigma$, and~$\sigma\tau'$ is "small"

It remains to prove $\rho(C[T])=\rho(C'[\rename\tau(T)])$.

Recall that $C[T]$ is "unfold-equivalent" to\footnote{If $Q_i$ is of rank $0$, it can be considered as a "piece" of rank $1$ with unused exit.} $\Context(Q_0,Q_1,\dots, Q_m)[T]$. Since $\Alg$ is "unfold-invariant", they have same image under $\rho$.
By the "algebra" axioms, replacing each $Q_j$ by $Q_{\tau'(\tau(j))}$, does not change the evaluation under the algebra. Identifying copies with the same representative~$\tau'(\tau(j))$ gives a "folding" of the resulting $A$-system, so also preserves its value. The resulting wiring sends exits with the same~$\tau$-class to the same successor (or directly back to~$T$ in the $\varepsilon$ case), and discard unreachable vertices. We therefore have shown that~$C'[\rename\tau(T)]$ has same value under $\rho$ as~$C[T]$.
\end{proofof}

\begin{lemma}\label{lemma:small-yield-decomposition}
Let~$S$ be a closed "($\alphabet$-set-system)-set-system", and let $T,\delta,(\sigma_t)_t$ satisfy the four conditions of \Cref{lemma:yield-composition}, with $\tau$ being the empty map.
There are $T',\delta',(\sigma'_t)_t$ satisfying the same conditions for~$S$, with every~$\sigma'_t$ "small" and $\rho(\flatten(T'))=\rho(\flatten(T))$.
\end{lemma}
\begin{proof}
By induction on the number of vertices~$u$ whose map~$\sigma_u$ is not "small". If there are none, there is nothing to do.
Otherwise replace the subsystem~$T(u)$ by a hole and flatten to obtain a "closed context"~$C[\hole_m]$, where $m=\rk(T(u))$. Apply \Cref{lemma:context-smallification} with~$\sigma_u$ to obtain~$\tau,\tau'$.
In the outer "system"~$T$, replace $T(u)$ by $\rename\tau(T(u))$, give direction~$j$ at~$u$ its former successor in direction~$\tau'(j)$, and retain only the vertices reachable from the "initial vertex". Denote this "system-system" by~$T'$. The chosen vertex~$u$ remains reachable, since the path to its first visit is unchanged. Its local subsystem is a "system" since~$\tau$ is surjective. The preceding lemma gives $\rho(\flatten(T'))=\rho(\flatten(T))$.

Restrict~$\delta$ and all other local maps to the retained vertices, and put $\sigma'_u=\sigma_u\tau'$. To check the local "yield" at~$u$, choose a witness $U\ssfolding T(u)$ and a "morphism@@ss" $\rename{\sigma_u}(U){\ssmorphism} S(\delta(u))$. Surjective renaming gives a {"folding"} $\rename\tau(U)\ssfolding\rename\tau(T(u))$, and
\[
\rename{\sigma'_u}(\rename\tau(U))=\rename{\sigma_u}(U).
\]
Thus the required local "yield" exists. An edge in direction~$j$ of~$u$ still maps to an edge of~$S$, since $\sigma'_u(j)=\sigma_u(\tau'(j))$. The other conditions are inherited. The map at~$u$ is now small, so we can apply the induction hypothesis to conclude.
\end{proof}

\begin{lemma}\label{lemma:profile-equivalence-closed}
If~$S,S'$ are closed "($\alphabet$-set-system)-set-systems" of the same shape, with "small-profile-equivalent" labels at every vertex, then $\flatten(S)$ and $\flatten(S')$ are "profile-equivalent".
\end{lemma}
\begin{proofof}{lemma:profile-equivalence-closed}
Let $(a,\emptymap)\in\profile(\flatten(S))$. Decompose a representing "yield" by {\Cref{lemma:yield-decomposition-converse}} and apply \Cref{lemma:small-yield-decomposition}. This gives a closed "system-system"~$T$ with $\rho(\flatten(T))=a$ and local "yields" $(T(t),\sigma_t)$ whose maps are all small.
"Small-profile-equivalence@small-profile-equivalence" supplies local "yields" $(R_t,\sigma_t)$ of~$S'(\delta(t))$ with $\rho(R_t)=\rho(T(t))$. Replace the labels of~$T$ by~$R_t$. Composition in {\Cref{lemma:yield-composition}} gives a "yield" of~$\flatten(S')$, and the "algebra" axioms preserve its value~$a$. Symmetry gives equality of "profiles".
For "root-profiles" apply the same argument to $\fuproot(S)$ and $\fuproot(S')$, as in the proof of \Cref{statement:pe-congruence}.
\end{proofof}

\begin{corollary}\label{statement:spe-refines-ssc}
"Small-profile-equivalence@small-profile-equivalence" refines $\equiSset_L$.
\end{corollary}
\begin{proofof}{statement:spe-refines-ssc}
For "small-profile-equivalent" "set-systems"~$S,S'$ and any "closed set-context"~$C$, \Cref{lemma:profile-equivalence-closed} makes $C[S]$ and $C[S']$ "profile-equivalent". By \Cref{statement:prof-refines-language}, $\yieldss(\Yields(C[S]))\subseteq L\cup\Plant(L)$ if and only if $\yieldss(\Yields(C[S']))\subseteq L\cup\Plant(L)$. This is precisely $S\equiSset_L S'$.
\end{proofof}

Finally, for a small map $\sigma\colon\interval m\to\interval n$, we have $m\leqslant n\cdot b_A$. There are only finitely many such maps and finitely many values in each~$A_m$. Thus there are finitely many "small-profiles" and "small-root-profiles" in each rank. By \Cref{statement:spe-refines-ssc}, the "syntactic set-algebra" is "rankwise-finite", proving \Cref{lemma:algebra-to-set-algebra}.

\section{The automaton property}\label{sec:core}

In \Cref{subsection:extremal-elements}, we define "extremal accepted situations" and "extremal elements". Finite minimality then gives extremal replacements in every "rankwise-finite" "set-algebra" with an upward-closed "accepting set" (\Cref{lemma:extremal-replacement,cor:extremal-replacements}). The "automaton property" of \Cref{def:autprop} requires two further features: "extremal elements" decompose into extremal unary "direction projections", and acceptance of the resulting unary/nullary "set-systems" is determined by their paths. We establish these features for "rankwise-finite" "syntactic set-algebras", using an exact description of the accepting "contexts" of "extremal elements".

\subsection{Extremal accepted situations and extremal elements}
\label{subsection:extremal-elements}

\AP For a "set-algebra"~$\yAlg=(Y,\yaeval)$, denote $\intro*\Yw$ the "ranked alphabet" $Y$ from which all symbols of "rank" $2$ or above are removed.
We use its "extended set-algebra"~$\yAlg^+$ for the unary context pieces, with the order and evaluation conventions of \Cref{definition:extended-set-algebras}. The element filling the hole remains in the ordinary sort~$Y_k$.
Fix an "accepting set" $F\subseteq Y_0$. For a "closed@@ss" "$Y$-set-system"~$S$, we will denote its acceptance by $\intro*\Acc(S)$, meaning $\yaeval(S)\in F$.

\begin{definition}\AP\label{definition:extremal-accepted-situation}
An ""extremal accepted situation"" of rank~$k$ is a tuple $(P_0,\dots,P_k,\delta)\in (Y_1^+)^{k+1}\times Y_k$ such that
\[
\Acc\bigl(\Context(P_0,\dots,P_k)[\delta]\bigr)
\]
and the tuple is componentwise minimal for~$\yaleq$ among tuples satisfying this condition.
Explicitly, whenever $Q_0,\dots,Q_k\in Y_1^+$ and $\eta\in Y_k$ satisfy
\[
Q_i\yaleq P_i\quad(0\leqslant i\leqslant k),\qquad
\eta\yaleq\delta,\qquad
\Acc\bigl(\Context(Q_0,\dots,Q_k)[\eta]\bigr),
\]
we have $Q_i=P_i$ for every~$i$ and $\eta=\delta$.

An element~$\delta\in Y_k$ is an ""extremal element"" if it is the last component of some "extremal accepted situation". Both notions are relative to the fixed "accepting set"~$F$.
\end{definition}

\subsection{Extremal replacements}
\label{subsection:extremal-replacements}

Monotonicity of context evaluation (\Cref{lemma:context-evaluation-monotone}) and finite minimality (\Cref{lemma:choice-extremal-context}) allow an element in an accepting "context" to be replaced by an "extremal element" below it. This replacement principle holds in every "rankwise-finite" "set-algebra" with an upward-closed "accepting set".

We first record the monotonicity needed to compare "contexts" whose pieces have been decreased.
\begin{lemma}\label{lemma:context-evaluation-monotone}
Evaluation of "$Y$-set-systems" is monotone in their labels, with the shape fixed. Moreover, if $P_i,Q_i\in Y_1^+$ satisfy $P_i\yaleq Q_i$ for $0\leqslant i\leqslant k$, and $a\yaleq b$ in~$Y_k$, then
\[
\yaeval\bigl(\Context(P_0,\dots,P_k)[a]\bigr)
\yaleq
\yaeval\bigl(\Context(Q_0,\dots,Q_k)[b]\bigr).
\]
\end{lemma}
\begin{proof}
Let~$S,S'$ have the same shape and satisfy $S(v)\yaleq S'(v)$ at every vertex. Replace each label by the "set-system" $\atomic(S(v))\sssum\atomic(S'(v))$, and flatten. Its evaluation is~$\yaeval(S)$, since each substituted label evaluates to $S(v)\yameet S'(v)=S(v)$. There is a "morphism@@ss" from~$S'$ into this "flattening", sending each vertex to its second copy. The "morphism@@ss" axiom of "set-algebras" gives $\yaeval(S)\yaleq\yaeval(S')$.

For the second assertion, write $P_i=(p_i,E_i)$ and $Q_i=(q_i,F_i)$, as in \Cref{definition:extended-set-algebras}. Then $p_i\yaleq q_i$ and $E_i\supseteq F_i$. We can first increase the ordinary labels from $p_i$ to $q_i$,and the hole label, the first assertion guarantees that this makes the global evaluation increase. Then delete the initial, root, and edge contributions of the markers in $E_i\setminus F_i$. The resulting "set-system" maps into the preceding one by the identity on vertices, so the "morphism@@ss" axiom again increases the evaluation. This proves the required inequality.
\end{proof}

\begin{lemma}[choice of an extremal accepted situation]\label{lemma:choice-extremal-context}
Assume~$\yAlg$ is "rankwise-finite". For every accepted tuple $(P_0,\dots,P_k,a)\in(Y_1^+)^{k+1}\times Y_k$, there is an "extremal accepted situation" $(M_0,\dots,M_k,\delta)$ such that $M_i\yaleq P_i$ for every~$i$ and $\delta\yaleq a$.
\end{lemma}
\begin{proof}
Consider the set of accepted tuples $(Q_0,\dots,Q_k,b)$ satisfying $Q_i\yaleq P_i$ for all~$i$ and $b\yaleq a$. It is nonempty, since it contains the given tuple, and finite, since $Y_1^+$ and~$Y_k$ are finite. Any componentwise minimal tuple $(M_0,\dots,M_k,\delta)$ is an "extremal accepted situation", by \Cref{lemma:context-evaluation-monotone}.
\end{proof}

\begin{lemma}[extremal replacement]\AP\plabel{lemma:extremal-replacement}
Let~$\YAlg=(Y,\yaeval)$ be a "rankwise-finite" "set-algebra" and let $F\subseteq Y_0$ be an upward-closed "accepting set". For every "closed $Y$-set-context"~$C[\hole_k]$ and every $a\in Y_k$ such that $\Acc(C[a])$, there is an "extremal element"~$\delta\in Y_k$ satisfying
\[
\delta\yaleq a\qquad\text{and}\qquad\Acc(C[\delta]).
\]
\end{lemma}
\begin{proofof}{lemma:extremal-replacement}
Let~$C[\hole_k]$ be a "closed set-context", and let~$a\in Y_k$ satisfy $\Acc(C[a])$.
Let~$R_0,\dots,R_k$ be the "pieces@@context" of~$C$, and put $P_i=\yaeval^+(R_i)\in Y_1^+$.
By \Cref{lemma:context-decomposition,lemma:extended-context-evaluation}, for every~$b\in Y_k$,
\[
\yaeval(C[b])=\yaeval\bigl(\Context(P_0,\dots,P_k)[b]\bigr).
\]
In particular, $(P_0,\dots,P_k,a)$ is accepted. Apply \Cref{lemma:choice-extremal-context} to obtain an "extremal accepted situation" $(M_0,\dots,M_k,\delta)$ below it. Its last component~$\delta$ is extremal and satisfies $\delta\yaleq a$. By \Cref{lemma:context-evaluation-monotone},
\[
\yaeval\bigl(\Context(M_0,\dots,M_k)[\delta]\bigr)
\yaleq\yaeval\bigl(\Context(P_0,\dots,P_k)[\delta]\bigr)
 =\yaeval(C[\delta]).
\]
The left-hand side belongs to~$F$ by the choice of the extremal situation. Since~$F$ is upward closed, $\Acc(C[\delta])$, as required. The argument applies to every~$k$, including ranks~$0$ and~$1$.
\end{proofof}

\begin{corollary}[simultaneous extremal replacements]\label{cor:extremal-replacements}
Let~$\YAlg$ be "rankwise-finite" and let $F\subseteq Y_0$ be upward closed. A closed "$Y$-set-system"~$S$ satisfies $\Acc(S)$ if and only if there are "extremal elements" $\delta_v\yaleq S(v)$, one for each $v\in\V_S$, such that
\[
\Acc\bigl(S[v/\delta_v]_{v\in\V_S}\bigr).
\]
Here $S[v/\delta_v]_{v\in\V_S}$ has the same vertices, initial and root sets, and edges as~$S$, and label~$\delta_v$ at~$v$.
\end{corollary}
\begin{proof}
Enumerate the finitely many vertices of~$S$. At each vertex, regard the rest of the current "set-system" as a "closed set-context" and apply \Cref{lemma:extremal-replacement}. This decreases that label to an "extremal element" and preserves acceptance. The converse follows from \Cref{lemma:context-evaluation-monotone}. The argument also covers the empty "set-system".
\end{proof}

\subsection{The automaton property}
\label{subsection:automaton-property}

Extremal replacement reduces acceptance to "extremal elements". To construct an automaton, we further need to split these elements into unary directions and determine acceptance from paths. We call this combination the "automaton property".

\AP Write $0$ for the empty closed "set-system"; it has no yields. For $a\in Y_k$ and $1\leqslant i\leqslant k$, its $i$th ""direction projection"" is
\[
\operatorname{dir}_i(a):=\yaeval\bigl(a(0,\dots,0,x_1,0,\dots,0)\bigr)\in Y_1,
\]
where $x_1$ occurs in position~$i$. Concretely, keep the "initial vertex" labelled~$a$, its edge in direction~$i$ to~$x_1$, and no edges in the other directions. In rank~$k$, $\operatorname{dir}_i(a)(x_i)$ denotes the value obtained by renaming its single variable to~$x_i$.

\AP A ""regular branch"" of a closed "$\Yw$-set-system"~$G$ is a closed "system"~$B$ over~$\Yw$ admitting a "morphism@@ss" $B\ssmorphism G$. Such a system is a finite chain ending at a nullary vertex or a lasso, and represents respectively a finite initial-to-leaf path or an ultimately periodic infinite path from an "initial vertex" of~$G$. We say that it is accepted if $\Acc(B)$.

\begin{definition}\AP\label{def:autprop}
Let~$\YAlg=(Y,\yaeval)$ be a "rankwise-finite" "set-algebra" and $F\subseteq Y_0$ an "accepting set". The pair $(\YAlg,F)$ has the ""automaton property"" if the following two conditions hold:
\begin{enumerate}
\item (Direction decomposition) For every "extremal element" $\delta\in Y_k$ with $k\geqslant1$, each $\operatorname{dir}_i(\delta)$ is extremal in~$Y_1$, and
\[
\delta=\bigmeet_{i=1}^k\operatorname{dir}_i(\delta)(x_i).
\]
\item (Path acceptance) Let~$G$ be a closed "$\Yw$-set-system" with no "root vertices", with every vertex reachable from an "initial vertex", and with at least one outgoing edge at every unary vertex. If every unary label is an "extremal element", then $\Acc(G)$ if and only if every "regular branch" of~$G$ is accepted.
\end{enumerate}
\end{definition}

The first condition supplies unary transition labels. The second condition allows acceptance of the resulting "set-systems" to be checked along their "regular branches". No direction decomposition is required in rank~$0$: nullary elements serve as leaf labels.

\subsection{Establishing the automaton property for syntactic set-algebras}

For the remainder of this section, let~$L$ be a "language of $\alphabet$-regular-trees" recognised by a "rankwise-finite" "algebra", and let~$\YAlg$ be its "syntactic set-algebra", with "accepting set" $F\subseteq Y_0$. Equality of recognising values implies indistinguishability in every "closed context", so the ordinary "syntactic algebra" of~$L$ is "rankwise-finite". It is also "unfold-invariant", since~$L$ is "unfold-invariant" and "unfolding" is a congruence for "contexts". Thus \Cref{lemma:algebra-to-set-algebra} makes~$\YAlg$ "rankwise-finite", and \Cref{lemma:synt-recognition} gives upward closure of~$F$.

We use two consequences of the syntactic definition: planting a closed "set-system" preserves acceptance, and
\begin{equation}\label{eq:syntactic-context-test}
a\yaleq b\quad\Longleftrightarrow\quad
\text{for every closed set-context $C$, $\Acc(C[a])$ implies $\Acc(C[b])$.}
\end{equation}
These statements also hold for "set-systems" and "contexts" labelled in~$Y$: choose alphabet-labelled representatives of their labels and flatten, then apply \Cref{lemma:syntorder,lemma:synt-recognition}. The formal unary markers are expanded before doing so.

\begin{lemma}\label{lemma:extremal-context-filter}
Let~$(P_0,\dots,P_k,\delta)$ be an "extremal accepted situation". For all $Q_0,\dots,Q_k\in Y_1^+$,
\[
\Acc\bigl(\Context(Q_0,\dots,Q_k)[\delta]\bigr)
\quad\Longleftrightarrow\quad
P_i\yaleq Q_i\quad(0\leqslant i\leqslant k).
\]
\end{lemma}
\begin{proof}
We will start by defining auxiliary objects, and then prove the forward implication. The reverse implication is monotonicity from \Cref{lemma:context-evaluation-monotone}.
Put $M=\Context(P_0,\dots,P_k)$, and write~$h$ for its hole vertex, as illustrated in \Cref{fig:extremal-context}.
\begin{figure}[H]
\centering
\def\svgwidth{.45\linewidth}
\begingroup%
  \makeatletter%
  \providecommand\color[2][]{%
    \errmessage{(Inkscape) Color is used for the text in Inkscape, but the package 'color.sty' is not loaded}%
    \renewcommand\color[2][]{}%
  }%
  \providecommand\transparent[1]{%
    \errmessage{(Inkscape) Transparency is used (non-zero) for the text in Inkscape, but the package 'transparent.sty' is not loaded}%
    \renewcommand\transparent[1]{}%
  }%
  \providecommand\rotatebox[2]{#2}%
  \newcommand*\fsize{\dimexpr\f@size pt\relax}%
  \newcommand*\lineheight[1]{\fontsize{\fsize}{#1\fsize}\selectfont}%
  \ifx\svgwidth\undefined%
    \setlength{\unitlength}{173.27171746bp}%
    \ifx\svgscale\undefined%
      \relax%
    \else%
      \setlength{\unitlength}{\unitlength * \real{\svgscale}}%
    \fi%
  \else%
    \setlength{\unitlength}{\svgwidth}%
  \fi%
  \global\let\svgwidth\undefined%
  \global\let\svgscale\undefined%
  \makeatother%
  \begin{picture}(1,0.49652913)%
    \lineheight{1}%
    \setlength\tabcolsep{0pt}%
    \put(0,0){\includegraphics[width=\unitlength,page=1]{nfcontext.pdf}}%
    \put(0.21122453,0.21597132){\color[rgb]{0,0,0}\makebox(0,0)[lt]{\lineheight{1.25}\smash{\begin{tabular}[t]{l}$P_0$\end{tabular}}}}%
    \put(0,0){\includegraphics[width=\unitlength,page=2]{nfcontext.pdf}}%
    \put(0.79974469,0.0875688){\color[rgb]{0,0,0}\makebox(0,0)[lt]{\lineheight{1.25}\smash{\begin{tabular}[t]{l}$P_k$\end{tabular}}}}%
    \put(0,0){\includegraphics[width=\unitlength,page=3]{nfcontext.pdf}}%
    \put(0.79974469,0.26936262){\color[rgb]{0,0,0}\makebox(0,0)[lt]{\lineheight{1.25}\smash{\begin{tabular}[t]{l}$P_2$\end{tabular}}}}%
    \put(0.79974469,0.3732458){\color[rgb]{0,0,0}\makebox(0,0)[lt]{\lineheight{1.25}\smash{\begin{tabular}[t]{l}$P_1$\end{tabular}}}}%
    \put(0.4863,0.22805){\makebox(0,0){$h$}}%
  \end{picture}%
\endgroup%

\caption{The context $M=\Context(P_0,\dots,P_k)$. The square is the hole~$h$ of rank~$k$; filling it with~$\delta$ gives $M[\delta]$. Throughout these figures, dashed boxes denote whole set-systems; the hole and the vertex labelled~$\delta$ have solid outlines. Internal roots of the pieces and contributions of formal markers are omitted.}
\label{fig:extremal-context}
\end{figure}
Expand the formal markers in $M[\delta]$, so that~$h$ is now labelled~$\delta$.
For $1\leqslant i\leqslant k$, we construct an ordinary unary "set-system"~$U_i$ as in \Cref{fig:extremal-construction}. To build $U_i$, we keep the vertices and edges of $M[\delta]$, take $h$ as the only "initial vertex", retain all its "root vertices", and additionally make all its former "initial vertices" roots. Finally add the edge $(h,i,x_1)$. Set
\[
D=\ssSum_{i=1}^k U_i(x_i),\qquad d=\yaeval(D).
\]
For $k=0$, instead let~$D$ be the closed "set-system" obtained in the same way, with "initial vertex"~$h$ and with no added "variable edge".

\begin{figure}[H]
\centering
\def\svgwidth{.96\linewidth}
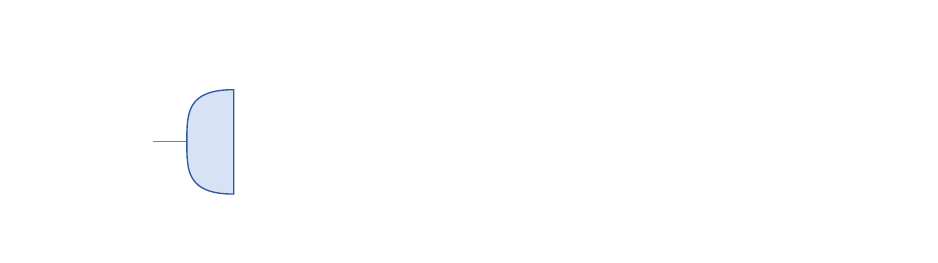
\caption{Construction of $U_i$ and $D$. In $U_i$, the extra choice in direction~$i$ leads to~$x_1$, and the former initial vertices represented by~$P_0$ become roots.}
\label{fig:extremal-construction}
\end{figure}
Intuitively, we can think of~$D$ as a "disjoint sum" of copies of $M[\delta]$, one for each direction, so it internalizes both the context $M$ and the extremal element~$\delta$. 

\underline{$M[d]$ is accepted}:

There is a "morphism@@ss" $\plant(M[D]){\ssmorphism}\plant(M[\delta])$, sending every copied vertex to its original. In particular, each added exit in direction~$i$ is connected to the "initial points" of~$P_i$, exactly as at~$h$. Root markers are preserved because all roots and former initials of~$M[\delta]$ were retained as roots in every copy. The "morphism@@ss" axiom of "set-algebras" and invariance of acceptance under planting give $\Acc(M[d])$.

\underline{$d=\delta$:}

Let $C$ be an arbitrary "set-context", of the form $C=\Context(Q_0,\dots,Q_k)$ and assume $\Acc(C[d])$. For $0\leqslant i\leqslant k$, define
\[
T_i:=\plant\bigl(\Context(P_0,\dots,P_{i-1},P_i\yameet Q_i,P_{i+1},\dots,P_k)[\delta]\bigr).
\]
Here the piece $P_i\yameet Q_i$ is realised by the disjoint sum of the expanded pieces, which has the required extended value. The cases $i>0$ and $i=0$ are illustrated in \Cref{fig:extremal-mixed-contexts}.
\begin{figure}[H]
\centering
\begin{minipage}[c]{.53\linewidth}
\def\svgwidth{.98\linewidth}
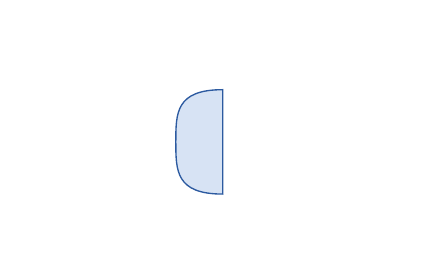
\end{minipage}\hfill
\begin{minipage}[c]{.43\linewidth}
\def\svgwidth{.98\linewidth}
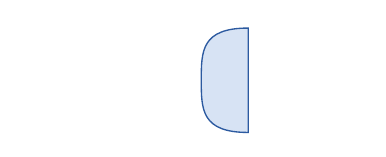
\end{minipage}
\caption{The set-systems $T_i$ for $i>0$ (left) and $T_0$ (right). In~$T_i$, direction~$i$ admits both~$P_i$ and~$Q_i$. In~$T_0$, both~$P_0$ and~$Q_0$ contribute roots. }
\label{fig:extremal-mixed-contexts}
\end{figure}
For every $0\leqslant i\leqslant k$, there is a "morphism@@ss" $T_i\ssmorphism\plant(C[D])$.
For $i>0$, use the copy~$U_i$ inside of $D$: its original edges supply~$P_i$ and its added exit supplies~$Q_i$. For $i=0$, use any copy~$U_j$ (or~$D$ itself when $k=0$), and the outer piece~$Q_0$ in $C$. Thus we have $\Acc(T_i)$ for each $i$, by the "morphism" axiom, since $\Acc(C[D])$ and $T_i\ssmorphism\plant(C[D])$. 
Minimality of $(P_0,\dots,P_k,\delta)$ implies $P_i\yameet Q_i=P_i$, hence $P_i\yaleq Q_i$ for every~$i$.

By monotonicity, these inequalities together with $\Acc(M[\delta])$ imply $\Acc(C[\delta])$. Since this holds for any context $C$ succh that $\Acc(C[d])$, and $\yaleq$ is the syntactic order, we have $d\yaleq\delta$. Together with $\Acc(M[d])$, minimality of $(P_0,\dots,P_k,\delta)$ gives $d=\delta$. 

\textbf{Conclusion with the forward implication of the lemma:} 

Assume $\Acc(\Context(Q_0,\dots,Q_k)[\delta])$. We have $d=\delta$ so $\Acc(\Context(Q_0,\dots,Q_k)[d])$. Therefore, the reasoning above can be made, allowing us to conclude that $P_i\yaleq Q_i$ for every~$i$.
\end{proof}

\begin{lemma}\label{lemma:extremal-directions}
For every "extremal element"~$\delta\in Y_k$ with $k\geqslant1$, we have
\[
\delta=\bigmeet_{i=1}^k\operatorname{dir}_i(\delta)(x_i).
\]
\end{lemma}
\begin{proof}
Choose an "extremal accepted situation" witnessing that~$\delta$ is an "extremal element". The construction in the proof of \Cref{lemma:extremal-context-filter} gives unary "set-systems"~$U_j$ with
$\delta=\yaeval(\ssSum_j U_j(x_j))$.
Projecting to direction~$i$ keeps~$U_i(x_i)$ and replaces every other summand~$U_j(x_j)$ by~$U_j[0]$, viewed as a rank-$k$ "set-system" with unused variables. Hence the meet of all these projections is the value of the original sum together with copies of such cut-off summands. Each $U_j[0]$ maps to~$U_j(x_j)$ by the identity on vertices. By the "morphism@@ss" axiom in \Cref{definition:set-algebras}, its value is above the value of the original sum, so these extra summands do not change the meet. We obtain $\bigmeet_{i=1}^k\operatorname{dir}_i(\delta)(x_i)=\yaeval(\ssSum_j U_j(x_j))=\delta$.
\end{proof}

\begin{lemma}\label{lemma:extremal-direction-is-extremal}
Every "direction projection" of an "extremal element" is extremal in~$Y_1$.
\end{lemma}
\begin{proof}
Let~$(P_0,\dots,P_k,\delta)$ be extremal, put $M=\Context(P_0,\dots,P_k)$ and $b_j=\operatorname{dir}_j(\delta)$, and fix~$i$. By \Cref{lemma:extremal-directions}, we can represent~$\delta$ inside~$M$ by $\ssSum_j b_j(x_j)$, as illustrated in \Cref{fig:extremal-expanded-context}.
\begin{figure}[H]
\centering
\begingroup
\definecolor{directionblue}{HTML}{2C5AA0}
\definecolor{directionfill}{HTML}{D7E3F4}
\definecolor{directionedge}{HTML}{808080}
\definecolor{directionchoice}{HTML}{6C5353}
\begin{tikzpicture}[
  x=1cm,y=1cm,
  every node/.style={font=\small},
  wire/.style={draw=directionedge,line width=.45pt},
  atom/.style={draw=directionblue,fill=directionfill,
    rounded corners=1.5pt,line width=.65pt,
    minimum width=7.5mm,minimum height=4.5mm,inner sep=1.5pt},
  piece/.style={atom,rounded corners=2pt,
    minimum width=12mm,minimum height=6mm,dash pattern=on 3pt off .8pt},
  choice/.style={diamond,draw=directionchoice,fill=directionchoice!15,
    line width=.65pt,minimum size=3.5mm,inner sep=0pt},
  initial/.style={circle,draw=black!75,fill=black!20,
    line width=.65pt,minimum size=2.6mm,inner sep=0pt},
  >=stealth
]
  \node[anchor=west] at (-.3,1.5) {$M[\delta]$};
  \node[initial] (minit) at (-.3,0) {};
  \node[piece] (pzero) at (.8,0) {$P_0$};
  \node[choice] (mentry) at (1.8,0) {};
  \node[atom] (bone) at (3,.6) {$b_1$};
  \node[atom] (bk) at (3,-.6) {$b_k$};
  \node[piece] (pone) at (4.65,.6) {$P_1$};
  \node[piece] (pk) at (4.65,-.6) {$P_k$};
  \foreach \x in {3,4.65} {
    \foreach \dy in {-.15,0,.15} {
      \fill[directionblue] (\x,\dy) circle[radius=.35mm];
    }
  }
  \draw[wire] (minit) -- (pzero);
  \draw[wire] (pzero) -- (mentry);
  \draw[wire] (mentry) -- (bone);
  \draw[wire] (mentry) -- (bk);
  \draw[wire] (bone) -- (pone);
  \draw[wire] (bk) -- (pk);
  \draw[wire,->] (pone.east)
    .. controls (6.1,1.85) and (1.8,1.85) .. (mentry.north);
  \draw[wire,->] (pk.east)
    .. controls (6,-1.85) and (1.8,-1.85) .. (mentry.south);
\end{tikzpicture}
\endgroup
\caption{A representation of $M[\delta]$ with~$\delta$ expanded into its unary summands $b_1(x_1),\dots,b_k(x_k)$.}
\label{fig:extremal-expanded-context}
\end{figure}
We form a closed unary "set-context"~$C[\hole_1]$ by replacing the $i$th summand in $M[\ssSum_j b_j(x_j)]$ by a unary hole, as illustrated in \Cref{fig:extremal-direction-context}. It is accepted when filled by~$b_i$. By \Cref{lemma:extremal-replacement}, there is an extremal $c\yaleq b_i$ with $\Acc(C[c])$. Let $d=c(x_i)\yameet\bigmeet_{j\neq i}b_j(x_j)$.
\begin{figure}[H]
\centering
\begingroup
\definecolor{directionblue}{HTML}{2C5AA0}
\definecolor{directionfill}{HTML}{D7E3F4}
\definecolor{directionedge}{HTML}{808080}
\definecolor{directionchoice}{HTML}{6C5353}
\begin{tikzpicture}[
  x=1cm,y=1cm,
  every node/.style={font=\small},
  wire/.style={draw=directionedge,line width=.45pt},
  atom/.style={draw=directionblue,fill=directionfill,
    rounded corners=1.5pt,line width=.65pt,
    minimum width=7.5mm,minimum height=4.5mm,inner sep=1.5pt},
  piece/.style={atom,rounded corners=2pt,
    minimum width=12mm,minimum height=6mm,dash pattern=on 3pt off .8pt},
  unaryhole/.style={draw=directionchoice,fill=white,line width=.65pt,
    minimum width=6mm,minimum height=6mm,inner sep=1pt},
  choice/.style={diamond,draw=directionchoice,fill=directionchoice!15,
    line width=.65pt,minimum size=3.5mm,inner sep=0pt},
  initial/.style={circle,draw=black!75,fill=black!20,
    line width=.65pt,minimum size=2.6mm,inner sep=0pt},
  >=stealth
]
  \node[anchor=west] at (-.3,2.1) {$C[\hole_1]$};
  \node[initial] (cinit) at (-.3,0) {};
  \node[piece] (pzero) at (.8,0) {$P_0$};
  \node[choice] (centry) at (1.8,0) {};
  \node[atom] (bone) at (3,1.2) {$b_1$};
  \node[unaryhole] (hi) at (3,0) {$h_i$};
  \node[atom] (bk) at (3,-1.2) {$b_k$};
  \node[piece] (pone) at (4.65,1.2) {$P_1$};
  \node[piece] (pi) at (4.65,0) {$P_i$};
  \node[piece] (pk) at (4.65,-1.2) {$P_k$};
  \foreach \x in {3,4.65} {
    \foreach \y in {.6,-.6} {
      \foreach \dy in {-.15,0,.15} {
        \fill[directionblue] (\x,{\y+\dy}) circle[radius=.35mm];
      }
    }
  }
  \draw[wire] (cinit) -- (pzero);
  \draw[wire] (pzero) -- (centry);
  \draw[wire] (centry) -- (bone);
  \draw[wire] (centry) -- (hi);
  \draw[wire] (centry) -- (bk);
  \draw[wire] (bone) -- (pone);
  \draw[wire] (hi) -- (pi);
  \draw[wire] (bk) -- (pk);
  \draw[wire,->] (pone.east)
    .. controls (6.1,2.55) and (1.8,2.55) .. (centry.north);
  \draw[wire,->] (pi.east)
    .. controls (7.4,3.4) and (1.3,3.4) .. (centry.north west);
    \draw[wire,->] (pk.east)
    .. controls (6,-2.5) and (1.8,-2.5) .. (centry.south);

  \node[anchor=west] at (6.4,2.1) {$d$};
  \node[initial] (dinit) at (6.4,0) {};
  \node[choice] (dentry) at (7.1,0) {};
  \node[atom] (dbone) at (8.4,1.2) {$b_1$};
  \node[atom] (dc) at (8.4,0) {$c$};
  \node[atom] (dbk) at (8.4,-1.2) {$b_k$};
  \node[anchor=west] (xone) at (9.7,1.2) {$x_1$};
  \node[anchor=west] (xi) at (9.7,0) {$x_i$};
  \node[anchor=west] (xk) at (9.7,-1.2) {$x_k$};
  \foreach \x in {8.4,9.9} {
    \foreach \y in {.6,-.6} {
      \foreach \dy in {-.15,0,.15} {
        \fill[directionblue] (\x,{\y+\dy}) circle[radius=.35mm];
      }
    }
  }
  \draw[wire] (dinit) -- (dentry);
  \draw[wire] (dentry) -- (dbone);
  \draw[wire] (dentry) -- (dc);
  \draw[wire] (dentry) -- (dbk);
  \draw[wire] (dbone) -- (xone);
  \draw[wire] (dc) -- (xi);
  \draw[wire] (dbk) -- (xk);
\end{tikzpicture}
\endgroup
\caption{The unary context~$C[\hole_1]$ (left) and the disjoint sum evaluating to~$d$ (right). The vertical dots indicate the omitted directions. Filling~$h_i$ with~$b_i$ restores the decomposition in \Cref{fig:extremal-expanded-context}; filling it with~$c$ replaces the $i$th summand. Internal roots and contributions of formal markers are omitted.}
\label{fig:extremal-direction-context}
\end{figure}
Then $d\yaleq\delta$ and $\Acc(M[d])$, so extremality gives $d=\delta$. Therefore, we have $b_i=\operatorname{dir}_i(\delta)=\operatorname{dir}_i(d)\yaleq\operatorname{dir}_i(c(x_i))=c$. We can conclude that $b_i=c$ is extremal.
\end{proof}

We now show that acceptance for an extremal element is determined by acceptance in each direction: 

\begin{lemma}[behavior of extremal elements]\label{lemma:restore-extremal-successors}
Let~$H$ be a closed "$Y$-set-system", and let~$v$ be a unary vertex with an "extremal element"~$b$ as its label and a nonempty set~$E_v$ of outgoing edges. For each~$e\in E_v$, let~$H_e$ be obtained from $H$ by keeping only $e$ as outgoing edge of $v$. If $\Acc(H_e)$ for every~$e\in E_v$, then $\Acc(H)$.
\end{lemma}
\begin{proof}
\emph{Expressing acceptance in terms of context pieces.}
Replace~$v$ by a unary hole in~$H$ and in each~$H_e$. The resulting "set-contexts" can be decomposed into "pieces@@context"~$(Q_0,Q_1)$ and~$(Q_0,Q_{1,e})$ respectively. 
Recall that a "piece" $Q$ is a pair $(Q^{\circ},Q^{\bullet})$ of an ordinary "set-system" $Q^{\circ}$ and a set $Q^{\bullet}\subseteq\{\varepsilon,\mathsf r\}$ of formal markers. 

Set
\[
q_0:=\yaeval^+(Q_0),\qquad
q_1:=\yaeval^+(Q_1),\qquad
q_{1,e}:=\yaeval^+(Q_{1,e}).
\]
By \Cref{lemma:context-decomposition,lemma:extended-context-evaluation},
\begin{align*}
\yaeval(H)&=\yaeval\bigl(\Context(q_0,q_1)[b]\bigr),\\
\yaeval(H_e)&=\yaeval\bigl(\Context(q_0,q_{1,e})[b]\bigr).
\end{align*}

\emph{Computing the outgoing piece.}
The pieces~$Q_1$ and~$Q_{1,e}$ have the same vertices, labels, edges, and root points. Only their initial points differ: $Q_{1,e}$ starts at the target of~$e$, represented by~$x_1$ when~$e$ is a self-loop, whereas~$Q_1$ starts at all these targets. We claim that
\begin{equation}\label{eq:outgoing-piece-meet}
q_1=\bigmeet_{e\in E_v}q_{1,e}.
\end{equation}
Indeed, the map identifying corresponding copies of vertices is a "folding" from the disjoint sum of the ordinary parts~$Q_{1,e}^{\circ}$ onto~$Q_1^{\circ}$. It preserves the initial sets because their union is the initial set of~$Q_1^{\circ}$; edges and roots agree in every copy. Unfold-invariance and the evaluation axiom of \Cref{definition:set-algebras} therefore give
\[
\yaeval(Q_1^{\circ})=\bigmeet_{e\in E_v}\yaeval(Q_{1,e}^{\circ}).
\]
The formal markers of~$Q_1$ are likewise the union of those of the~$Q_{1,e}$. The meet formula in \Cref{definition:extended-set-algebras} now gives \eqref{eq:outgoing-piece-meet}.

\emph{Using extremality.}
Choose an "extremal accepted situation"~$(P_0,P_1,b)$ witnessing extremality of~$b$. The context evaluation identities above and \Cref{lemma:extremal-context-filter} give, for each~$e\in E_v$,
\[
\Acc(H_e)\quad\Longleftrightarrow\quad
P_0\yaleq q_0\ \text{ and }\ P_1\yaleq q_{1,e}.
\]
By hypothesis, these inequalities hold for every~$e$. Since~$E_v$ is nonempty and the meet is an infimum (\Cref{lemma:meet-is-meet}), \eqref{eq:outgoing-piece-meet} yields
\[
P_0\yaleq q_0,\qquad
P_1\yaleq\bigmeet_{e\in E_v}q_{1,e}=q_1.
\]
Another application of \Cref{lemma:extremal-context-filter} gives $\Acc(\Context(q_0,q_1)[b])$, which is equivalent to~$\Acc(H)$.
\end{proof}

This can now be lifted to a path criterion for acceptance of "set-systems" with "extremal elements" at unary vertices:
\begin{lemma}[path criterion]\label{lemma:extremal-path-criterion}
Let~$G$ be a closed "$\Yw$-set-system" with no "root vertices", with every vertex reachable from an "initial vertex", and with at least one outgoing edge at every unary vertex. Suppose every unary label is an "extremal element". Then $\Acc(G)$ if and only if every "regular branch" of~$G$ is accepted.
\end{lemma}
\begin{proof}
If $\Acc(G)$, then every "regular branch"~$B$ is accepted: by definition $B$ is a closed "$\Yw$-system" with a "morphism@@ss" $B\ssmorphism G$, and the "morphism@@ss" axiom of "set-algebras" gives $\yaeval(G)\yaleq\yaeval(B)$. The accepting set is upward closed, so $\Acc(B)$.

Conversely, suppose every "regular branch" of~$G$ is accepted. We prove by induction that every $H$ obtained from $G$ by retaining a nonempty subset of the outgoing edges at each unary vertex is accepted. 

The induction is on the number of additional edges per unary vertex:
\[
N(H):=\sum_{\substack{v\in\V_H\\\rk(H(v))=1}}
\bigl(|\Edges_H(v)|-1\bigr).
\]
If~$N(H)=0$, then~$H$ is a "folding" image of a disjoint sum of branches from its initial vertices and tails with no initial vertex; each summand maps into an accepted "regular branch" of~$G$, so the "morphism@@ss" axiom and closure of~$F$ under finite meets (\Cref{lemma:synt-recognition}) give $\Acc(H)$.

If~$N(H)>0$, choose a unary vertex~$v$ with at least two outgoing edges. For each outgoing edge~$e$, retain only~$e$ at~$v$ to obtain~$H_e$. Each~$H_e$ still belongs to the family under consideration and satisfies $N(H_e)<N(H)$, so it is accepted by induction. The label at~$v$ is extremal, so \Cref{lemma:restore-extremal-successors} gives $\Acc(H)$.

Taking~$H=G$ completes the proof.
\end{proof}

\begin{corollary}\label{cor:synt-aut-prop}
Let~$L$ be a "language of $\alphabet$-regular-trees" recognised by a "rankwise-finite" "algebra". Then its "syntactic set-algebra", with its "accepting set", has the "automaton property".
\end{corollary}
\begin{proof}
By the reduction to the ordinary "syntactic algebra" above and \Cref{lemma:algebra-to-set-algebra}, its "syntactic set-algebra" is "rankwise-finite". The first condition of \Cref{def:autprop} follows from \Cref{lemma:extremal-directions,lemma:extremal-direction-is-extremal}, and the second is \Cref{lemma:extremal-path-criterion}.
\end{proof}

\section{MSO, automata, bisimulation and \texorpdfstring{$\mu$}{mu}-calculus}\label{sec:mso}
\label{section:bisim-invariant}

In this section, we combine the different tools seen so far, and complete the proof of our main theorems.

We use standard definitions of the "Monadic Second-Order logic@monadic second-order logic" ("MSO") and $\mu$-calculus on "transition systems", together with a corresponding notion of "MSO" for our ranked "systems". We assume the reader has a good familiarity with "MSO".
Further definitions and notations can be found in~\cite{Thomas97}.
\AP We shall use finite ""signatures"" containing ""relational symbols"" only (i.e.\ no functional symbols). The notion of a ""structure"" over a given "signature" is the usual one. Given a sentence~$\varphi$ and a structure~$S$, we say that $S$ ""models"" $\varphi$, noted $S\models \varphi$, if $\varphi$ is true in~$S$.

In \Cref{subsection:transition-system}, we define "transition systems" and show their relationship with our ranked "systems".
In \Cref{subsection:MSO}, we interpret "MSO" on "transition systems" and their closed "system" encodings, and translate "MSO" into recognition by "rankwise-finite" "algebras".

In \Cref{subsection:algebra-to-automata}, we use the "automaton property" to translate algebraic recognition of "languages of regular-trees" into "unfold-automata".
Finally, in \Cref{subsection:bisimulation-invariant-case}, we use this result in the fragment of "bisimulation-invariant" languages to show that "bisimulation-invariant" "MSO" and $\mu$-calculus are equi-expressive.

\subsection{Transition systems as systems, and bisimulation}
\label{subsection:transition-system}

In this section, we define the notion of "transition systems", which is the classical one, and state how these objects can be encoded in "$\alphabetTS$-systems", where $\alphabetTS$ is a suitably defined "ranked alphabet". We shall also see how to define a notion of ``"bisimulation"'' over "$\alphabetTS$-systems" that corresponds to the usual definition of "bisimilarity" (see \Cref{lemma:bisimulation-is-bisimulation}).

Let us fix a finite set~$\Pred$ of unary predicates.

\AP A ""transition system"" $(V,i,T,\gamma)$ consists of a finite\footnote{In this work, as for "systems", "transition systems" are finite, unless explicitly stated otherwise.} set of ""vertices@@ts""~$V$, an ""initial vertex@@ts""~$i$, a binary relation~$T$ called the ""transition relation@@ts"", and a labelling function $\gamma:V\to 2^\Pred$. As before, we assume all "vertices@@ts" to be reachable from the "initial vertex@@ts".

We now explain how to encode "transition systems" as "systems" as defined in this document.

\AP 
Let $\intro*\alphabetTS$ be the "ranked alphabet" which has an element $\nu_n$ of "rank"~$n$ for all ""$\Pred$-valuation"" $\nu\in2^\Pred$, and all~$n\in\N$. 
	We can see "transition systems" as "$\alphabetTS$-systems" as follows.
Given a "closed@@system" "$\alphabetTS$-system"~$S$, its ""decoding@@ts"" is the "transition system"~$\intro*\tsdecode(S)$ with the same vertices and initial vertex, valuation~$\gamma(v)=\nu$ when~$S(v)=\nu_n$, and $T(u,v)$ if~$(u,d,v)$ is an edge for some~$d$. Conversely, $S$ is called an ""encoding@@ts"" of~$\tsdecode(S)$. The decoding is unique, while several non-isomorphic closed "$\alphabetTS$-systems" may encode the same "transition system", since directions may be reordered or duplicated. Every "transition system" admits an "encoding@@ts" by assigning one direction to each distinct successor.

\AP In this work, we define the ""bisimilarity"" relation as the least equivalence relation over "closed $\alphabetTS$-systems" that contains "unfold-equivalence@@ss" and the relation ``"encoding@@ts" the same "transition system"''. The following lemma states why this is consistent with the standard terminology.
\begin{lemma}\AP\label{lemma:bisimulation-is-bisimulation}
	Two closed "$\alphabetTS$-systems" are "bisimilar" if and only if their "decodings@@ts" are "bisimilar" in the usual sense.
\end{lemma}
\begin{proof}
	Let us recall the following definition of ``usual "bisimulation"''.
	Two "transition systems"~$(V,i,T,\gamma)$ and~$(V',i',T',\gamma')$ of respective vertex sets~$V,V'$ are ``""usual-bisimilar""'' if there exists a relation $R\subseteq V\times V'$ such that $R(i,i')$, and for all $R(x,x')$:
	\begin{itemize}
	\item $\gamma(x)=\gamma'(x')$,
	\item For all~$y$ such that $T(x,y)$ there exists $y'\in V'$ such that $T'(x',y')$ and $R(y,y')$.
	\item For all~$y'$ such that $T'(x',y')$ there exists $y\in V$ such that $T(x,y)$ and $R(y,y')$.
	\end{itemize}
	Let now~$S,S'$ be closed "$\alphabetTS$-systems". We have to show that $\tsdecode(S)$ and $\tsdecode(S')$ are "usual-bisimilar" if and only if $S$ and $S'$ are "bisimilar" according to our definition.

	\textit{First direction.}
	Assume that $\tsdecode(S)=\tsdecode(S')$, then clearly $\tsdecode(S)$ and $\tsdecode(S')$ are "usual-bisimilar", witnessed by the identity relation.
	If $S'$ is an "unfolding@@ss" of~$S$, witnessed by a "folding@@ss" $\eta\colon S'\ssfolding S$, then the graph of~$\eta$ is a usual "bisimulation" between their decodings: preservation of labels gives the same valuations, and local surjectivity gives the successor conditions. Thus "unfold-equivalent@@ss" systems have "usual-bisimilar" decodings, using a common "system" unfolding from \Cref{lemma:common-system-unfolding}. By transitivity, "bisimilar"~$S,S'$ have "usual-bisimilar" decodings.
	
	\textit{Converse direction.}
	It is sufficient to show, given two "usual-bisimilar" "transition systems"~$\T=(V,i,T,\gamma)$ and $\T'=(V',i',T',\gamma')$, that there exist two "bisimilar" "$\alphabetTS$-systems"~$S$ and~$S'$ that respectively "encode@@ts" $\T$ and $\T'$.
	
	We first claim that the result holds when the relation~$R\subseteq V\times V'$ that witnesses that~$\T$ and~$\T'$ are "usual-bisimilar" is the graph of a map $\eta$ from~$V$ to~$V'$.
	Let~$n$ be the maximal ""out-degree@@ts"" of a "vertex@@ts"~$v\in V$.
	We define~$S'$ to have the same vertices, and "initial vertex@@ts" as~$\T'$.
	For every vertex~$v'\in V'$ with valuation~$\nu$ and distinct successors $v'_1,\dots,v'_{n'}$ in~$\T'$ (in a fixed order), set $S'(v'):=\nu_{nn'}$ and let its successor in direction~$i+n'j$ be~$v'_i$, for $i\in\interval{n'}$ and $0\leqslant j<n$.
 	Clearly, $\tsdecode(S')=\T'$.

	We now define $S$ to have the same vertices and "initial vertex@@ts" as~$\T$.
	For all vertices~$v\in V$ we set $S(v):=S'(\eta(v))$.
	Now, let~$v'_1,\dots,v'_{n'}$ (chosen in the same order as above) be the successor vertices of $\eta(v)$ in $\T'$, and for each $i\in \interval{n'}$, let~$v_{i,0},\dots,v_{i,k_i-1}$ be the successors of~$v$ in $\T$ that are mapped by~$\eta$ to~$v'_i$.
	By definition of "bisimulation", we have $k_i\geq 1$, and hence $v_{i,0}$ is well defined.
	For $i\in\interval{n'}$ and $0\leqslant j<n$, define the successor of~$v$ in direction~$i+n'j$ to be~$v_{i,j}$ if~$j<k_i$, and~$v_{i,0}$ otherwise.
	Note that by choice of~$n$, $k_i\leqslant n$, and as a consequence, all successors of~$v$ in~$\T$ are also successors of~$v$ in~$S$.
	The image under~$\eta$ of this successor is~$v'_i$, the successor of~$\eta(v)$ in the same direction.
	Consequently, $\tsdecode(S)=\T$ and~$\eta$ is a "morphism@@ss" $S\ssmorphism S'$. Since both are "systems", \Cref{remark:system-morphisms} makes~$\eta$ a "folding@@ss", so~$S$ is an "unfolding@@ss" of~$S'$.
	This concludes the proof of the claim.

	A simple example of this construction is presented below:
	\begin{center}
		\scalebox{.8}{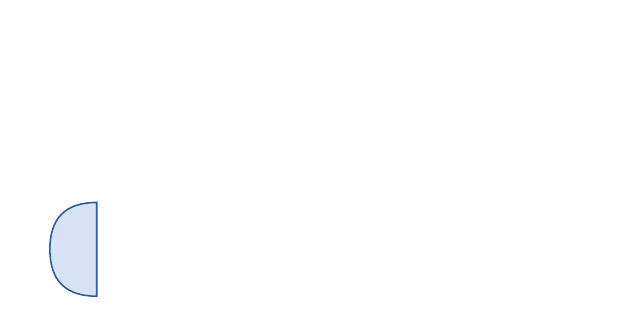}
	\end{center}
	
	Let us assume now that~$R\subseteq V\times V'$ is a witness that $\T=(V,i,T,\gamma)$ and $\T'=(V',i',T',\gamma')$ are "usual-bisimilar".
	On~$R$, put $T_R=\{((x,x'),(y,y'))\in R^2\mid T(x,y),~T'(x',y')\}$ and $\gamma_R(x,x')=\gamma(x)=\gamma'(x')$. Let~$\T_R$ be the restriction of this labelled graph to the pairs reachable from~$(i,i')$, with~$(i,i')$ as its initial vertex. This is a "transition system" satisfying our reachability convention.
	Each coordinate projection witnesses a usual "bisimulation" from~$\T_R$ to~$\T$ or~$\T'$: the successor conditions for~$R$ still hold after restriction, since a successor of a reachable pair is reachable.
	The functional case therefore gives closed "$\alphabetTS$-systems" $S,S_R,S'_R,S'$ with $\tsdecode(S)=\T$, $\tsdecode(S_R)=\tsdecode(S'_R)=\T_R$, $\tsdecode(S')=\T'$, and "foldings@@ss" $S_R\ssfolding S$ and $S'_R\ssfolding S'$.
	As a consequence~$S$ and $S'$ are "bisimilar".
\end{proof}

\subsection{From monadic second-order logic to algebras}
\label{subsection:MSO}




\AP
""Monadic second-order logic"" (or simply \reintro*"MSO") is the extension of first-order logic with ""set variables"" $X,Y,Z,\dots$, the possibility to write existential and universal ""set quantifiers"", and a membership predicate $x\in Y$ expressing that an element belongs to a set.

In order to define "MSO" over (transition) "systems", we therefore need to define the atomic predicates that will be used in the logic. Let us start by presenting the canonical definition for "MSO" over "transition systems".




\paragraph*{MSO on classical "transition systems".}
Let~$\T=(V,i,T,\gamma)$ be a "transition system" over the fixed finite set~$\Pred$ of unary predicates. We interpret "MSO" over the relational signature containing a unary symbol~$I$ for the "initial vertex@@ts"~$i$, a binary symbol~$T$ for the "transition relation@@ts", and a unary symbol~$p$ for each~$p\in\Pred$, interpreted as $\{v\in V\mid p\in\gamma(v)\}$.
So the grammar of "MSO" on "transition systems" is given by:
\[
\varphi ::= x\in X \mid I(x) \mid  p(x) \mid T(x,y) \mid \neg \varphi \mid \varphi_1 \lor \varphi_2 \mid \exists x.\varphi \mid \exists X.\varphi
\]
The usual syntactic sugar $\land,\Rightarrow,\forall x,\forall X$ is defined as expected.

Satisfaction of an MSO-formula $\varphi$ in $\T$ is the standard Tarski semantics: assignments give values to first-order variables $x,y,\dots$ in $V$ and to second-order variables $X,Y,\dots$ as subsets of $V$, atomic formulas are evaluated using the relations above, and quantifiers range over elements/subsets of $V$.

\begin{example}\label{example:MSO}
If $p\in\Pred$, the formula $\exists x.p(x)$ states that there is a vertex where $p$ holds. Recall that all vertices are reachable from the "initial vertex@@ts".
\end{example}

Now, with \Cref{theorem:bisim-invariant-finite} as target result, we define the "MSO" atomic predicates
on "$\alphabetTS$-systems" in a way that reflects the structure of the encoded "transition system".

\paragraph*{MSO on "$\alphabetTS$-systems".}
In order to be able to reproduce \Cref{example:MSO} in the setting of "systems", we need to ensure some form of uniformity across ranks. If we were to naively define "MSO" on "$\alphabet$-systems" using the full $\alphabet$ as signature (which would be infinite), no finite formula could capture this simple example, as one would need to use symbols of arbitrary high arity in a unique formula.
See \Cref{remark:non-uniformMSO} for more details on the non-uniform "MSO" setting.
On the other hand, "$\alphabetTS$-systems" have this uniformity built-in, allowing us to rely on the underlying set of predicates. We can now define "MSO" on "$\alphabetTS$-systems", using the same signature as for "transition systems".
For a "closed@@system" "$\alphabetTS$-system"~$S$, define $S\models\varphi$ by $\tsdecode(S)\models\varphi$, using the same vertex and set assignments for any free logical variables. Directions in~$S$ are forgotten by this decoding.

\begin{lemma}\label{lemma:MSO-is-recognisable}
	Given an "MSO-sentence"~$\varphi$, the "language of systems"~$\{S\text{ closed "$\alphabetTS$-system"}\mid S\models\varphi\}$ is "recognisable" by a "rankwise-finite" "algebra".
\end{lemma}
\begin{proof}


	For simplicity, we will use here only second-order variables of "MSO", via the following grammar:
\[
\varphi ::= X\subseteq Y \mid I(X) \mid p(X) \mid T(X,Y) \mid\neg \varphi \mid \varphi_1 \lor \varphi_2 \mid \exists X.\varphi
\]
Here $I(X)$ means that the initial vertex belongs to~$X$, $p(X)$ means that $p$ holds at every vertex in~$X$, and $T(X,Y)$ means that $T(x,y)$ holds for some $x\in X$ and $y\in Y$. On singleton sets these predicates agree with their first-order counterparts.

This pure second-order "MSO" is equivalent to the variant with both first and second-order variables. Indeed, it is straightforward to translate the second-order atomic predicates to their first-order counterparts, and conversely, a formula with first-order variables could be replaced with one in the pure second-order fragment, by replacing each first-order variable $x$ with a second-order variable $X$ constrained by the predicate $\mathit{Singleton}(X):=X\neq\emptyset\wedge\forall Y.(Y\subseteq X\Rightarrow (Y=\emptyset\lor Y=X))$, where $X=\emptyset$ is syntactic sugar for $\forall Y. X\subseteq Y$.

We can therefore restrict ourselves to second-order variables in the following.
\AP
To encode valuations of free set variables, we add marking predicates to the input alphabet. For $k\in\N$, a ""$k$-marked system"" is a closed system over~$\alphabetTS_k$, the "ranked alphabet" obtained from~$\alphabetTS$ by adding unary predicates~$\alpha_1,\dots,\alpha_k$. The predicate~$\alpha_i$ records membership in the set assigned to~$X_i$. Thus a "$k$-marked system" determines the truth value of a formula with free variables among~$X_1,\dots,X_k$, and a $0$-marked system is a closed "$\alphabetTS$-system".

We show that for any formula $\varphi$, with $k$ free variables (for any $k\in\N$), the set $\{S\text   { "$k$-marked system"}\mid S\models\varphi\}$ is "recognisable" by a "rankwise-finite" "algebra". We proceed by induction on $\varphi$.

The base case $I(X_1)$ is recognised by a "rankwise-finite" "algebra" $\Alg=(A,\aeval)$, where $A_n=\{0,1\}$ for all $n\in\Nats$, and $\aeval(S)$ is the label of the "initial vertex" of~$S$. The "recognising morphism" is the "morphism@@algebra" determined by sending input letters containing~$\alpha_1$ to~$1$ and the others to~$0$, in the corresponding rank. Its "accepting set" is~$\{1\}$.

For $X_1\subseteq X_2$, use the "rankwise-finite" "algebra" of \Cref{ex:aexists-alg}, mapping a letter to~$1$ exactly when~$\alpha_1$ holds and~$\alpha_2$ does not, and to~$0$ otherwise. The "accepting set"~$\{0\}$ tests the absence of such a vertex, as in \Cref{ex:reco-exists-lang}. The predicate~$p(X_1)$ is treated in the same way, with~$p$ in place of~$\alpha_2$.

For the base case $T(X_1,X_2)$, we have to design a "rankwise-finite" "algebra" that detects the presence of an edge from some vertex in $X_1$ to some vertex in $X_2$. This can be done in the same spirit as the previous case, with an "algebra" $\mathcal{A}=(A,\aeval)$, where $A_n=(\powerset(\interval n)^2\times\{0,1\})\uplus\{\top\}$.
An element $(E,F,b)\in A_n$ consists of two sets $E,F\subseteq \interval n$ and a boolean $b\in\{0,1\}$. Here~$E$ records the indices of reachable exit variables~$x_j$, $F$ records those with an incoming edge from a vertex carrying~$\alpha_1$, and~$b$ records whether the "initial vertex" carries~$\alpha_2$. The element~$\top$ records the presence of an edge from a vertex carrying~$\alpha_1$ to a vertex carrying~$\alpha_2$, in any direction. The function~$\aeval$ maintains this information under "flattening". For instance, for an "$A$-system"~$S$, $\aeval(S)=\top$ when either:
\begin{itemize}
\item $\top$ can be reached in $S$ from the "initial vertex", following at each vertex $v_i$ with label $(E_i,F_i,b_i)$ a direction belonging to $E_i$, or
\item following such a path leads to some $v$ labelled $(E,F,b)$, with some $j\in F$ such that the $j$-successor of $v$ is labelled by $(E',F',1)$. This corresponds to the case where an edge from $\alpha_1$ to $\alpha_2$ is created by connecting $v$ to $v'$ along direction $j$.
\end{itemize}

The induction cases $\varphi_1\vee\varphi_2$, $\neg\varphi$, and $\exists X.\varphi$ correspond to union, complement, and "projection" along the rank-preserving map that forgets the marking predicate for~$X$. To put two formulas over the same set of free logical variables, we add unused marking predicates by "cylindrification". All these operations on "languages of systems" preserve recognition by a "rankwise-finite" "algebra", by \Cref{statement:recognition-closure}.
\end{proof}

\begin{remark}\label{remark:non-uniformMSO}
In the setting of binary trees that is more usual in automata theory,  \Cref{lemma:MSO-is-recognisable} can be adapted by restricting the alphabet to symbols of rank $2$ and by adding directional edges predicates $T_1,T_2$, matching the theory of "MSO" with two successors.
Thus, every "MSO"-definable language of infinite binary trees, restricted to
"regular trees@regular-tree", is recognisable by a "rankwise-finite" "algebra".
\end{remark}


\AP 
 The sentence~$\psi$ is ""bisimulation-invariant"" if $S\models\psi$ is equivalent to~$S'\models\psi$ for all "bisimilar" closed "$\alphabetTS$-systems"~$S,S'$. By \Cref{lemma:bisimulation-is-bisimulation}, this agrees with invariance under usual "bisimulation" of their decodings. Since our "systems" are finite, "bisimulation-invariant" means over finite transition systems unless stated otherwise. In particular, its "language of systems" is invariant under "unfold-equivalence@@ss", and hence is a "language of regular-trees".


\subsection{From algebra to automata in the unfold-invariant setting}
\label{subsection:algebra-to-automata}

We define "unfold-automata" and their accepting runs (\Cref{definition:unfold-automaton}), and prove that "rankwise-finite" algebraic recognition can be translated into such automata, preserving the semantics for "languages of regular-trees" (\Cref{thm:alg2aut}). The construction uses the "automaton property" of "rankwise-finite" "syntactic set-algebras".

Let~$\Sigma$ be a "ranked alphabet", and put $\Sigma_{>0}=\bigcup_{n>0}\Sigma_n$. For the application to "transition systems", we take~$\Sigma=\alphabetTS$.

We first treat "languages of regular-trees", that is, "languages of systems" invariant under "unfold-equivalence@@ss". The additional assumption of "bisimulation-invariance" will be added in \Cref{subsection:bisimulation-invariant-case}.

An "unfold-automaton" recognises closed "regular-trees" through runs on finite "system" representatives:
\begin{definition}\label{definition:unfold-automaton}
	An ""unfold-automaton"" is a tuple $\aut=(H,\Sigma,\Delta,\Omega)$ where
	\begin{itemize}
		\item $H=H_1\uplus H_0$ is a finite "ranked set" of unary and nullary elements (they roughly correspond to the states of the automaton, but label transitions and leaves rather than vertices),
		\item $\Sigma$ is the "ranked alphabet" of the automaton,
		\item $\Delta=(\Delta_+,\Delta_0)$ is the transition relation, with $\Delta_+:\Sigma_{>0}\to \powerset((H_1)^+)$ where for each $a\in\Sigma_{>0}$, we have $\Delta_+(a)\subseteq (H_1)^{\rk(a)}$, and $\Delta_0:\Sigma_0\to \powerset(H_0)$,
		\item $\Omega\subseteq (H_1)^*H_0\cup(H_1)^\omega$ is an acceptance condition whose finite-word part is regular and whose infinite-word part is $\omega$-regular.
	\end{itemize}
	
\end{definition}

\begin{figure}[H]
  \centering
\begin{tikzpicture}[
  x=1cm,y=1cm,
  every node/.style={font=\small},
  vertex/.style={circle,draw=black!75,fill=black!5,
    minimum size=7mm,inner sep=1pt,line width=.6pt},
  edge/.style={-{Stealth[length=2mm]},line width=.6pt},
  type/.style={fill=white,inner sep=2pt}
]
  \node[vertex] (a) at (0,0) {$a$};
  \draw[edge] (a) -- node[type,pos=.55] {$t_1$} (1.8,1);
  \draw[edge] (a) -- node[type,pos=.55] {$t_2$} (1.8,0);
  \draw[edge] (a) -- node[type,pos=.55] {$t_3$} (1.8,-1);
  \foreach \i/\y in {1/1,2/0,3/-1} {
    \node[right,font=\scriptsize] at (1.8,\y) {\(\i\)};
  }
\end{tikzpicture}
  \caption{A transition $(t_1,t_2,t_3)\in\Delta_+(a)$ for $a\in\Sigma_3$.}
  \label{fig:unfold-automaton-transition}
\end{figure}

We will simply note $\Delta(a)$ for either $\Delta_+(a)$ or $\Delta_0(a)$ depending on the rank of $a$.

Such automata will be used to recognise a set of "regular-trees" in the following way. 

For a closed "system"~$S$, a ""leaf"" is a vertex of rank~$0$. A "branch" is a maximal path from the "initial vertex", either infinite or ending at a leaf, as in \Cref{ex:reco-all-paths}. We represent branches by sequences of edges, retaining their directions.

A run of an "unfold-automaton"~$\aut$ on a closed "regular-tree"~$r$ is a labelling~$\rho$ of the edges and leaves of a representative "system"~$S\in r$ by elements of~$H_1$ and~$H_0$, respectively, consistent with~$\Delta$:
\begin{itemize}
\item for each vertex~$s\in\V_S$ of rank~$k>0$, let~$e_i$ be its unique outgoing edge in direction~$i$. We require $(\rho(e_1),\dots,\rho(e_k))\in\Delta_+(S(s))$. The targets of these edges need not be distinct.
\item for each leaf~$s\in\V_S$, we require $\rho(s)\in\Delta_0(S(s))$.
\end{itemize}

For an infinite "branch"~$\pi=e_1e_2\dots$, put $\rho(\pi)=\rho(e_1)\rho(e_2)\dots$. For a branch~$\pi=e_1\dots e_n$ ending at a leaf~$s$, put $\rho(\pi)=\rho(e_1)\dots\rho(e_n)\rho(s)$. The run is accepting if $\rho(\pi)\in\Omega$ for every branch~$\pi$, and~$r$ is accepted if it has an accepting run. We also call finite and ultimately periodic branches (as edge sequences) "regular branches". They are represented by chains and lassos; when their labels are algebra elements, these are the "regular branches" used in \Cref{def:autprop}.

Choosing a run of~$\aut$ on~$r$ amounts to choosing a representative "system"~$S\in r$, then an element of~$\Delta(a)$ at each vertex labelled~$a$. The word acceptance condition~$\Omega$ checks the resulting branches.

\begin{lemma}[regular branches]\label{lemma:unfold-aut-regular-branches}
Let~$\aut=(H,\Sigma,\Delta,\Omega)$ be an "unfold-automaton". A run~$\rho$ on a representative "system"~$S$ is accepting if and only if $\rho(\pi)\in\Omega$ for every "regular branch"~$\pi$ of~$S$. Consequently, $\aut$ accepts a closed "regular-tree"~$r$ if and only if there are a representative~$S\in r$ and a run on~$S$ such that all "regular branches" are accepting.
\end{lemma}
\begin{proof}
The forward implication is immediate. Conversely, suppose every "regular branch" is accepting. All finite branches are regular, so it remains to consider infinite branches. Since~$\Omega\cap H_1^\omega$ is $\omega$-regular, its complement in~$H_1^\omega$ is recognised by a finite B\"uchi word automaton~$\mathcal{A}$. If an infinite branch of~$S$ is rejecting, an accepting run of~$\mathcal{A}$ on its label word gives an infinite path in the finite product of~$S$ and~$\mathcal{A}$, where an edge~$e$ of~$S$ updates~$\mathcal{A}$ by reading~$\rho(e)$. This path visits an accepting product vertex infinitely often. A finite prefix reaching such a vertex, followed by repetition of a nonempty closed walk through it, is again accepting. Its projection to~$S$ is an ultimately periodic branch whose label word is rejected by~$\Omega$, contradicting the hypothesis. The final assertion follows from the definition of acceptance.
\end{proof}

We are now ready to state the main result of this section:

\begin{theorem}\label{thm:alg2aut}
	Let~$L$ be a "language of regular-trees" over~$\Sigma$ recognised by a "rankwise-finite" "algebra". Then~$L$ can be recognised by an "unfold-automaton".
\end{theorem}

To prove the theorem, let~$\YAlg=(Y,\yaeval)$ be the "syntactic set-algebra" of~$L$, with "accepting set"~$F\subseteq Y_0$. By \Cref{lemma:synt-recognition,cor:synt-aut-prop}, it is "rankwise-finite", $F$ is upward closed, and $(\YAlg,F)$ has the "automaton property". We use simultaneous extremal replacements (\Cref{cor:extremal-replacements}), direction decomposition (\Cref{lemma:extremal-directions,lemma:extremal-direction-is-extremal}), and the path criterion (\Cref{lemma:extremal-path-criterion}) to construct~$\aut=(H,\Sigma,\Delta,\Omega)$.

Take $H_1=Y_1$ and $H_0=Y_0$, the unary and nullary elements of~$\YAlg$.

\begin{figure}[H]
	\centering
	\scalebox{.7}{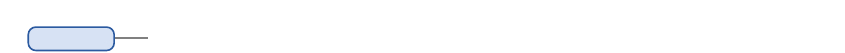}
	\caption{~~~~\textbf{Left}: Evaluating $a_1b_1c_1d_0$ of $Y_1^*Y_0$~~~~~~~~~~~\textbf{Right}: Evaluating $a_1e_1^\omega$ of $Y_1^\omega$}\label{fig:omega}
\end{figure}

Let us define the acceptance condition $\Omega\subseteq Y_1^\omega\cup Y_1^*Y_0$.
For a word~$w\in Y_1^*Y_0$, view~$w$ as a closed "system" over~$\Yw$, a chain ending at its nullary letter (left of \Cref{fig:omega}), and put $w\in\Omega$ exactly when $\Acc(w)$.
The evaluation axiom of "set-algebras" (\Cref{definition:set-algebras}) makes unary composition associative and its action on nullary elements compatible with composition. Hence, word evaluation can be computed in a finite semigroup with elements from $Y_1\uplus Y_0$, so this finite-word language is regular.

For infinite branches, put $V_b=\{w\in (Y_1)^+:\yaeval(w)=b\}$, where a nonempty finite word is viewed as a rank-$1$ "system": a unary chain whose final edge leads to~$x_1$. Unary composition makes~$Y_1$ a finite semigroup; we write~$ae$ for the product of~$a$ and~$e$. Thus, each language~$V_b$ is regular. Define
\[
\Omega\cap Y_1^\omega
=\bigcup_{\substack{a,e\in Y_1\\ae=a,\ e^2=e\\\yaeval(ae^\omega)\in F}}
V_a(V_e)^\omega,
\]
where $ae^\omega$ denotes the closed "system" consisting of an initial $a$-vertex followed by a loop at an $e$-vertex, as on the right of \Cref{fig:omega}. This is an $\omega$-regular language. Ramsey's theorem gives a factorisation of every infinite word into a nonempty prefix of value~$a$ and nonempty blocks of value~$e$, with $ae=a$ and $e^2=e$.

On an ultimately periodic word, the condition agrees with evaluation of its unary lasso. Indeed, refine a homogeneous factorisation by retaining infinitely many boundaries at the same position in the period. Grouping blocks keeps the values~$a,e$ unchanged; repeating one grouped block yields the original ultimately periodic word. The evaluation axiom and invariance under "unfold-equivalence@@ss" therefore identify its value with $\yaeval(ae^\omega)$. Conversely, cutting a lasso into a sufficiently long prefix and idempotent powers of its period gives such a factorisation. Thus, on finite and ultimately periodic words, $\Omega$ tests acceptance of the corresponding closed unary/nullary "systems".

Let~$h\colon\Sigma\to Y$ be the "map of ranked sets" given by $h(a)=\setsyntmorph{\atomic(a)}_L$, the syntactic value of the input letter~$a$. We define the transition relation $\Delta$ of the automaton $\aut$ as follows. 
\[
\begin{split}
\Delta_+(a)&=\{(\operatorname{dir}_1(\delta),\dots,\operatorname{dir}_k(\delta)):\\
&\hspace{3em}\delta\in Y_k\text{ is an "extremal element" and }\delta\yaleq h(a)\}
\quad(k=\rk(a)>0),\\
\Delta_0(a)&=\{\delta\in Y_0:\delta\text{ is an "extremal element" and }\delta\yaleq h(a)\}.
\end{split}
\]
Each transition is witnessed by an "extremal element" below the syntactic value of the input letter. Its "direction projections" label the outgoing edges, and nullary "extremal elements" label the leaves. For every $a$, the set $\Delta(a)$ is finite, as a subset of $Y_1^k$ or $Y_0$.

\begin{lemma}\label{lemma:extremal-aut-completeness}
If a closed "regular-tree"~$r$ over~$\Sigma$ belongs to~$L$, then~$\aut$ has an accepting run on every representative~$S\in r$.
\end{lemma}
\begin{proof}
Fix a representative "$\Sigma$-system"~$S\in r$. Its relabelling~$\slift h(S)$ by evaluations of letters in $Y$ is accepted. By \Cref{cor:extremal-replacements}, choose "extremal elements" $\delta_v\yaleq h(S(v))$ simultaneously so that the relabelled system~$D=S[v/\delta_v]_{v\in\V_S}$ is accepted. For every non-leaf~$v$, replace~$\delta_v$ by
\[
\ssSum_{i=1}^{\rk(S(v))}\operatorname{dir}_i(\delta_v)(x_i),
\]
and apply "flattening", retaining the nullary labels at leaves. The result is a closed "$\Yw$-set-system"~$G$. By \Cref{lemma:extremal-directions} and the evaluation axiom, $\yaeval(G)=\yaeval(D)$, so $\Acc(G)$.

We now describe a run $\rho$ of~$\aut$ on~$S$. For each non-leaf vertex~$v$ with outgoing edges $e_1,\dots,e_k$, choose $\rho(e_i)=\operatorname{dir}_i(\delta_v)$ for $i=1,\dots,k$. For $v$ a leaf, we choose $\rho(v)=\delta_v$.

Every "regular branch" of~$S$, labelled by~$\rho$, gives a "regular branch"~$B$ of~$G$. Its "morphism@@ss" $B\ssmorphism G$ and upward closure of~$F$ give $\Acc(B)$, so its label word belongs to~$\Omega$. By \Cref{lemma:unfold-aut-regular-branches}, the run~$\rho$ is accepting.
\end{proof}

\begin{lemma}\label{lemma:extremal-aut-soundness}
If a closed "regular-tree"~$r$ over~$\Sigma$ is accepted by~$\aut$, then $r\in L$.
\end{lemma}
\begin{proof}
By \Cref{lemma:unfold-aut-regular-branches}, choose a run~$\rho$ on a representative "$\Sigma$-system"~$S\in r$ such that all "regular branches" are accepting. At each vertex choose an "extremal element"~$\delta_v\yaleq h(S(v))$ witnessing its transition, and let~$D=S[v/\delta_v]_{v\in\V_S}$. Replace each non-leaf label of~$D$ by the disjoint sum of its "direction projections" and apply "flattening", keeping~$\delta_v$ at leaves. The resulting closed "$\Yw$-set-system"~$G$ has no "root vertices". Every vertex is reachable from an "initial vertex", and every unary vertex has at least one outgoing edge, because~$S$ is a "system". Its unary labels are "extremal elements" by \Cref{lemma:extremal-direction-is-extremal}.

The "regular branches" of~$G$ represent exactly the "regular branches" of~$S$ labelled by~$\rho$. They are accepted by the definition of~$\Omega$, so \Cref{lemma:extremal-path-criterion} gives $\Acc(G)$. By \Cref{lemma:extremal-directions} and the evaluation axiom, $\yaeval(D)=\yaeval(G)\in F$. Finally, $\delta_v\yaleq h(S(v))$ and monotonicity (\Cref{lemma:context-evaluation-monotone}) give $\yaeval(D)\yaleq\yaeval(\slift h(S))$. Upward closure of~$F$ yields $\Acc(\slift h(S))$, and \Cref{lemma:synt-recognition} gives $r\in L$.
\end{proof}

Together, \Cref{lemma:extremal-aut-completeness,lemma:extremal-aut-soundness} prove \Cref{thm:alg2aut}.

\begin{remark}\label{remark:run-unfold}
	An arbitrary "unfold-automaton" may need to "unfold" an input "system" to be able to label it with an accepting run. However, the "unfold-automaton" constructed in \Cref{thm:alg2aut} has an accepting run on every representative of each accepted "regular-tree". This can be thought of as a normal form for "unfold-automata", made possible by the flexibility of their accepting condition. Intuitively, for "unfold-automata" in this normal form, an accepting run need not be more complicated than the input regular tree. For an example of an "unfold-automaton" that does not have this property, consider $H_1=\{a,b\}$ and $\Omega=(ab)^\omega$. Then, the "system" consisting of a single-vertex self-loop cannot be directly accepted, but its "unfolding" consisting of a $2$-cycle can.
\end{remark}


\subsection{The bisimulation-invariant case}
\label{subsection:bisimulation-invariant-case}

We now work over~$\alphabetTS$ and show that "bisimulation-invariance" turns algebraic recognition into $\mu$-calculus definability.
\begin{theorem}\label{thm:alg2mu}
Let~$L$ be a "bisimulation-invariant" "language of regular-trees" recognised by a "rankwise-finite" "algebra". Then~$L$ can be defined by a $\mu$-calculus formula.
\end{theorem}

Fix~$L$ as in the theorem. We first put the "unfold-automaton" into a form that allows directions to be duplicated. Write $\vec b^{\,m}=(b_1^m,\dots,b_n^m)$, where~$b_i^m$ denotes~$m$ consecutive copies of~$b_i$. An "unfold-automaton"~$\aut=(H,\Sigma,\Delta,\Omega)$ is a ""clone-automaton"" if
\[
\vec b\in\Delta(\nu_n)\quad\Longrightarrow\quad
\vec b^{\,m}\in\Delta(\nu_{nm})
\qquad(n,m\geqslant1).
\]

We will call ""clone"" of a transition $\vec b$ any transition of the form~$\vec b^{\,m}$ for some~$m\geqslant1$. 
\begin{lemma}\label{lemma:bisim-aut}
There exists a "clone-automaton" recognising~$L$. Moreover, it has an accepting run on every representative of each "regular-tree" in~$L$.
\end{lemma}
\begin{proof}
Let~$\aut=(H,\Sigma,\Delta,\Omega)$ be the "unfold-automaton" constructed in \Cref{thm:alg2aut} from the "syntactic set-algebra"~$\YAlg$ of~$L$. We define the following auxiliary notations:
\[
\operatorname{compr}_k(\vec b)
=\left(\bigmeet_{j=1}^{k}b_{(i-1)k+j}\right)_{i=1}^{n}
\qquad\bigl(\vec b\in Y_1^{nk}\bigr).\]
\[
\begin{aligned}
\alpha(\nu,n,k)&=\{\operatorname{compr}_k(\vec b):\vec b\in\Delta(\nu_{nk})\},\\
\beta(\nu,n)&=\{\vec c\in Y_1^n:\vec c\in\alpha(\nu,n,k)\text{ for infinitely many }k\}.
\end{aligned}
\]

Intuitively, $\alpha(\nu,n,k)$ is the set of ``compressed'' transitions of~$\aut$ at~$\nu_{nk}$, and $\beta(\nu,n)$ is the set of ``eventual'' compressed transitions. 

We define the wanted "clone-automaton"~$\aut'=(H,\Sigma,\Delta',\Omega)$, with
\[
\Delta'(\nu_N)=\bigcup_{N=nm}\{\vec c^{\,m}:\vec c\in\beta(\nu,n)\}
\quad(N>0),
\qquad \Delta'_0=\Delta_0.
\]

The idea is that every transition of~$\aut'$ is a "clone" of an eventual compressed transition of~$\aut$, so we can recover a behavior that was already in $\aut$, provided that the order of directions can be permuted. This condition will be guaranteed by the assumption that $L$ is "bisimulation-invariant".

Let $\rho'$ be an accepting run of~$\aut'$ on some~$S\in r$. Let~$v$ be a non-leaf vertex of $S$, and $\vec c^{\,m}$ be the transition of $\rho'$ at $ v$, with~$\vec c\in\beta(\nu,n)$. There exists~$k\geqslant m$ and~$\vec b\in\Delta(\nu_{nk})$ with~$\operatorname{compr}_k(\vec b)=\vec c$. We build a new "system" $T$ by relabelling~$v$ with~$\nu_{nk}$, and sending its new direction~$(i-1)k+j$ to its old successor in direction~$(i-1)m+\min(j,m)$. This reroutes directions in order to use a group of $k$ transitions for each group of $m$ successors, overloading the last successor of each group when needed. We then build a run $\rho$ as well, by using~ transition $b_i$ for direction $i$. The leaves are left unchanged. The following example has~$n=3$, $m=2$, and~$k=3$:
\begin{center}
\scalebox{.9}{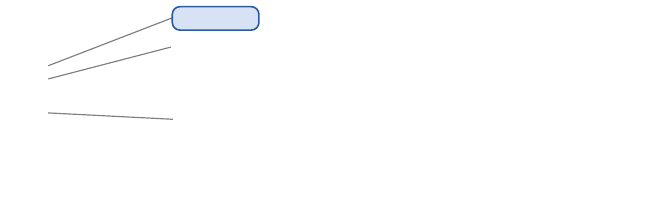}
\end{center}

\underline{$L(\aut')\subseteq L$:}

By performing these replacements simultaneously, we obtain a "system"~$T$ and a run~$\rho$ of~$\aut$ on $T$. Every "regular branch" of~$T$ projects to a "regular branch" of~$S$ with pointwise smaller labels, since~$c_i\yaleq b_{(i-1)k+j}$. E.g. in the above example, $c_1=b_1\meet b_2\meet b_3$. Monotonicity (\Cref{lemma:context-evaluation-monotone}) and upward closure of~$F$ therefore give acceptance of the new chain or lasso, so its label word belongs to~$\Omega$. By \Cref{lemma:unfold-aut-regular-branches}, the run~$\rho$ is accepting. The replacements preserve the decoded successor relation, so~$S$ and~$T$ are "bisimilar". Soundness of~$\aut$ and "bisimulation-invariance" of~$L$ give~$r\in L$.
\smallskip

\underline{$L\subseteq L(\aut')$:}

Let~$r\in L$ and $S\in r$ be a representative "$\alphabetTS$-system". For every~$\nu_n$ occurring at a non-leaf of~$S$, all sufficiently large~$k$ satisfy~$\alpha(\nu,n,k)\subseteq\beta(\nu,n)$: each tuple outside~$\beta(\nu,n)$ occurs for only finitely many~$k$, and~$Y_1^n$ is finite. Let us fix one~$k$ such that for all $\nu_n$ appearing in $S$, we have $\alpha(\nu,n,k)\subseteq\beta(\nu,n)$.

Let~$S^{(k)}$ be obtained by duplicating each edge of~$S$ into~$k$ consecutive edges with the same target. Each label $\nu_n$ is therefore changed to $\nu_{nk}$. Its decoding equals that of~$S$, so its "regular-tree" belongs to~$L$. By \Cref{lemma:extremal-aut-completeness}, there is an accepting run $\rho$ of $\aut$ on~$S^{(k)}$. Let~$b_{v,i,j}$ be the transition of $\rho$ on copy~$j$ of direction~$i$ at~$v$, and~$\ell_v$ for the label of a leaf. We build a run $\rho'$ of $\aut'$ on~$S$, by labelling direction~$i$ at~$v$ by
\[
c_{v,i}=\bigmeet_{j=1}^{k}b_{v,i,j},
\]
and keep~$\ell_v$ at leaves. By the choice of~$k$, each non-leaf transition lies in~$\beta(\nu,n)\subseteq\Delta'(\nu_n)$.

To check acceptance, replace each non-leaf~$v$ of~$S$ by~$\ssSum_i c_{v,i}(x_i)$, keep~$\ell_v$ at leaves, and flatten to obtain~$G$, the "$\Yw$-set-system" corresponding to the run~$\rho'$ on $S$, as in the proofs of Lemmas \ref{lemma:extremal-aut-completeness} and \ref{lemma:extremal-aut-soundness}. Expanding each~$c_{v,i}$ into the disjoint sum of the~$b_{v,i,j}$ recovers~$G_k$, the "$\Yw$-set-system" corresponding to $S^{(k)}$, because all copies of an edge have the same target.
Every unary label~$b_{v,i,j}$ of~$G_k$ is extremal by \Cref{lemma:extremal-direction-is-extremal}. Its "regular branches" correspond to those of~$S^{(k)}$ labelled by~$\rho$, and are therefore accepted. Moreover, $G_k$ has no "root vertices", every vertex is reachable from an "initial vertex", and every unary vertex has at least one outgoing edge. Thus \Cref{lemma:extremal-path-criterion} gives~$\Acc(G_k)$.
The evaluation axiom gives
\[
\yaeval(G)=\yaeval(G_k)\in F.
\]
Every "regular branch"~$B$ of~$G$ has a "morphism@@ss"~$B\ssmorphism G$, so~$\yaeval(G)\yaleq\yaeval(B)$ and~$\Acc(B)$. Thus, every "regular branch" of the compressed run~$\rho'$ satisfies~$\Omega$. By \Cref{lemma:unfold-aut-regular-branches}, $\rho'$ is an accepting run of~$\aut'$ on~$S$, thus $S\in L(\aut')$.
\end{proof}
\medskip

\noindent\textbf{From "clone-automata" to $\mu$-calculus formulas.}

Let~$\aut=(H,\Sigma,\Delta,\Omega)$ be the "clone-automaton" just constructed. We translate it into an automaton of \cite[Theorem~7]{JaninW96}, working directly on finite "transition systems".
For~$\nu\in2^\Pred$, put
\[
B_+(\nu)=\{\{b_1,\dots,b_n\}:(b_1,\dots,b_n)\in\Delta(\nu_n),\ n>0\}.
\]
This set is finite, since it is a subset of~$\powerset(H_1)$.
Since $\Omega$ is $\omega$-regular, there is a deterministic parity automaton~$\D$ over alphabet~$H$ with finite state set~$Q$, initial state~$q_0$, transition function~$\delta_\D$, a parity condition~$\Omega_\D$ recognising~$\Omega\cap H_1^\omega$. It also has final states~$F_\D$ recognising~$\Omega\cap H_1^*H_0$ among finite words.

For~$B=\{b_1,\dots,b_n\}\in B_+(\nu)$ and~$p_i=\delta_\D(q,b_i)$, we define the formula:
\[
\theta_{q,B}=\exists x_1,\dots,x_n.
\bigl(p_1(x_1)\land\dots\land p_n(x_n)
\land\forall z.(p_1(z)\lor\dots\lor p_n(z))\bigr).
\]
The modal automaton~$\C$ has state set~$Q$, initial state~$q_0$, parity condition~$\Omega_\D$, and transitions
\[
\delta(q,\nu)=\bigvee_{B\in B_+(\nu)}\theta_{q,B}
\ \lor\!
\bigvee_{\substack{b_0\in\Delta_0(\nu_0)\\\delta_\D(q,b_0)\in F_\D}}
\forall z.\bot.
\]
In the notation of \cite{JaninW96}, take~$\Sigma_p=\Pred$ and~$\Sigma_r=\{*\}$.

Its acceptance game \cite[Definition~6]{JaninW96} starts at~$(q_0,i)$ on a "transition system" with initial vertex~$i$. At~$(q,s)$ with valuation~$\nu$, Player~I chooses a marking~$m:Q\to\powerset(\mathit{succ}(s))$ satisfying~$\delta(q,\nu)$. Player~II chooses~$(p,t)$ with~$t\in m(p)$, which becomes the next position. A player unable to move loses; infinite plays are decided by~$\Omega_\D$. The automaton accepts if Player~I has a winning strategy.

\begin{lemma}\label{lemma:modal-automaton-translation}
For every closed "$\alphabetTS$-system"~$S\in r$, the automaton~$\C$ accepts~$\tsdecode(S)$ if and only if~$r\in L$.
\end{lemma}
\begin{proof}
Suppose~$\C$ accepts~$\tsdecode(S)$. Fix a positional winning strategy for Player~I, by positional determinacy of parity games \cite{EmersonJutla91}. We construct a "system"~$S'$ with initial vertex~$(q_0,i_S)$, keeping only positions reached by the edges described below.

At a reached non-leaf position~$(q,s)$ with valuation~$\nu$, let~$m$ be the chosen marking and let~$B\in B_+(\nu)$ witness a satisfied disjunct. Let~$\vec c=(c_1,\dots,c_k)\in\Delta(\nu_k)$ with~$\{c_1,\dots,c_k\}=B$. We define~$N=|\mathit{succ}(s)|$ and~$p_j=\delta_\D(q,c_j)$. From the formula $\theta_{q,B}$, we know that each~$m(p_j)$ is nonempty, and by the rules of the game it has at most~$N$ elements. We label~$(q,s)$ by~$\nu_{kN}$ and choose ~$\vec c^{\,N}\in\Delta(\nu_{kN})$ as transition of $\aut$ on this vertex. We route its $j$th block of~$N$ directions onto $\{(p_j,t):t\in m(p_j)\}$, using every target at least once. See \Cref{fig:marking-to-clone-run} for an example of this construction.

\begin{figure}[H]
  \centering
\begingroup
\definecolor{markingblue}{HTML}{2C5AA0}
\definecolor{markingbrown}{HTML}{9B5C24}
\begin{tikzpicture}[
  x=1cm,y=1cm,
  every node/.style={font=\small},
  source/.style={circle,draw=black!75,fill=black!5,
    minimum size=8mm,inner sep=1pt,line width=.6pt},
  target/.style={draw=black!65,fill=black!3,rounded corners=1.5pt,
    minimum width=12mm,minimum height=6mm,inner sep=2pt,line width=.6pt},
  wire/.style={-{Stealth[length=1.8mm]},line width=.65pt},
  first/.style={wire,draw=markingblue},
  second/.style={wire,draw=markingbrown},
  annotation/.style={fill=white,inner sep=1.5pt,font=\scriptsize}
]
  \node at (1.2,3) {$S$: marking at $s$};
  \node[source,label=above:{$\nu_3$}] (s) at (0,0) {$s$};
  \node[target] (t1) at (2.4,1.8) {$t_1$};
  \node[target] (t2) at (2.4,0) {$t_2$};
  \node[target] (t3) at (2.4,-1.8) {$t_3$};
  \foreach \i in {1,2,3} {
    \draw[wire,draw=black!65] (s) -- (t\i);
  }
  \node[align=center] at (1.2,-2.8) {
    $m(p_1)=\{t_1,t_2\}$\\[2pt]
    $m(p_2)=\{t_2,t_3\}$
  };

  \draw[wire,draw=black!65] (3.6,0) -- (4.9,0);
  \node[font=\scriptsize] at (4.25,.45) {$k=2,\ N=3$};

  \node at (8,3) {$S'$: cloned transition};
  \node[source,label=above:{$\nu_6$}] (qs) at (5.8,0) {$(q,s)$};
  \node[target,draw=markingblue,fill=markingblue!6]
    (p1t1) at (10,2.1) {$(p_1,t_1)$};
  \node[target,draw=markingblue,fill=markingblue!6]
    (p1t2) at (10,.7) {$(p_1,t_2)$};
  \node[target,draw=markingbrown,fill=markingbrown!6]
    (p2t2) at (10,-.7) {$(p_2,t_2)$};
  \node[target,draw=markingbrown,fill=markingbrown!6]
    (p2t3) at (10,-2.1) {$(p_2,t_3)$};

  \draw[first] (qs) -- node[annotation,pos=.55,above,sloped]
    {$1:c_1$} (p1t1);
  \draw[first] (qs) to[bend left=8]
    node[annotation,pos=.8,above,sloped] {$2:c_1$} (p1t2);
  \draw[first] (qs) to[bend right=8]
    node[annotation,pos=.85,below,sloped] {$3:c_1$} (p1t2);
  \draw[second] (qs) to[bend left=8]
    node[annotation,pos=.85,above,sloped] {$4:c_2$} (p2t2);
  \draw[second] (qs) to[bend right=8]
    node[annotation,pos=.8,below,sloped] {$5:c_2$} (p2t2);
  \draw[second] (qs) -- node[annotation,pos=.55,below,sloped]
    {$6:c_2$} (p2t3);
  \node at (8,-2.8) {$\vec c^{\,3}=(c_1,c_1,c_1,c_2,c_2,c_2)$};
\end{tikzpicture}
\endgroup
  \caption{From a marking to a cloned transition, with $k=2$, $N=3$, and $p_j=\delta_\D(q,c_j)$.}
  \label{fig:marking-to-clone-run}
\end{figure}

At a reached leaf position with valuation~$\nu$, the satisfied leaf disjunct supplies~$b_0\in\Delta_0(\nu_0)$ with~$\delta_\D(q,b_0)\in F_\D$. We keep the vertex label~$\nu_0$ and run label~$b_0$ in the new "system" $S'$. 

The projection~$(q,s)\mapsto s$ from $S'$ to $S$ preserves valuations and sends every constructed edge to an original edge. The universal clause in~$\theta_{q,B}$ ensures that every original successor is covered. Hence this projection witnesses a usual "bisimulation" between the decodings, and~$S'$ is "bisimilar" to~$S$ by \Cref{lemma:bisimulation-is-bisimulation}.

Every branch of~$S'$ follows the winning strategy, and its edge labels update~$\D$ to the recorded states. Infinite branches satisfy~$\Omega_\D$; finite branches satisfy~$F_\D$ after reading their final label~$b_0$. Thus the run of~$\aut$ on~$S'$ is accepting. Since~$\aut$ recognises~$L$, and~$L$ is "bisimulation-invariant", we obtain~$r\in L$.

Conversely, suppose~$r\in L$. By \Cref{lemma:bisim-aut}, choose an accepting run~$\rho$ of~$\aut$ on~$S$ itself. At a non-leaf position~$(q,s)$, use the marking
\[
m(p)=\{t:\text{some edge }e\text{ from }s\text{ to }t
\text{ satisfies }\delta_\D(q,\rho(e))=p\}.
\]
It satisfies the disjunct determined by the support of the transition at~$s$. Every response can be realised by an edge of~$S$ with the corresponding update of~$\D$. Thus each play follows a branch labelled by~$\rho$, and infinite plays satisfy~$\Omega_\D$. At a reached leaf~$s$, acceptance of that finite branch gives~$\delta_\D(q,\rho(s))\in F_\D$. The empty marking is therefore legal and leaves Player~II unable to move. This is a winning strategy, so~$\C$ accepts~$\tsdecode(S)$.
\end{proof}

By \cite[Theorem~7]{JaninW96}, the automaton~$\C$ is equivalent to a $\mu$-calculus formula. We obtain \Cref{thm:alg2mu}, and together with \Cref{lemma:MSO-is-recognisable}, we obtain our main result of equivalence between bisimulation-invariant MSO and $\mu$-calculus for finite transition systems (\Cref{theorem:bisim-invariant-finite}).

\subsection*{Comments on effectivity}

The well-known upward direction of \Cref{theorem:jw}, and \Cref{theorem:bisim-invariant-finite}, which consists in translating a "$\mu$-calculus sentence" into an equivalent "MSO-sentence" is purely syntactic, and effective. 

For the other direction, let us first note that, given an "MSO-sentence", it is undecidable whether it is "bisimulation-invariant" (over all "transition systems", or over finite "transition systems"), as observed in~\cite{CiardelliOtto20} already for First-Order Logic.

Consequently, the algorithm that we provide has the following characteristics:
\begin{description}
\item[Input] An "MSO-sentence" $\Phi$.
\item[Output] A $\mu$-calculus sentence $\Psi$ such that, if $\Phi$ is "bisimulation-invariant" over finite "transition systems", then $\Psi$ and $\Phi$ coincide on all finite "transition systems".
\end{description}

\begin{proof}[Proof (Sketch)]

Let $\Phi$ be the input "MSO" formula.
Let $d=|\Phi|$ be the size of~$\Phi$.
Let us assume $\Phi$ to be "bisimulation-invariant" over finite "transition systems".
Then the property it defines is "recognised" by a "rankwise-finite" "algebra" by \Cref{lemma:MSO-is-recognisable}.
The induction in that proof uses finite products and rankwise powersets (\Cref{statement:recognition-closure}), giving a computable bound~$N(d,k)$ on the size of each sort~$A_k$. Passing to the ordinary "syntactic algebra" cannot increase these sizes. Counting the "small-profiles" in \Cref{lemma:algebra-to-set-algebra} then gives a computable bound~$M(d)$ on both~$|Y_0|$ and~$|Y_1|$ for the "syntactic set-algebra" used in \Cref{thm:alg2aut}.

The acceptance condition~$\Omega$ constructed there is determined by the finite multiplication table on~$Y_1$, its action on~$Y_0$, the accepting set~$F$, and the values~$\yaeval(ae^\omega)$ for~$a,e\in Y_1$. Hence the size of a deterministic parity word automaton with final states recognising~$\Omega$ has a computable bound in terms of~$M(d)$. For each valuation~$\nu$, the modal automaton~$\C$ has at most~$2^{|Y_1|}$ non-leaf disjuncts, each mentioning at most~$|Y_1|$ states, and at most~$|Y_0|$ leaf disjuncts. Thus its size is effectively bounded, as is the size~$N'(d)$ of the equivalent $\mu$-calculus formula obtained by \cite[Theorem~7]{JaninW96}. Although~$\beta$ is defined using infinitely many~$k$, only the finite sets of supports~$B_+(\nu)$ enter~$\C$; this size bound does not require computing~$\beta$.

Pushing further, by the effective finite model property of the $\mu$-calculus \cite[Thm 15]{BradfieldWalukiewicz2018}, there exists an effective bound $N''(d)$ (exponential in $N'(d)$) such that two $\mu$-calculus sentences of size at most~$N'(d)$ are equivalent if and only if they are equivalent over all "transition systems" of size at most~$N''(d)$.
Hence, a possible algorithm is as follows: try all $\mu$-calculus sentences of size at most~$N'(d)$, and for each of them, check whether it is equivalent to $\Phi$ on all "transition systems" of size at most~$N''(d)$. If yes, choose $\Psi$ to be this $\mu$-calculus sentence. If no such $\mu$-calculus sentence is discovered, output any sentence.
Under the bisimulation-invariance assumption, some candidate~$\Psi_0$ of size at most~$N'(d)$ agrees with~$\Phi$ on all finite "transition systems". Consequently, any selected~$\Psi$ agrees with~$\Psi_0$ on all systems of size at most~$N''(d)$, and hence everywhere by choice of~$N''(d)$. This proves the promised correctness of the algorithm.
\end{proof}

\section*{Conclusion}
\label{section:conclusion}

Our notion of  "systems" is equivalent to notions of tree "algebras" developed elsewhere in the literature.
The notion of "set-systems" is a kind of extended powerset construction on it, expanded with "root vertices", something that is new to the best of our knowledge, though close to more classical powerset constructions. One originality here is mainly that all operations are performed while keeping the structures "folded", and only finitely "unfolding" them when necessary.
More crucial is the argument that transforms "algebras" to "set-algebras", which circumvents the impossibility to have the powerset distribute over the monad of an algebraic theory that has non-linear identities. We are studying further this categorical aspect.
It would be interesting to lift our result to the counting setting. It is known \cite[Thm 4.2.1.3]{JaninHDR} that an MSO-definable property that is invariant under counting "bisimulation" can always be defined in the counting $\mu$-calculus\footnote{also known as graded $\mu$-calculus}. Our proof is not directly adaptable to this setting: one would need additional arguments yielding a counting bound, in the spirit of the "small-profiles" from \Cref{subsection:small-profile}.
Another interesting direction is to see to which extent an algebraic approach can be used to bypass automata for deciding the MSO-theory of infinite trees, i.e.\ a purely algebraic proof of Rabin's theorem.
\smallskip

\textbf{Acknowledgements.}
We thank Achim Blumensath for helpful discussions, proofreading this work, and finding a major mistake in a previous version of this paper (now solved). We also thank Sebastian Enqvist and Yde Venema for helping improve the presentation of a proof, and Veeti Ahvonen for pointing us towards \cite[Thm 4.2.1.3]{JaninHDR}, the counting variant of \Cref{theorem:jw}.

\bibliography{bibliobisimmso}
\end{document}